\documentclass[a4paper,11pt]{article}
\pdfoutput=1 % if your are submitting a pdflatex (i.e. if you have
\usepackage{jheppub} % for details on the use of the package, please
\usepackage[T1]{fontenc} % if needed

\usepackage{amssymb}

\title{\boldmath $t\bar{t}b\bar{b}$ at the LHC:
 On the size of corrections and $b$-jet definitions}

\author[a]{Giuseppe Bevilacqua,}
\author[b]{Huan-Yu  Bi,}
\author[c]{Heribertus Bayu Hartanto,}
\author[d]{Manfred  Kraus,}
\author[b]{Michele Lupattelli}
\author[b]{and Malgorzata  Worek}

\affiliation[a]{MTA-DE Particle Physics Research Group, University of
  Debrecen, H-4010 Debrecen, \\PBox 105, Hungary}
\affiliation[b]{ Institute for Theoretical Particle Physics
and Cosmology, RWTH Aachen University, \\D-52056 Aachen, Germany}
\affiliation[c]{Cavendish Laboratory, University of Cambridge,
J.J. Thomson Avenue, Cambridge CB3 0HE, United Kingdom} 
\affiliation[d]{Physics Department, Florida State University,
Tallahassee, FL 32306-4350, USA}

\emailAdd{giuseppe.bevilacqua@science.unideb.hu}
\emailAdd{bihy@physik.rwth-aachen.de}
 \emailAdd{hbhartanto@hep.phy.cam.ac.uk}
 \emailAdd{mkraus@hep.fsu.edu}
 \emailAdd{lupattelli@physik.rwth-aachen.de}
 \emailAdd{worek@physik.rwth-aachen.de}

 \abstract{
We report on the calculation of the next-to-leading order QCD
corrections to the production of a $t\bar{t}$ pair in association with
two heavy-flavour jets. We concentrate on the di-lepton $t\bar{t}$ decay
channel at the LHC with $\sqrt{s}=13$ TeV.  The computation is based
on  $pp \to e^+ \nu_e\, \mu^-\bar{\nu}_\mu\, b\bar{b} \,b\bar{b}$
matrix elements and includes all resonant and non-resonant diagrams,
interferences and off-shell effects of the top quark and the $W$ gauge
boson. As it is customary for such studies, results are presented in
the form of inclusive and differential fiducial cross sections. We
extensively investigate the dependence of our results upon variation
of renormalisation and factorisation scales and parton distribution
functions in the quest for an accurate estimate of the theoretical
uncertainties. We additionally study the impact of the contributions
induced by the bottom-quark parton density.  Results presented here
are particularly relevant for measurements of $t\bar{t}H(H\to
b\bar{b})$ and the determination of the Higgs coupling to the top
quark. In addition, they might be used  for precise measurements of
the top-quark fiducial cross sections and to investigate top-quark
decay modelling at the LHC. }

\dedicated{\rm TTK-21-16,  P3H-21-029, CAVENDISH-HEP-21/08}


\begin{document} 
\maketitle
\flushbottom

% =============================================
%
\section{Introduction}
\label{sec:introduction}
%
% =============================================

\begin{figure}[t!]
  \begin{center}
  \includegraphics[trim=0 737 0 0, width=\textwidth]{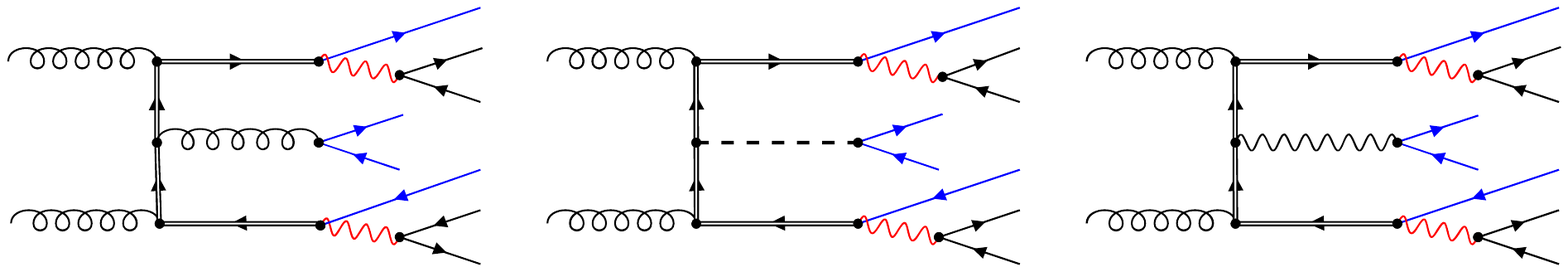}
  \end{center}
\caption{\label{fig:fd-ttbb} \it Representative tree level Feynman
diagrams for the QCD $pp\to t\bar{t}b\bar{b}$ production process as
well as for 
$pp\to t\bar{t}H(H\to b\bar{b})$ and $pp \to t\bar{t}Z (Z\to
b\bar{b})$ that lead to the same $e^+ \nu_e \, \mu^- \bar{\nu}_\mu \,
b\bar{b}\, b\bar{b}$ final state respectively at ${\cal
O}(\alpha_s^4\alpha^4)$ and ${\cal O}(\alpha_s^2\alpha^6)$.  The
double line indicates the top (anti-top) quark, the blue line
corresponds to bottom (anti-bottom) quarks, whereas the $W$ gauge
boson is depicted in red. Furthermore, the dashed line represents the
SM Higgs boson.}
\end{figure}

The discovery of the Higgs boson at the LHC was only the start of the
wide program, the main purpose of which is to identify the properties
and couplings of this recently discovered particle. In the Standard
Model (SM) the Higgs boson couples to the fundamental fermions
via the Yukawa interaction with a coupling strength that is
proportional to the fermion mass. Probing the coupling of the Higgs
boson to the top quark, the heaviest observed particle in the SM,
comprises a crucial test of the consistency of the Higgs
sector. Furthermore, the Top-Yukawa coupling, denoted as $Y_t$,
might be used to constrain various models of physics beyond the SM
(BSM) that very often predict a different coupling strength than the
SM one. The latter is expected to be close to unity.  Indirect
constraints on the coupling between the top quark and the Higgs boson
are available from processes including virtual top quark loops. Here
the best example comprises Higgs boson production through $gg$
fusion. On the other hand, the $Y_t$ coupling can be probed directly
in the associated production of the Higgs boson with the $t\bar{t}$
pair, the process which has recently been observed by both the ATLAS
and CMS collaborations \cite{Sirunyan:2018hoz,Aaboud:2018urx}.
Even though $pp\to t\bar{t}H$ contributes only about $1\%$ to the
total $pp \to H$ production cross section, it offers a very
distinctive signature. For the Higgs boson with the observed mass
value the dominant decay mode is $H\to b\bar{b}$ with the branching
ratio of $58\%$ \cite{deFlorian:2016spz}.  This decay mode is
additionally sensitive to the coupling of the Higgs boson to the
bottom quark but it is not easily accessible experimentally.
Nevertheless, both ATLAS and CMS reported searches for $t\bar{t}H$
production in the $b\bar{b}$ decay channel of the Higgs boson
\cite{Sirunyan:2018mvw,Aaboud:2017rss}. The main experimental
challenge for this channel is the correct identification of the
candidates for the Higgs boson decay from the so-called combinatorial
background.  The latter is responsible for a substantial smearing of
the Higgs boson peak in the $b\bar{b}$ invariant mass spectrum. Further
challenges include the possibility of misidentification of light jets
with $b$-jets and problems with the control of various SM backgrounds,
see Figure \ref{fig:fd-ttbb} for examples of Feynman diagrams for 
$t\bar{t}H(H\to b\bar{b})$ production and processes that lead to the same
$t\bar{t}b\bar{b}$ final state\footnote{All Feynman diagrams in this
paper were produced with the help of the \textsc{FeynGame} program
\cite{Harlander:2020cyh}.}. With the help of $b$-jet tagging
algorithms as well as boosted top quarks and Higgs boson it is
possible to isolate the contribution of the $t\bar{t}H(H\to b\bar{b})$
process from the most general reducible background represented by
$t\bar{t}jj$ production and from the irreducible $Z$-peak background
\cite{Plehn:2009rk}. Nevertheless, the $pp\to t\bar{t}H \to
t\bar{t}b\bar{b}$ process suffers greatly from the
$t\bar{t}b\bar{b}$ background, that is the most important irreducible
background process for this SM Higgs boson channel.  In addition,
searches for four top-quark production $(t\bar{t}t\bar{t})$ are also
affected by the QCD $t\bar{t}b\bar{b}$ background
\cite{Aaboud:2018jsj,Sirunyan:2017tep}. Consequently, measurements of
$t\bar{t}H(H\to b\bar{b})$ and $t\bar{t}t\bar{t}$ at the LHC would
benefit considerably from a better understanding of the QCD
$t\bar{t}b\bar{b}$ production process and from the
improved modelling of top-quark decays.
%
% =============================================
% =============================================

The $pp\to t\bar{t}b\bar{b}$ process is also interesting from the
theoretical point of view. This is due to the presence of two very
different and distinctive scales, the top-quark mass $m_t$ and the
$b$-jet transverse momentum. The former governs $t\bar{t}$ production
and subsequent top-quark decays, the latter describes the two $b$-jets
coming from the $g\to b\bar{b}$ splitting. However, this rather
straightforward picture is not adequate anymore once the contributions
from off-shell top quarks and $W$ gauge bosons are consistently taken
into account. Away from the $t\bar{t}$ threshold and for large values
of $M(b\bar{b})$ various other mechanisms start to play a non
negligible part, that makes the $t\bar{t}b\bar{b}$ process truly
multi-scale production.

Calculations at NLO in QCD are available for  stable top quarks
already for some time \cite{Bredenstein:2008zb,
Bredenstein:2009aj,Bevilacqua:2009zn,Bredenstein:2010rs,Worek:2011rd,
Bevilacqua:2014qfa}. They suffer from large uncertainties in the
choice of factorisation and renormalisation scales that are of the
order of $33\%$ and large NLO corrections of the order of $77\%$.  The
latter is mainly due to the $gg$ initial state. Indeed, NLO QCD
corrections to the subprocess initiated by $q\bar{q}$ annihilation
only are small with the ${\cal K}$-factor of the order of ${\cal K}=
\sigma^{\rm NLO}/\sigma^{\rm LO}\approx 2.5\%$ whereas the NLO
theoretical error due to scale variations is at the $17\%$ level
\cite{Bredenstein:2008zb}.  Unfortunately, the large scale variation
and the size of the corrections themselves for the full $pp$ process
imply that a full NNLO study would be indispensable. As the latter
will remain out of reach in the nearest future, it seems that, as
already suggested in Ref.  \cite{Bredenstein:2010rs}, additional
kinematic restrictions to the $b\bar{b}$ system, e.g.  $M(b\bar{b})
\gtrsim 100$ GeV or $p_T(b\bar{b}) \gtrsim 200$ GeV as motivated by
the studies of the SM Higgs boson with the mass of $m_H=125$ GeV, must
be introduced in order to reduce large theoretical uncertainties and
higher-order QCD corrections.  Alternatively, a veto on extra jet
radiation might be carried out to achieve the same goals. In
Ref. \cite{Buccioni:2019plc} even calculations of NLO QCD corrections
to the $pp\to t\bar{t}b\bar{b}j$ process were presented.  For
integrated NLO cross sections scale uncertainties at the level of
$25\%$ were obtained, whereas, the usual ${\cal K}$-factor was
calculated to be $1.45$.  Having an additional resolved jet present
allowed to investigate in detail the modelling of recoil effects in
$t\bar{t}b\bar{b}$ production. We note here that, generally speaking,
the $pp \to t\bar{t}b\bar{b}j$ process at NLO entails information that
can be used to gain some insights into the perturbative convergence of
the inclusive $t\bar{t}b\bar{b}$ cross section beyond NLO.

Computing higher-order QCD corrections to processes with stable top
quarks, no matter how technically complex they might be, can only give
us a general idea of the size of the NLO corrections.  Such theoretical
predictions cannot provide a reliable description of the top-quark
decays and are not sufficient to detail QCD radiation pattern for this
process. Thus, for more realistic studies radiative top-quark decays
are needed.  The first step in this direction was achieved, for the 
$t\bar{t}b\bar{b}$ production process, by matching the $t\bar{t}b\bar{b}$
matrix elements, calculated with massless or massive $b$ quarks at NLO
in QCD, to parton-shower (PS) programs
\cite{Kardos:2013vxa,Cascioli:2013era,
Garzelli:2014aba,Bevilacqua:2017cru}. In these studies, however,
top-quark decays have been either completely omitted or performed
within the \textsc{Pythia} MC framework. Either way, these predictions
did not include $t\bar{t}$ spin correlations. On the other hand, in
Ref. \cite{Jezo:2018yaf} NLO matrix element calculations matched to a
parton-shower (NLO+PS) simulations of $pp\to t\bar{t}b\bar{b}$ were
presented in the four-flavour scheme for stable top quarks and with
(LO) spin correlated top-quark decays. In general, it was shown that
matching and shower uncertainties were very small for this process
once $m_b\ne 0$ was considered.  Moreover, neither ${\cal K}$-factor
nor the size of scale uncertainties was substantially changed by PS
effects. Actually, the scale dependence uncertainties  even increased
slightly, reaching $40\%$.  The experimental measurements of $pp$
production cross sections for $t\bar{t}b\bar{b}$ have been carried out
by both ATLAS and CMS \cite{Sirunyan:2017snr,Aaboud:2018eki,
Sirunyan:2020kga}.  The measured inclusive fiducial cross sections
generally exceed theoretical predictions for $t\bar{t}b\bar{b}$ as
provided by the already mentioned NLO + PS simulations.  Nevertheless,
they are compatible within the total uncertainties.  Very recently,
the computation of NLO QCD corrections to $t\bar{t}b\bar{b}$
production in the di-lepton top-quark decay channel was reported in
Ref. \cite{Denner:2020orv}.  Specifically, higher order $\alpha_s$
corrections to the $e^+ \nu_e \, \mu^- \bar{\nu}_\mu \, b\bar{b}\,
b\bar{b}$ final state at ${\cal O}(\alpha_s^4\alpha^4)$ were
calculated for the LHC energy of $\sqrt{s}=13$ TeV. In
Ref. \cite{Denner:2020orv} they have shown that at the level of cross
section the NLO QCD corrections to $t\bar{t}b\bar{b}$ were close to
$100\%$. Furthermore, at the differential level on top of this overall
shift, moderate shape distortions up to $25\%$ were obtained.

The goal of the paper is manifold. Due to the complexity of the
process we will present an independent computation of the complete NLO
QCD corrections to the off-shell production of $t\bar{t}b\bar{b}$ in
the di-lepton top-quark decay channel. In the first step, we use the
same scale choice and the SM input parameters as well as phase-space
cuts to confirm the results presented in Ref. \cite{Denner:2020orv} at
the integrated and at the differential level. Our second goal is to
extend this analysis to other  selected renormalisation and
factorisation scale choices and to different PDF sets. By using the
error PDF sets we will additionally study the internal NLO PDF
uncertainties. Furthermore, a stability test of LO and NLO fiducial
cross sections with respect to the $b$-jet transverse momentum cut
will be performed. Afterwards additional cuts will be introduced to
examine their impact on the size of the ${\cal K}$-factor and the
theoretical uncertainties due to scale dependence. Another  goal of
the paper is to investigate the initial state bottom-quark
contributions and their impact on the integrated and differential
$t\bar{t}b\bar{b}$ cross sections. To this end two approaches, the so
called {\it charge-blind} and {\it charge-aware} heavy-flavour jet
tagging, will be introduced.

The paper is organised as follows. In Section \ref{sec:methods} we
briefly describe the \textsc{Helac-Nlo} framework that we use for the
calculation and discuss the cross-checks that have been performed. Our
theoretical setup for LO and NLO QCD results is given in Section
\ref{sec:setup}. Phenomenological results for the integrated and
differential fiducial $t\bar{t}b\bar{b}$ cross sections are discussed
in detail in Section \ref{sec:ttbb-integ} and Section
\ref{sec:ttbb-diff}.  The initial state bottom-quark contribution is
examined in Section \ref{sec:ttbb-initial-b}. The comparison with
theoretical predictions presented in Ref. \cite{Denner:2020orv} is
performed in Section \ref{sec:comparison}. Finally, in Section
\ref{sec:sum} our results for the $t\bar{t}b\bar{b}$ production
process are summarised.

%=============================================
%
\section{Description of the calculation and its validation}
\label{sec:methods}
%
% =============================================

% =============================================
\begin{figure}[t!]
  \begin{center}
  \includegraphics[trim=0 509 0 0, width=\textwidth]{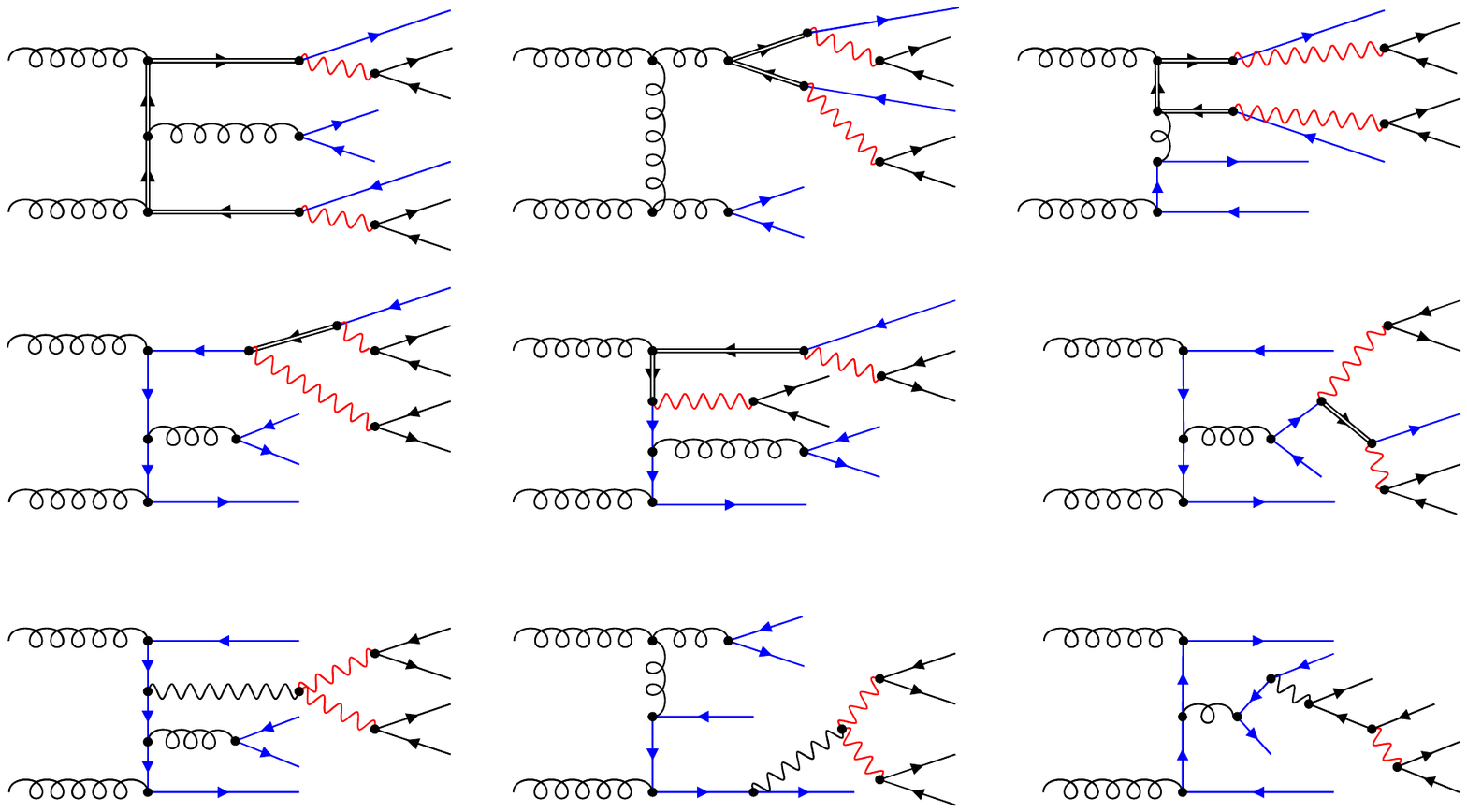}
  \end{center}
\caption{\label{fig:fd} \it Representative tree level Feynman diagrams
for the $pp\to e^+ \nu_e \, \mu^- \bar{\nu}_\mu \, b\bar{b}\, b\bar{b}
+X$ process at ${\cal O}(\alpha_s^4\alpha^4)$. Diagrams with two, only
one and no top-quark resonances are presented. The double line
indicates the top (anti-top) quark, the blue line corresponds to the bottom
(anti-bottom) quark whereas  the $W$ gauge boson is depicted in red. Also
shown is a diagram that contributes to the finite $W$ width
corrections. }
\end{figure}
% =============================================

We consider the fully realistic $e^+ \nu_e\, \mu^- \bar{\nu}_\mu\,
b\bar{b}\, b\bar{b}+X$ final state. We consistently take into account
resonant and non-resonant top-quark contributions and all interference
effects among them. In addition, non-resonant and off-shell effects
due to the finite $W$ gauge boson width are included. Due to their
insignificance we neglect flavor mixing. A few examples of Feynman
diagrams contributing to the leading order process at ${\cal
O}(\alpha_s^4\alpha^4)$ are presented in Figure \ref{fig:fd}. They are
shown for the dominant $gg$ partonic subprocess. NLO QCD corrections
are calculated with the help of the \textsc{Helac-Nlo} Monte Carlo (MC)
program \cite{Bevilacqua:2011xh}. This is the first computation of
a $2\to 6$ process (the decay products of the $W$'s are not
counted, because they do not couple to colour charged states) carried
out within this framework. Even though \textsc{Helac-Nlo} has already
been employed for the calculations of NLO QCD corrections to
$t\bar{t}+X$, $X=j,\gamma,Z(\to \nu\nu),W^\pm(\to \ell \nu)$ with full
top quark off-shell effects included
\cite{Bevilacqua:2015qha,Bevilacqua:2016jfk,Bevilacqua:2018woc,
Bevilacqua:2019cvp,Bevilacqua:2020pzy}, these processes were at most $2
\to 5$ processes from the QCD point of view. For the $gg \to e^+
\nu_e\, \mu^- \bar{\nu}_\mu\, b\bar{b}\, b\bar{b}$ partonic reaction
there are $3904$ LO diagrams. For each $q\bar{q} \to e^+ \nu_e\, \mu^-
\bar{\nu}_\mu\, b\bar{b}\, b\bar{b}$ subprocess, where $q$ stands for
$u\,d,c,s$, we have $930$ LO diagrams. The calculation of the LO scattering
amplitudes is performed within the \textsc{Helac-Dipoles} package
\cite{Czakon:2009ss}.  The phase-space integration is performed and
optimised with the help of \textsc{Parni} \cite{vanHameren:2007pt} and
\textsc{Kaleu} \cite{vanHameren:2010gg}. The produced top quarks are
unstable particles, thus, the inclusion of their decays is performed in
the complex-mass scheme \cite{Denner:1999gp,Denner:2005fg,
Bevilacqua:2010qb,Denner:2012yc}. It fully respects gauge invariance
and is straightforward to apply.  The resonant
electroweak vector bosons are also treated in the complex-mass scheme.
%
% =============================================
\begin{figure}[t!]
  \begin{center}
  \includegraphics[trim= 18 720 0 0, width=\textwidth]{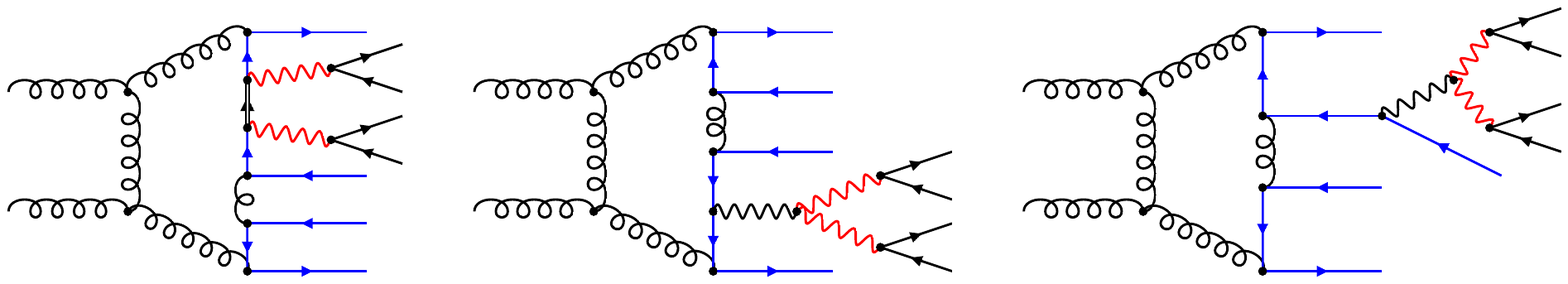}
\end{center}
\vspace{1cm}
  \caption{\label{fig:fd-1loop} \it  Examples of octagon-, heptagon-
and hexagon-type of one-loop diagrams contributing to the $pp\to e^+
\nu_e \, \mu^- \bar{\nu}_\mu \, b\bar{b}\, b\bar{b} +X$ process.  The
double line indicates the top (anti-top) quark, the blue line
corresponds to bottom (anti-bottom) quarks whereas the $W$ gauge boson
is  depicted in red. }
\end{figure}
% =============================================
%

We compute the virtual corrections using \textsc{Helac-1Loop}
\cite{vanHameren:2009dr} and \textsc{CutTools} \cite{Ossola:2007ax}.
Specifically, one-loop amplitudes are generated with
\textsc{Helac-1Loop} and are further reduced at the integrand level
using the so-called OPP method \cite{Ossola:2006us} as implemented in
\textsc{CutTools}.  The most complicated one-loop diagrams in our
calculations are octagons (8-point integrals). In the $gg$ channel
they involve tensor integrals up to rank six. In Table
\ref{tab:one-loop} the number of one-loop Feynman diagrams, that
corresponds to each type of correction for the dominant $gg$ partonic
subprocess, is provided (see examples in Figure
\ref{fig:fd-1loop}). They are obtained with the help of the
\textsc{Qgraf} program \cite{Nogueira:1991ex} as \textsc{Helac-Nlo}
does not employ Feynman diagrams for the amplitude calculations. We
have cross-checked our results with the publicly available
general-purpose MC program \textsc{MadGraph5}${}_{-}$\textsc{aMC@Nlo}
\cite{Alwall:2014hca}. Specifically, we compared numerical values of 
the one-loop virtual corrections for a
few phase-space points for $gg$ and $q\bar{q}$ $(q=u,d,b)$ partonic
subprocesses. Virtual corrections come from the interference of the
tree-level and one-loop amplitudes, summed over all colors and spins,
and for $N_F= 5$ massless quark flavours. We compared the finite parts
along with the coefficients of the poles in $\epsilon$. Additionally,
coefficients for color and spin summed results for the ${\cal
I}$-operator were cross checked between \textsc{Helac-1Loop} and
\textsc{Helac-Dipoles}. We have found perfect agreement in each
case. At the one loop level the appearance of a non-zero top-quark
width in the propagator requires the evaluation of scalar integrals
with complex masses, which is supported by the \textsc{OneLOop}
program \cite{vanHameren:2010cp}, used for the evaluation of the
integrals. The preservation of gauge symmetries by this approach is
explicitly checked by studying the Ward identities up to the one-loop
level. For $gg$ subprocess we perform this test for every phase-space
point. Quadruple precision is used to recompute events which fail
the gauge invariance check. For the $q\bar{q}$ subprocess we use the
so-called scale test \cite{Badger:2010nx}, which is based on momentum
rescaling. Since we know how the recalculated amplitude scales when
the momenta are rescaled, it is possible to compare the two
results. Also in this case the test is performed for each
phase-space point. For the failed points the amplitude is recomputed
using higher precision. In addition, reweighting techniques,
helicity and colour sampling methods are used in order to
optimise the performance of the \textsc{Helac-Nlo} framework.
%
% =============================================
\begin{table}[t!]
\begin{center}
\begin{tabular}{cc}
\hline \hline  \\ [-0.4cm]
  {One-loop correction type}
  &{Number of Feynman diagrams} \\[0.2cm]
  \hline \hline  \\ [-0.4cm]
 {Self-energy}       & 93452   \\ [0.2cm]
 {Vertex}               & 88164   \\ [0.2cm]
 {Box-type}           & 49000   \\ [0.2cm]
 {Pentagon-type}  & 25876   \\ [0.2cm]
 {Hexagon-type}   & 11372   \\[0.2cm]
{Heptagon-type}   & 3328   \\ [0.2cm]
{Octagon-type}     & 336   \\ [0.2cm]
  \hline \hline \\[-0.4cm]
{Total number}      & 271528    \\[0.2cm]
   \hline \hline
 \end{tabular}
\end{center}
\caption{\label{tab:one-loop} \it
The number of one-loop Feynman diagrams for the dominant subprocess
$gg \to e^+\nu_e \, \mu^- \bar{\nu}_\mu \, b\bar{b} \, b\bar{b}$ at
${\cal O}(\alpha_s^5 \alpha^4)$ split by loop topology. The Higgs
boson exchange contributions are not considered and the
Cabibbo-Kobayashi-Maskawa mixing matrix is kept diagonal.}
\end{table}
% =============================================
\begin{table}[h!]
  \begin{center}
    \begin{tabular}{cccc}
      \hline\hline\\[-0.4cm]
      Partonic  &Number of &Number of&Number of\\
      Subprocess &Feynman diagrams &CS Dipoles&NS Subtractions\\[0.2cm]
  \hline\hline \\ [-0.4cm]
  $gg  \to   e^+\nu_e\,
  \mu^-\bar{\nu}_\mu\, b\bar{b} \, b\bar{b}\, g$ & $41 364$ & 90 &
                                                                   18\\[0.2cm]
   $q\bar{q}  \to   e^+\nu_e\,
      \mu^-\bar{\nu}_\mu\, b\bar{b} \, b\bar{b}\, g$ &  $9576$
                           & 50 & 10\\[0.2cm]
  $g q\to   e^+\nu_e\,
      \mu^-\bar{\nu}_\mu\, b\bar{b} \, b\bar{b} \, q$
                &   $9576$& 50 & 10\\[0.2cm]
   $g \bar{q} \to   e^+\nu_e\,
      \mu^-\bar{\nu}_\mu\, b\bar{b} \, b\bar{b}\, \bar{q}$ &
                                                             $9576$& 50& 10\\[0.2cm]
       \hline \hline
\end{tabular}
\end{center}
\caption{\label{tab:real-emission}  \it The list of partonic
  subprocesses contributing to the subtracted real emission at
${\cal O}(\alpha_s^5 \alpha^4)$  for the $pp \to e^+\nu_e \, \mu^-
\bar{\nu}_\mu \, b\bar{b} \, b\bar{b}+X$ process where $q = u, d,
  c, s$. Also shown are the number of Feynman diagrams, as well as the
 number of Catani-Seymour and Nagy-Soper subtraction terms that
  correspond to these partonic subprocesses.}
\end{table}
% =============================================

To compute the real corrections we isolate the singularities from the
soft or the collinear parton emissions via subtraction methods for NLO
QCD calculations. We employ \textsc{Helac-Dipoles}, which implements
the dipole formalism of Catani and Seymour
\cite{Catani:1996vz,Catani:2002hc} for arbitrary helicity eigenstates
and colour configurations of the external partons and the Nagy-Soper
subtraction scheme \cite{Bevilacqua:2013iha}, which makes use of
random polarisation and colour sampling of the external partons.  Two
independent subtraction schemes allow us to cross check the
correctness of the real emission part of the calculation in an even
more robust way. We use a phase-space restriction on the contribution
of the subtraction terms as originally proposed in
Ref. \cite{Nagy:1998bb,Nagy:2003tz,Bevilacqua:2009zn} for the
Catani-Seymour scheme and in Ref. \cite{Czakon:2015cla} for the
Nagy-Soper one. We consider two extreme choices to cross check the
independence of the results on this parameter. All partonic
subprocesses that are taken into account for the real emission
contributions are listed in Table \ref{tab:real-emission}, together
with the number of the corresponding Feynman diagrams, the number of
Catani-Seymour dipoles and Nagy-Soper subtraction terms. In each case,
there are five times fewer terms in the Nagy-Soper subtraction scheme
compared to the Catani-Seymour approach. The difference corresponds to
the total number of possible spectators in the process under scrutiny,
which are relevant only in the Catani-Seymour case.

Our theoretical predictions are stored in the form of modified Les
Houches Event Files \cite{Alwall:2006yp} and ROOT Ntuples
\cite{Antcheva:2009zz}. Inspired by the ideas proposed in
Ref. \cite{Bern:2013zja} each ``event'' is stored with supplementary
matrix element and PDF information. Ntuples contain unweighted
events, that helps to keep the storage as small as possible. With the goal of
optimising the performance of the unweighting procedure, the
so-called partial unweighting is implemented in  the
\textsc{Helac-Nlo} software, see
e.g. \cite{Bevilacqua:2016jfk} for more details.  Ntuples allow us to
obtain theoretical predictions for different scale choices and PDF
sets. Thus, for example the error PDF sets can easily be employed to
calculate the internal NLO PDF uncertainties.  Furthermore, any
infrared-safe (IR-safe) observable can be generated, ranges and bin
sizes can be adjusted while no additional time consuming running of
the code is required.

% =============================================
%
\section{LHC setup}
\label{sec:setup}
%
% =============================================

We consider the $pp\to e^+\nu_e\, \mu^-\bar{\nu}_\mu \, b\bar{b}\,
b\bar{b} +X$ process at NLO in QCD for LHC Run II energy of
$\sqrt{s}=13$ TeV.  Specifically, $\alpha_s$ corrections to the
born-level process at ${\cal O}(\alpha_s^4\alpha^4)$ are
evaluated. Different lepton generations are used to avoid virtual
photon singularities stemming from $\gamma \to \ell^+ \ell^-$. We do
not consider $\tau$ leptons as they are usually studied separately at
the LHC due to their very rich and complex decay pattern.  We keep the
Cabibbo-Kobayashi-Maskawa (CKM) mixing matrix diagonal. We take LO and
NLO NNPDF3.1 PDF sets \cite{Ball:2017nwa} as the default PDF sets. In
both cases they are obtained with $\alpha_s(m_Z)=0.118$. However, we
will show results for other PDF sets, specifically for CT18
\cite{Hou:2019efy} and MMHT14 \cite{Harland-Lang:2014zoa}. The
difference between various PDF sets originate, among others, from the
choice of the data used and the theoretical assumptions made for the
global fit. Consequently, it is desirable to see theoretical
predictions also for different PDF sets. The running of the strong
coupling constant $\alpha_s$ with two-loop (one-loop) accuracy at NLO
(LO) is provided by the LHAPDF interface \cite{Buckley:2014ana} with
$N_F = 5$. The SM input parameters for our calculations are given by
\begin{equation}
\begin{array}{lll}
 G_{ \mu}=1.16638 \cdot 10^{-5} ~{\rm GeV}^{-2}\,, &\quad \quad \quad
                                                     \quad \quad
&   m_{t}=173 ~{\rm GeV} \,,
\vspace{0.2cm}\\
 m_{W}= 80.351972  ~{\rm GeV} \,, &
&\Gamma^{\rm NLO}_{W} =  2.0842989  ~{\rm GeV}\,, 
\vspace{0.2cm}\\
  m_{Z}=91.153481   ~{\rm GeV} \,, &
&\Gamma^{\rm NLO}_{Z} = 2.4942664~{\rm GeV}\,.
\end{array}
\end{equation}
All other partons, including bottom quarks, and leptons are treated as
massless particles. The top quark width is treated as a fixed
parameter throughout this work and its value is evaluated for
$\alpha_s(\mu_R=m_t)$. The top quark width at LO has been
computed based on the formulas from Ref. \cite{Jezabek:1988iv},
whereas the NLO QCD value has been obtained upon applying the relative
QCD corrections given in Ref.  \cite{Basso:2015gca} to the LO
width. For our set of input parameters we have 
\begin{equation}
\begin{array}{lll}
  \Gamma_{t}^{\rm LO} =  1.443303   ~{\rm GeV}\,, &
 \quad \quad \quad \quad\quad &
  \Gamma_{t}^{\rm NLO} =  1.3444367445 ~{\rm GeV}\,.
\end{array}
\end{equation}
Since we treat bottom quarks as massless partons there are no diagrams
with Higgs boson exchange at tree level. The Higgs boson contribution
appears only at NLO through closed fermion loops involving top
quarks. We checked that this contribution amounts to $0.3\%$ of the
total NLO cross section for our setup and is therefore
negligible. Given its very small impact we decided to neglect the
Higgs-boson contribution, which is equivalent to taking the $m_H \to
\infty$ limit. The electromagnetic coupling $\alpha$ is calculated
from the Fermi constant $G_\mu$ ($G_\mu-$scheme) via
\begin{equation}
\alpha_{G_\mu}=\frac{\sqrt{2}}{\pi} \,G_F \,m_W^2  \,\sin^2\theta_W \,,
~~~~~~~~~~~{\rm
  where} ~~
\sin^2\theta = 1-\frac{m_W^2}{m_Z^2}\,,
\end{equation}
where $G_\mu$ is measured in the muon decay. In the $G_\mu-$scheme
electroweak corrections related to the running of $\alpha$ are taken
into account. The evaluation of the residual theoretical uncertainties
due to the not yet calculated higher order corrections is based on the
exploration of the cross section dependence on the renormalisation
scale, $\mu_R$, and on the factorisation scale, $\mu_F$. For a given
definition of the $\mu_0$ scale, judiciously chosen to absorb the large
logarithmic corrections that appear at higher orders, we set
$\mu_R=\mu_F=\mu_0$ but vary the two scales independently in the range
\begin{equation}
  \frac{1}{2} \, \mu_0  \le \mu_R\,,\mu_F \le  2 \,  \mu_0\,,
\end{equation}
with the following additional condition 
\begin{equation}
\frac{1}{2} \le \frac{\mu_R}{\mu_F} \le  2 \,.
\end{equation}
This means that none of the ratios $\mu_F/\mu_0$, $\mu_R/\mu_0$ and
$\mu_F/\mu_R$ can be larger than two or smaller than one-half. In this
way we avoid having in the perturbative expansion logarithms of
arguments larger than a chosen amount, regardless of its
arbitrariness. In practice, it comes down to considering the following
pairs
\begin{equation}
\label{scan}
\left(\frac{\mu_R}{\mu_0}\,,\frac{\mu_F}{\mu_0}\right) = \Big\{
\left(2,1\right),\left(0.5,1  
\right),\left(1,2\right), (1,1), (1,0.5), (2,2),(0.5,0.5)
\Big\} \,.
\end{equation}
By searching for the minimum and maximum of the resulting cross
sections one obtains an uncertainty band. The narrower the band is,
the smaller the higher order corrections are expected to be.  The
variation of the cross section with respect to the scale choice is
unphysical. It is just a reflexion of the truncation of the
perturbative series. Indeed, if the cross sections are known to all
orders, they will not exhibit this dependence. The scale variation is,
thus, by no means a rigorous way to estimate the theoretical
uncertainty. At best, it might only give an indication of the true
uncertainty. For $\mu_0$ we consider two cases. The mass of the
heaviest particle appearing in the process is typically considered to
be a natural scale choice.  Thus, for our fixed scale setting we
choose $\mu_0= m_t$. Integrated fiducial cross sections are mostly
influenced by final state production relatively close to the
$t\bar{t}$ threshold, which justifies our choice. However,
differential cross sections extend up to energy scales that are much
larger than the threshold. To this end we define the dynamical scale
$\mu_0=H_T/3$. The latter is given by
\begin{equation}
   H_T=
 p_T(b_1) + p_T(b_2) + p_T(b_3) + p_T(b_4) + p_T(e^+) + p_T(\mu^-) +
 p_T^{miss}\,,
\end{equation}
where $b_i$, $i=1,\dots,4$ stands for the four $b$-jets and
$p_{T}^{miss}$ is the missing transverse momentum built out of the two
neutrinos $(\nu_e , \bar{\nu}_\mu)$. This particular choice is
especially suitable for the calculations with off-shell top-quark
contributions as it carries no information about the underlying 
top-quark resonant history. Thus, the top-quark reconstruction is not
attempted here. The PDF uncertainties together with the scale
dependence are the two most important ingredients of the theoretical
error on the predictions of the cross sections. All final-state
partons with pseudo-rapidity $|\eta| <5$ are recombined into jets with
separation $R=0.4$ in the rapidity-azimuthal-angle plane via the
IR-safe {\it anti}$-k_T$ jet algorithm
\cite{Cacciari:2008gp}. Furthermore, we use the
four-momentum recombination scheme. In the first part of the paper
contributions induced by the bottom quarks  are neglected
and we require exactly four $b$-jets as well as two charged
leptons. In the second part of the paper, when the
bottom quarks are going to be included in the initial state, we will
replace the requirement of having exactly four $b$-jets with another
one, namely, we will ask for at least four $b$-jets in the final
state. We put no restriction on the kinematics of the extra light jet (if
resolved) and on the missing transverse momentum. All final state
particles and $b$-jets  have to fullfil the following criteria, which we
consider to be very inclusive selection cuts
\begin{equation}
\begin{array}{ll l l } p_{T} ({\ell})>20 ~{\rm GeV}\,, & \quad\quad
\quad\quad\quad & p_{T} (b)>25 ~{\rm GeV}\,, \vspace{0.2cm}\\
|y(\ell)|<2.5\,,&& |y(b)|<2.5 \,,
\end{array}
\end{equation}
where $\ell=\mu^-,e^+$.

% =============================================
%
\section{Integrated fiducial cross sections}
\label{sec:ttbb-integ}
% 
% =============================================

With the input parameters and cuts specified in Section
\ref{sec:setup}, we arrive at the following predictions for the fixed
scale choice  $\mu_R=\mu_F=\mu_0=m_t$ and the default PDF sets
\begin{equation}
  \begin{split}
\sigma^{\rm LO}_{pp\to e^+ \nu_e \mu^-
  \bar{\nu}_\mu b\bar{b} b\bar{b}} \, ({\rm NNPDF3.1},
\mu_0=m_t)&=6.998 ^{\, +4.525 \, (65\%)}_{\, -2.569 \,(37\%)}
\, {\rm [scales]}\, {\rm fb}\,,\\
\sigma^{\rm NLO}_{pp\to e^+ \nu_e \mu^-
  \bar{\nu}_\mu  b\bar{b} b\bar{b}} \, ({\rm NNPDF3.1},
\mu_0=m_t)&=13.24^{\, +2.33 \, (18\%)}_{\, -2.89 \, (22\%)}\, {\rm [scales]}
\, {}^{\,+0.19\,(1\%)}_{\,-0.19\, (1\%)} \, {\rm [PDF]} \,  {\rm fb}\,.
\end{split}
\end{equation}
At the central scale $\mu_0 = m_t$, the $gg$ channel dominates the
total LO $pp$ cross section by $94\%$, followed by the $q\bar{q}
+\bar{q}q$ channels with $6\%$. The full $pp$ cross section receives
positive and large NLO QCD corrections of $89\%$. The theoretical
uncertainties resulting from scale variation taken in a very
conservative way as a maximum of the lower and upper bounds are $65\%$
at LO and $22\%$ at NLO. Therefore, by going from LO to NLO we have
reduced the theoretical error by a factor of $3$. In the case of truly
asymmetric uncertainties sometimes it is more appropriate to
symmetrise the errors. After symmetrisation the scale uncertainty at
LO is $51\%$ and at NLO does not change substantially, i.e. it is
reduced down to $20\%$. The ${\cal K}$-factor that we have obtained
${\cal K}=1.89$, is defined as the ratio of NLO to LO cross
sections. In our case both LO and NLO integrated fiducial cross
sections are calculated for LO and NLO PDF sets as obtained with
$\alpha_s(m_Z) = 0.118$. Had we used the LO NNPDF3.1 PDF set
with $\alpha_s(m_Z)=0.130$ our LO prediction would rather be
\begin{equation}
\sigma^{\rm LO}_{pp\to e^+ \nu_e \mu^-
  \bar{\nu}_\mu b\bar{b} b\bar{b}} \, ({\rm NNPDF3.1},
\mu_0=m_t) =9.151^{\,+6.512 \, (71\%)}_{\,
  -3.546\,(39\%)} \, {\rm [scales]}\, {\rm fb} \,.
\end{equation}
This would result in the ${\cal K}$-factor of ${\cal K}=1.45$. Both
findings are correct and reflect the different dependence on the scale
of LO and NLO cross sections.  Indeed, the LO cross section is much
more sensitive to the variation of scales and can change more
rapidly than the NLO one. Independently, we can conclude at this point
that NLO QCD corrections are large and indispensable to correctly
describe the process at hand.

Another source of the theoretical error comes from the
parameterization of the NNPDF3.1 PDF set. These uncertainties are due to
experimental errors in the various data that are used in the
fits. They do not, however, take into account additional systematics
coming from the underlying assumptions that enter the parametrisation
of different PDF sets. The latter cannot simply be quantified within a
given scheme. Therefore, we also provide NLO QCD results for two other
PDF sets, namely CT18NLO and MMHT2014. We use the corresponding
prescription from each group to provide the $68\%$ confidence level
(C.L.) PDF uncertainties \footnote{The CT18 NLO errors are rescaled
since they are orginally provided at $90\%$ C.L.}. Both CT18NLO PDFs
and MMHT2014 PDFs include a central set as well as respectively $N=58$
and $N=50$ error sets in the Hessian representation.  The NNPDF3.1 PDF
set uses the MC sampling method in
conjunction with neural networks. In that case the PDF uncertainties are
obtained using the replicas method with a set of $N = 100$ MC PDF
members.  The internal NNPDF3.1 PDF uncertainties are very small,
at the level of $1\%$ only.  Our findings for CT18NLO and MMHT2014 PDF
sets are, on the other hand, given by 
\begin{equation}
  \begin{split}
\sigma^{\rm NLO}_{pp\to e^+ \nu_e \mu^-
  \bar{\nu}_\mu b\bar{b}b\bar{b}} \, ({\rm CT18NLO},
\mu_0=m_t)&=  12.85^{+2.27\,(18\%)}_{-2.78 \, (22\%)}\, {\rm
  [scales]}\, {}^{+0.43 \,
  (3\%)}_{-0.39\, (3\%)} \, {\rm [PDF]} \, {\rm fb}\,,\\
\sigma^{\rm NLO}_{pp\to e^+ \nu_e \mu^-
  \bar{\nu}_\mu b\bar{b}b\bar{b}} \, ({\rm MMHT2014},
\mu_0=m_t)&=  13.12^{+2.31\, (18\%)}_{-2.86\, (22\%)}\, {\rm
  [scales]}\, {}^{+0.40\, (3\%)}_{-0.36\, (3\%)} \, {\rm [PDF]}\, 
{\rm fb}\,.
\end{split}
\end{equation}
We can see that for CT18NLO and MMHT2014 the internal PDF
uncertainties are slightly larger, i.e. of the order of $3\%$. These
uncertainties are nevertheless similar in size to the differences between
NLO QCD results obtained with various PDF sets. Indeed, the relative
difference between CT18NLO and NNPDF3.1 is $3\%$, whereas for 
MMHT2014 and NNPDF3.1 we have  instead $1\%$. We additionally
present LO predictions for various LO PDF sets. Due to the lack of LO
CT18 PDF sets we went back to the older version and used instead the
two LO CT14 PDF sets \cite{Dulat:2015mca}. More specifically, we
employ CT14lo, CT14llo and LO MMHT2014 with
$\alpha_s(m_Z)=0.118,0.130$ and $0.135$ respectively. The larger the
value of $\alpha_s(m_Z)$ the larger the resulting LO cross section
would be. The latter goes to the denominator of the ${\cal K}$-factor.
Our findings confirm this pattern and can be summarised as follows
\begin{equation}
  \begin{split}
\sigma^{\rm LO}_{pp\to e^+ \nu_e \mu^-
  \bar{\nu}_\mu b\bar{b} b\bar{b}} \, ({\rm CT14 lo},
\mu_0=m_t)&= 7.098^{\, +4.454 \, (63\%)}_{\, -2.561 \, (36\%)}
\, {\rm [scales]}\,
{\rm fb}\,,\\
\sigma^{\rm LO}_{pp\to e^+ \nu_e \mu^-
  \bar{\nu}_\mu b\bar{b} b\bar{b}} \, ({\rm CT14 llo},
\mu_0=m_t)&= 9.407^{\, +6.380 \, (68\%)}_{\, - 3.558\, (38\%)}
\, {\rm [scales]}\,
{\rm fb}\,,\\
\sigma^{\rm LO}_{pp\to e^+ \nu_e \mu^-
  \bar{\nu}_\mu b\bar{b} b\bar{b}} \, ({\rm MMHT2014},
\mu_0=m_t)&= 10.670^{\, +7.813 \, (73\%)}_{\, -4.205 \, (39\%)}
\, {\rm [scales]}\,
{\rm fb}\,.\\
\end{split}
\end{equation}
In turn, we receive the following spread of the ${\cal
K}$-factor values $1.81, 1.37, 1.23$ respectively.  Depending on
the LO PDF set employed the NLO QCD corrections to
$pp \to t\bar{t}b\bar{b}$ production in the di-lepton top quark decay
channel range from $89\%$ to $23\%$. On the other hand, the NLO
theoretical error is completely dominated by the scale dependence and
is consistently at the $22\%$ level (at the $20\%$ level after
symmetrisation) independently of the PDF set used.
%
% =============================================
\begin{figure}[t!]
  \begin{center}
\includegraphics[width=0.49\textwidth]{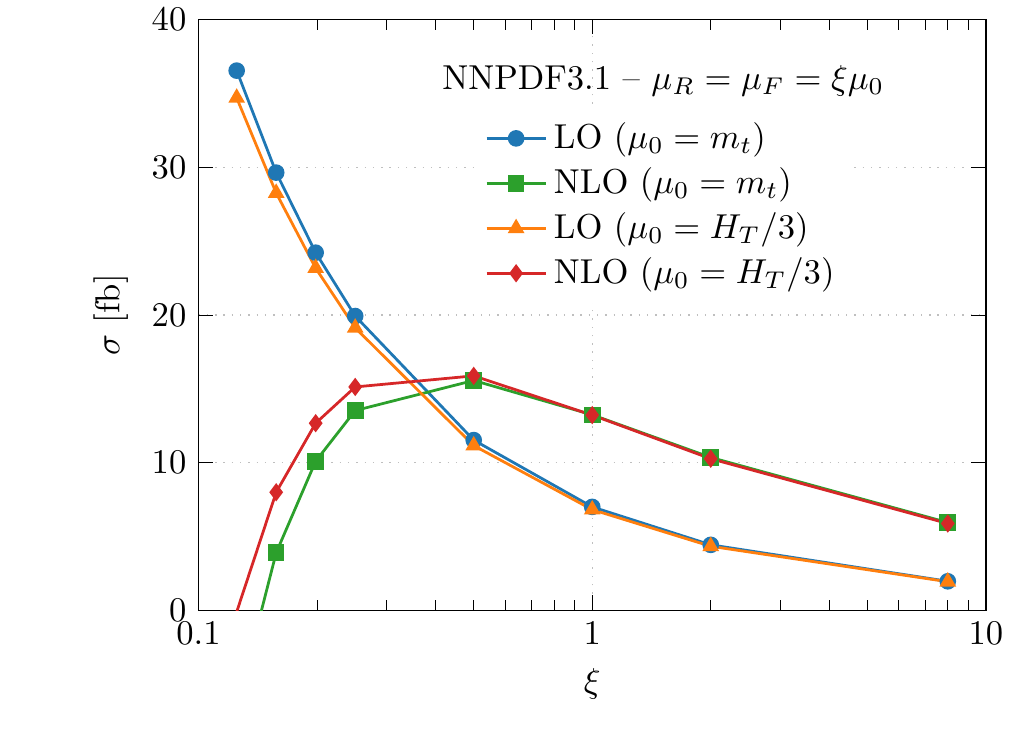}
\\
\includegraphics[width=0.49\textwidth]{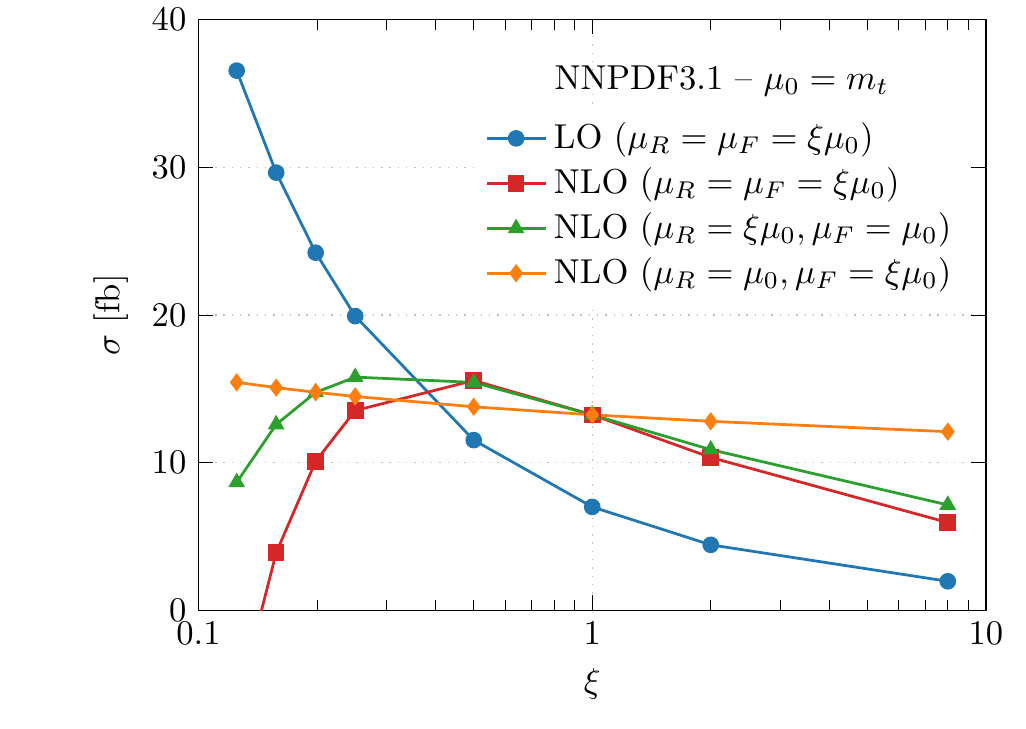}
\includegraphics[width=0.49\textwidth]{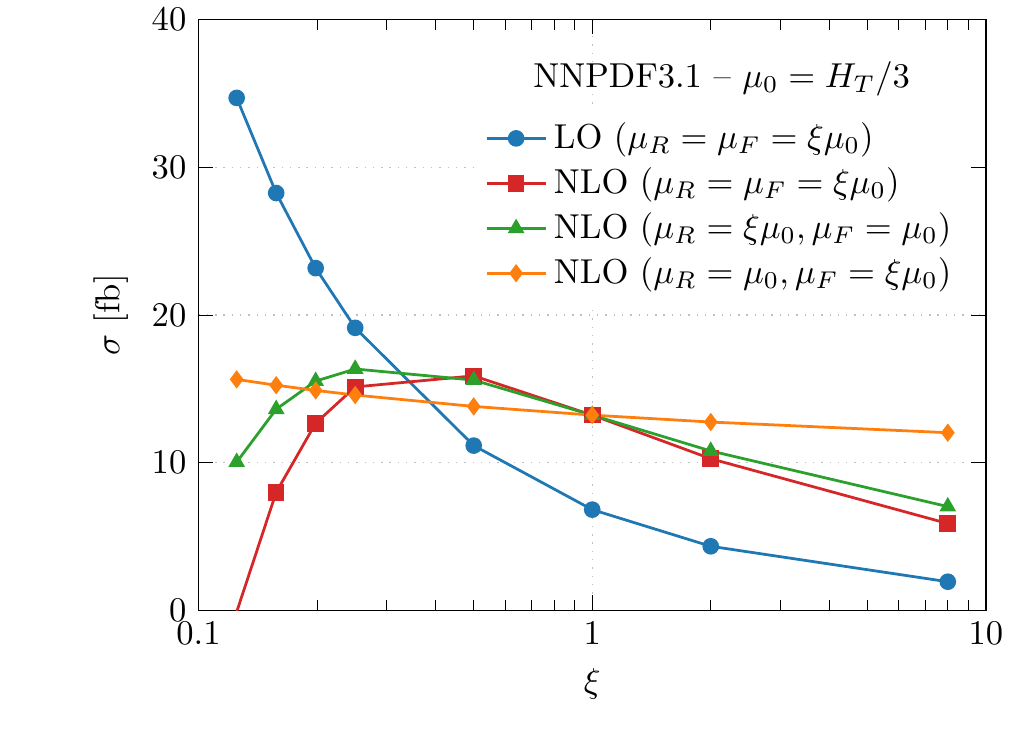}
  \end{center}
\caption{\label{fig:scale_dep} \it Scale dependence of the LO and NLO
integrated fiducial cross sections for the $pp\to e^+ \nu_e \,\mu^-
\bar{\nu}_\mu\, b\bar{b}\,b\bar{b}+X$ production process at the LHC
with $\sqrt{s} = 13$ TeV. Renormalisation and factorisation scales are
set to the common value $\mu_R=\mu_F=\mu_0$ with $\mu_0=m_t$ and
$\mu_0= H_T /3$. The LO and NLO NNPDF3.1 PDF sets are employed.
For each case of $\mu_0$ also shown is the variation of $\mu_R$ with
fixed $\mu_F$ and the variation of $\mu_F$ with fixed $\mu_R$.}
\end{figure}
% =============================================
%

For the dynamical scale setting $\mu_R=\mu_F=\mu_0
=H_T/3$ our results read
\begin{equation}
  \begin{split}
\sigma^{\rm LO}_{pp\to e^+ \nu_e \mu^-
  \bar{\nu}_\mu b\bar{b}b\bar{b}} \, ({\rm NNPDF3.1},
\mu_0=H_T/3)&=6.813^{\, +4.338 \, (64\%)}_{\, -2.481 \,(36\%)}\,{\rm [scales]}
\, {\rm fb}\,,\\
\sigma^{\rm NLO}_{pp\to e^+ \nu_e \mu^-
  \bar{\nu}_\mu b\bar{b}b\bar{b}} \,  ({\rm NNPDF3.1},
\mu_0=H_T/3)&=13.22^{\,+2.66 \,(20\%)}_{\,-2.95 \, (22\%)}\, {\rm [scales]} \,
{}^{\, +0.19\, (1\%)}_{\, -0.19\, (1\%)}\, {\rm [PDF]} \, {\rm fb}\,.
\end{split}
\end{equation}
Our NLO QCD findings for CT18NLO and MMHT2014 PDF sets
are given by

\begin{equation}
  \begin{split}
\sigma^{\rm NLO}_{pp\to e^+ \nu_e \mu^-
  \bar{\nu}_\mu b\bar{b}b\bar{b}} \, ({\rm CT18NLO},
\mu_0=H_T/3)&=  12.81^{\, +2.58\,(\, 20\%)}_{\, -2.84\, (\, 22\%)}\, {\rm
  [scales]}\, {}^{\, +0.42\, (3\%)}_{\, -0.38\, (3\%)}\, {\rm [PDF]} \, {\rm fb}\,,\\
\sigma^{\rm NLO}_{pp\to e^+ \nu_e \mu^-
  \bar{\nu}_\mu b\bar{b}b\bar{b}} \, ({\rm MMHT2014},
\mu_0=H_T/3)&=  13.10^{\, +2.64\,(20\%)}_{\, -2.92\, (22\%)}\, {\rm
  [scales]}\, {}^{\, +0.40\, (3\%)}_{\, -0.35\,(3\%)}\, {\rm [PDF]}\, 
{\rm fb}\,.
\end{split}
\end{equation}
As expected, theoretical predictions for $\mu_0=H_T/3$ for
the integrated fiducial cross sections are very similar to the results
for $\mu_0=m_t$.  We show these results for completeness and because
we will employ the former scale choice at the differential level,
where it plays a crucial role.

In Figure \ref{fig:scale_dep}, the graphical presentation of the
behaviour of LO and NLO cross sections upon varying the central value
of $\mu_R$ and $\mu_F$ by a factor of $\xi \in \{0.125, . . . , 8\}$
is shown for the NNPDF3.1 PDF sets. Both cases $\mu_0=m_t$ and
$\mu_0=H_T/3$ are depicted, that allowed us to compare the two
scales. For the sake of completeness, in Figure \ref{fig:scale_dep} we
present again the scale dependence of the LO and NLO integrated cross
sections for each case of $\mu_0$ separately.  Also shown is the
variation of $\mu_R$ with the fixed value of $\mu_F$ and the variation
of $\mu_F$ with fixed $\mu_R$. We can note that from the point of view
of the integrated fiducial cross sections each scale is a valid choice
that might be used in phenomenological applications.  We
can also add that in the range $\xi\in \left\{0.5,1,2 \right\}$ the
scale variation is driven by the changes in $\mu_R$. In other words,
should we vary $\mu_R$ and $\mu_F$ up and down by a factor of $2$
around $\mu_0$ simultaneously instead of independently the scale
dependence uncertainties would not change significantly. 
Furthermore, not only the choice of $\mu_0$ but also the value of
$\xi$ is important. The latter if not properly selected can introduce
large higher order effects and even result in negative (unphysical)
NLO cross sections as can be observed in Figure \ref{fig:scale_dep}
for $\xi \lesssim 0.25$. Consequently, such small values should not be
selected to set the scale for this process.

%=============================================
%
\subsection{Stability test of NLO fiducial cross sections}
%
%=============================================

%
% =============================================
\begin{table}[t!]
\begin{center}
  \begin{tabular}{ccccccc}
    \hline \hline\\[-0.4cm]
   $p_{T}(b)$ & $\sigma^{\rm LO}$ [fb] & $\delta_{\rm scale}$ &
    $\sigma^{\rm NLO}$ [fb] & $\delta_{\rm scale}$ & $\delta_{\rm PDF}$
    &  ${\cal K}=\sigma^{\rm NLO}/\sigma^{\rm LO}$ \\[0.2cm]
    \hline\hline\\[-0.4cm]
    \multicolumn{7}{c}{$\mu_R=\mu_F=\mu_0=m_t$} \\ [0.2cm]
    \hline \hline \\ [-0.4cm]
    $25$ & $6.998$ & ${}^{+4.525~(65\%)}_{-2.569~(37\%)}$ & $13.24$ &
     ${}^{+2.33~(18\%)}_{-2.89~(22\%)}$
      & ${}^{+0.19~(1\%)}_{-0.19~(1\%)}$ & $1.89$\\[0.2cm] 
    $30$ & $5.113$ & ${}^{+3.343~(65\%)}_{-1.889~(37\%)}$ & $9.25$ &
      ${}^{+1.32~(14\%)}_{-1.93~(21\%)}$
      & ${}^{+0.14 \, (2\%)}_{-0.14 \, (2\%)}$ & $1.81$\\[0.2cm] 
    $35$ & $3.775$ & ${}^{+2.498~(66\%)}_{-1.401~(37\%)}$ & $6.57$ &
      ${}^{+0.79~(12\%)}_{-1.32~(20\%)}$
      & ${}^{+0.10\, (2\%)}_{-0.10 \, (2\%)}$ & $1.74$\\[0.2cm] 
   $40$ & $2.805$ & ${}^{+1.867~(67\%)}_{-1.051~(37\%)}$ & $4.70$ &
     ${}^{+0.46~(10\%)}_{-0.91~(19\%)}$
      & ${}^{+0.08 \, (2\%)}_{-0.08 \, (2\%)}$ & $1.68$\\ [0.2cm]
    \hline\hline\\[-0.4cm]
      \multicolumn{7}{c}{$\mu_R=\mu_F=\mu_0=H_T/3$} \\[0.2cm]
    \hline \hline \\[-0.4cm]
    $25$ & $6.813$ & ${}^{+4.338~(64\%)}_{-2.481~(36\%)}$ & $13.22$ &
    ${}^{+2.66~(20\%)}_{-2.95~(22\%)}$
    & ${}^{+0.19~(1\%)}_{-0.19~(1\%)}$ & $1.94$\\[0.2cm] 
    $30$ & $4.809$ & ${}^{+3.062~(64\%)}_{-1.756~(37\%)}$ & $9.09$ &
    ${}^{+1.66~(18\%)}_{-1.98~(22\%)}$
    & ${}^{+0.16 \, (2\%)}_{-0.16 \, (2\%)}$ & $1.89$\\[0.2cm] 
    $35$ & $3.431$ & ${}^{+2.191~(64\%)}_{-1.256~(37\%)}$ & $6.37$ &
    ${}^{+1.07~(17\%)}_{-1.36~(21\%)}$
    & ${}^{+0.11\, (2\%)}_{-0.11 \, (2\%)}$ & $1.86$\\[0.2cm] 
    $40$ & $2.464$ & ${}^{+1.582~(64\%)}_{-0.901~(37\%)}$ & $ 4.51$ &
    ${}^{+0.72~(16\%)}_{-0.95~(21\%)}$
      & ${}^{+0.09 \, (2\%)}_{-0.09\, (2\%)}$ & $1.83$\\ [0.2cm]
   \hline \hline   
\end{tabular}
\end{center}
\caption{\label{tab:integrated} \it
  LO and NLO integrated fiducial cross sections for the $pp\to
e^+\nu_e\, \mu^-\bar{\nu}_\mu\, b\bar{b} \,b\bar{b} +X$ process at the
LHC with $\sqrt{s}=13$ TeV. Results are evaluated using
$\mu_R=\mu_F=\mu_0$ with $\mu_0=m_t$ and $\mu_0=H_T/3$. LO and NLO
NNPDF3.1 PDF sets are used. We display results for four different
values of the $p_T(b)$ cut.  Also given are the theoretical
uncertainties coming from scale variation $(\delta_{scale})$ and PDFs
$(\delta_{\rm PDF})$.  In the last column the ${\cal K}$-factor is
shown.}
\end{table}
% =============================================

In Table \ref{tab:integrated} we show the integrated fiducial cross
sections at LO and NLO for different cuts on the transverse momentum of
the $b$-jet. Theoretical uncertainties coming from scale variation are
denoted as $\delta_{\rm scale}$, whereas the internal NNPDF3.1 PDF
uncertainties are labeled as $\delta_{\rm PDF}$.  Also shown is the ${\cal
K}$-factor. Results are evaluated using $\mu_R=\mu_F=\mu_0$ with
$\mu_0=m_t$ and for the LO and NLO NNPDF3.1 PDF sets. We observe a
very stable behaviour of systematics when varying the $p_T(b)$ cut,
within the $25-40$ GeV range. Specifically, $\delta_{\rm scale}$ is
consistently of the order of $20\%$ and  the PDF uncertainties
are small for each value of the $p_T(b)$ cut. The size of NLO QCD
corrections is reduced from $89\%$ to $68\%$. This  reduction of
$21\%$  is well within the NLO uncertainties for this process.
Results are not changed when they are generated for the dynamical
scale choice. Theoretical predictions at LO and NLO for $\mu_0=H_T/3$
are also given in Table \ref{tab:integrated}. For the nominal value of
the $p_T(b)$ cut the ${\cal K}$-factor is slightly larger, i.e.  ${\cal
K}=1.94$. It is reduced down to ${\cal K}=1.83$ for $p_T(b)>40$
GeV. The $11\%$ difference is again well within $\delta_{\rm
scale}$. We can conclude that the perturbative expansion in $\alpha_s$
for the process at hand is not spoiled by the appearance of large
logarithms, thus, under excellent theoretical control.

%=============================================
%
\subsection{Additional cuts and comparison with ATLAS results}
% 
% =============================================

In the following we will examine the behaviour of the integrated cross
section upon adding additional cuts, i.e. making the available phase
space for the $2\to 8$ process more exclusive. We will show results
for $\mu_R=\mu_F=\mu_0=m_t$ only. We have checked, however, that the
LO and NLO theoretical predictions for $\mu_0=H_T/3$ are very
similar. In the first step we increase $p_T(\ell)$ from $20$ GeV to
$25$ GeV. Furthermore,  we introduce an extra cut, i.e.  the
separation between the $b$-jet and the charged lepton in the
rapidity-azimuthal-angle plane of $\Delta R(\ell b)>
0.4$. The NLO results with these cuts read
\begin{equation}
\sigma^{\rm NLO}_{pp\to e^+ \nu_e \mu^-
  \bar{\nu}_\mu b\bar{b}b\bar{b}} \,  ({\rm NNPDF3.1},
\mu_0=m_t)=9.70^{\, +1.66 \, (17\%)}_{\, -2.10\, (22\%)}
\, {\rm fb}\,.
\end{equation}
The corresponding ${\cal K}$-factor is ${\cal K}=1.88$. In the next
step  another cut has been added, i.e.  the separation between
charged leptons of $\Delta R(\ell\ell)> 0.4$. We
report the following NLO cross section for this  case
\begin{equation}
\sigma^{\rm NLO}_{pp\to e^+ \nu_e \mu^-
  \bar{\nu}_\mu b\bar{b}b\bar{b}} \,  ({\rm NNPDF3.1},
\mu_0=m_t)=9.52^{\, +1.62 \, (17\%)}_{\, -2.06\, (22\%)}
\, {\rm fb}\,.
\end{equation}
Finally, we put the restriction on the missing transverse momentum of
$p_T^{miss} > 50$ GeV. In this  case we report
\begin{equation}
\sigma^{\rm NLO}_{pp\to e^+ \nu_e \mu^-
  \bar{\nu}_\mu b\bar{b}b\bar{b}} \,  ({\rm NNPDF3.1},
\mu_0=m_t)=6.72^{\, +1.14 \, (17\%)}_{\, -1.46\, (22\%)}
\, {\rm fb}\,.
\end{equation}
These two additional conditions for $\Delta R(\ell\ell)$ and
$p_T^{miss}$ do not affect the ${\cal K}$-factor. It remains 
at the ${\cal K}=1.88$ level. We also notice that the theoretical
uncertainties due to scale dependence are remarkably stable for all
three cases. A few comments are in order. The first selection that we
have applied, namely 
\begin{equation}
\begin{array}{ll l l } p_{T} ({\ell})>25 ~{\rm GeV}\,, & \quad\quad
\quad\quad\quad & p_{T} (b)>25 ~{\rm GeV}\,, \vspace{0.2cm}\\
  |y(\ell)|<2.5\,,&& |y(b)|<2.5 \,, \vspace{0.2cm}\\
  \Delta R(b b) > 0.4\,, &&\Delta R(\ell b)> 0.4\,,
\end{array}
\end{equation}
with the $anti-k_T$ jet algorithm and the radius parameter  $R=0.4$,
is very close to the phase-space volume in which the fiducial
$t\bar{t}b\bar{b}$ cross section has recently been measured by the
ATLAS collaboration in the $e\mu$ top-quark decay channel
\cite{Aaboud:2018eki}. In that paper, among others,
$\sigma_{t\bar{t}b\bar{b}}$ was determined by requiring exactly one
electron and one muon (with opposite charges) and at least four
$b$-jets. We shall label this final state as $e\mu\, + \,4b$ in the
following. The measured cross section, after the estimated contributions
from $t\bar{t}H, t\bar{t}W^\pm$ and $t\bar{t}Z$ productions were
subtracted, was compared with the theoretical predictions for the
$t\bar{t}b\bar{b}$ process. The experimental result was found to be
higher than the SM prediction but still compatible within the quoted 
uncertainties. In detail, the measured ATLAS result is given by
\begin{equation}
\sigma^{\rm ATLAS}_{e\mu \, +\,  4b}
= (25 \,  \pm 6.5) \,  {\rm fb}\,,
\end{equation}
where $6.5$ fb $(26\%)$ is the total uncertainty for this measurement,
i.e. statistical plus systematical uncertainty.  In Table
\ref{tab:comparison} we show various theoretical predictions for
$t\bar{t}b\bar{b}$ production that have been used in
Ref. \cite{Aaboud:2018eki} to compare with the ATLAS data. In
particular, results shown there have been generated with the following
MC frameworks: \textsc{Sherpa+OpenLoops} \cite{Cascioli:2013era},
\textsc{PowHel+Pythia 8}
\cite{Kardos:2013vxa,Garzelli:2014aba,Bevilacqua:2017cru} and
\textsc{MadGraph5}${}_{-}$\textsc{aMC\@Nlo+Powheg-Box+Pythia 8}
\cite{Alwall:2014hca}, that are commonly used by the ATLAS and CMS
experimental collaborations. In all cases massive $b$-quarks and
four-flavour PDF sets have been employed. The second \textsc{PowHel}
result has been obtained for massless $b$-quarks with the five-flavour
PDF set.  All theoretical results are based on the NLO matrix element
calculations for the on-shell $t\bar{t}b\bar{b}$
process.  Let us note here, that the measured inclusive
fiducial cross section exceed theoretical predictions for the
$t\bar{t}b\bar{b}$ process. However, they are all compatible within
the given uncertainties. For comparison, we also add our result multiplied by a
factor of $2$ to account for the two decay channels $e^+\mu^-$ and
$e^-\mu^+$.    We can observe that the $\sigma^{\rm
\textsc{Helac-Nlo} \,(5FS)}_{e\mu+4b}$ result is closer to the 
$\sigma^{\rm ATLAS}_{e\mu+4b}$ one than other theoretical  predictions
given in Table \ref{tab:comparison}. Both results agree very well within
the quoted uncertainties. Specifically, the agreement of $0.7\sigma$
has been obtained\,\footnote{We note
that the scale choice used in Ref. \cite{Aaboud:2018eki} is different
than our dynamical scale setting $\mu_R=\mu_F=\mu_0=H_T/3$. The four
theoretical predictions from Ref. \cite{Aaboud:2018eki} employ $\mu_R=
\prod_{i=t,\bar{t},b,\bar{b}} E_{T,i}^{1/4}$, where $E_{T,i}$ refers
to the transverse energy. The factorisation scale is set to
$\mu_F=\frac{1}{2} \sum_{i=t,\bar{t},b,\bar{b},j} E_{T,i}$. Due to the
stable top quarks, that appear in the definition of $\mu_R$ and
$\mu_F$, we cannot use the exact same scale settings.  We can,
however, mimic it as close as possible by using the reconstructed
top-quark momenta. Had we used this approach our NLO QCD results would
rather be $\sigma^{\rm \textsc{Helac-Nlo}}_{e\mu\, + \,4b}= (20.3 \pm
4.2)$ fb instead of $\sigma^{\rm \textsc{Helac-Nlo}}_{e\mu\, +
  \,4b}=(19.4 \pm 4.2)$ fb.}.  A dedicated comparison
between \textsc{Helac-Nlo} and predictions obtained with the help of
various NLO matrix element calculations matched to parton shower
programs would be in order to understand the source of the spread
among these theoretical results. We leave such studies for the near
future. We note that in the experimental result
as well as in the theoretical predictions used in
Ref. \cite{Aaboud:2018eki} also leptonic $\tau^\pm$ decays were
included, i.e.  $e^\pm$ and $\mu^\pm$ from $\tau^\pm$ decays were
incorporated into the analysis. This is not the case for the
\textsc{Helac-Nlo} simulations. Nonetheless, we can roughly estimate
the size of the missing contribution by using the NLO fiducial cross
section for $pp\to \tau^+\nu_\tau\, \tau^-\bar{\nu}_\tau\, b\bar{b}
\,b\bar{b} +X$ multiplied by the corresponding branching ratio for the
leptonic $\tau^+\tau^-$ decays. Since our selection cuts are very
inclusive this estimate should not be very far away from the true
result. Assuming the following branching ratios ${\cal BR} (\tau^- \to
\mu^-\bar{\nu}_\mu \nu_\tau)=17.39\%$ and ${\cal BR} (\tau^- \to
e^-\bar{\nu}_e \nu_\tau)=17.82\%$ \cite{Zyla:2020zbs} we can estimate
that the \textsc{Helac-Nlo} result should be increased by about $0.6$
fb. Thus, the final theoretical result is rather given by
\begin{equation}
\sigma^{\rm
  \textsc{Helac-Nlo}}_{e\mu\, + \,4b}= (20.0 \pm 4.3)\, {\rm fb}\,,
\end{equation}
which is only $0.6\sigma$ away from the ATLAS measured cross
section. We conclude this part by saying that the precision of the
measurement is still slightly lower than that of the theoretical prediction
as obtained with the help of the \textsc{Helac-Nlo} MC program.  In
the former case the experimental total error is at the $26\%$ level. It
comprises statistical and systematical uncertainties. In the latter
case the estimated theoretical error is $22\%$. It 
combines  scale dependence and PDF uncertainties.
%
% =============================================
\begin{table}[t!]
\begin{center}
\begin{tabular}{lcc}
\hline \hline  \\ [-0.4cm]
 Theoretical predictions 
  & \quad \quad \quad &  $\sigma_{e\mu + 4b}$ [fb] \\[0.2cm]
  \hline \hline  \\ [-0.4cm]
 \textsc{Sherpa+OpenLoops}   (4FS)  &  & $17.2\pm 4.2$   \\[0.2cm] 
  \textsc{Powheg-Box+Pythia 8} (4FS)  &  & 16.5\\[0.2cm]
  \textsc{PowHel+Pythia 8} (5FS)  &  & 18.7\\[0.2cm]
  \textsc{PowHel+Pythia 8} (4FS)  &  &  18.2\\[0.2cm]
  \hline \hline  \\ [-0.4cm]
  \textsc{Helac-Nlo} (5FS)  &  &  $19.4 \pm 4.2$ \\[0.2cm]
   \hline \hline
 \end{tabular}
\end{center}
\caption{\label{tab:comparison} \it Predicted fiducial cross section
results for the $pp \to t\bar{t}b\bar{b}$ process in the $e\mu$ top
quark decay channel at the LHC with $\sqrt{s}=13$ TeV. Results for the
case of at least four $b$-jets in the final state is shown. Except
from the \textsc{Helac-Nlo} case all result are taken from
Ref. \cite{Aaboud:2018eki}.}
\end{table}
% =============================================
%

%=============================================
%
\section{Differential fiducial cross sections and
PDF uncertainties}
\label{sec:ttbb-diff}
% 
% =============================================

In addition to the normalization of the integrated fiducial cross
section higher-order QCD corrections can affect the shape of various
kinematic distributions. To quantify the size of these effects and to
investigate shape distortions that are introduced in this way we shall
test a variety of differential distributions for the $pp \to e^+ \nu_e
\, \mu^-\bar{\nu}_\mu \, b\bar{b}\, b\bar{b} +X$ process. All
observables are obtained for the $\mu_R=\mu_F=\mu_0=H_T/3$ scale
choice, NNPDF3.1 PDF sets and default cuts and parameters. Even though
we present results for the kinematic-dependent choice of $\mu_R$ and
$\mu_F$ only we have also studied differential predictions for the
kinematic-independent scale choice, i.e. for
$\mu_R=\mu_F=\mu_0=m_t$. We will use the latter findings to provide
relevant comments. However, in order not to lengthen the manuscript
unnecessarily, they will not be shown.  This is additionally justified
by the fact that at the differential level $\mu_0=m_t$ cannot describe
efficiently the multi-scale kinematics of the $t\bar{t}b\bar{b}$
process. The fixed scale setting leads to perturbative instabilities
in the TeV region, where large negative corrections are visible in the
tails of several distributions, see
e.g. Ref. \cite{Bevilacqua:2016jfk} for a detailed discussion of this
behaviour. We show in the upper plots the
absolute LO and NLO QCD predictions for the $pp\to e^+ \nu_e \,
\mu^-\bar{\nu}_\mu \, b\bar{b}\, b\bar{b}+X$ process.  The lower
panels display the differential ${\cal K}$-factors defined as ${\cal
K}=d\sigma^{\rm NLO}(\mu_0)/d\sigma^{\rm LO}(\mu_0)$. Additionally, we
provide the uncertainty bands from scale variation defined according
to ${\cal K}(\mu=\xi\mu_0)= d\sigma^{\rm NLO}(\mu=\xi
\mu_0)/d\sigma^{\rm LO}(\mu_0)$. The LO blue bands are given to
illustrate the relative scale uncertainty of the LO cross
section. They are defined according to $ d\sigma^{\rm
  LO}(\mu=\xi\mu_0)/d\sigma^{\rm LO}(\mu_0)$. The LO and NLO
uncertainty bands are obtained by performing a $7$-point scale
variation around the central value $\mu_0$. 

We note here that our analysis is not trying to identify the origin
of $b$-jets, i.e. it does not distinguish between the extra (prompt)
$b$-jets and $b$-jets that come from the top-quark decays. This is to
avoid the use of the reconstruction techniques for assigning $b$-jets
to a specific production process. We leave such topic  for future
studies  in which modeling uncertainties will be explored in
more detail.

\begin{figure}[t!]
  \begin{center}
\includegraphics[width=0.49\textwidth]{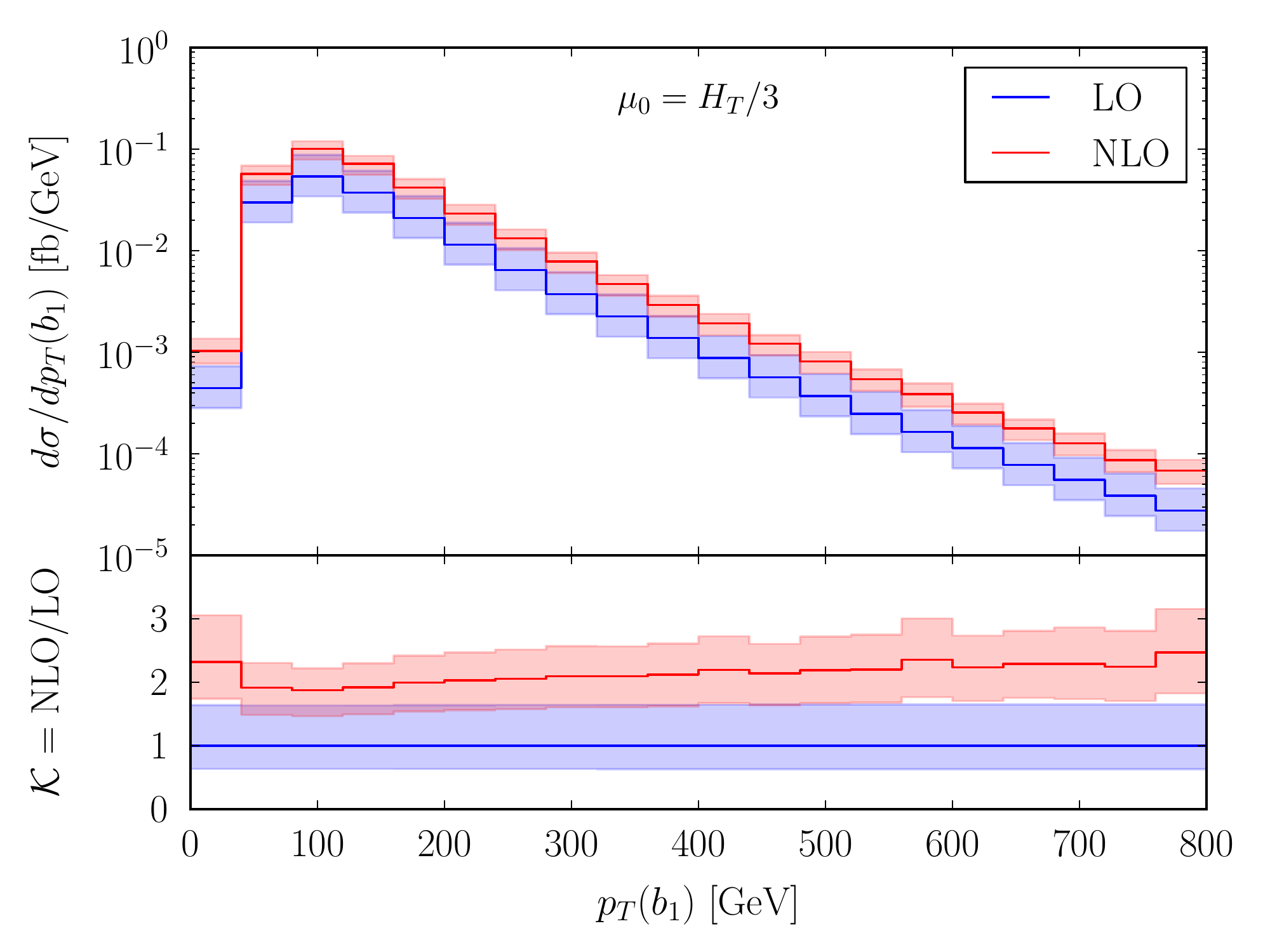}
\includegraphics[width=0.49\textwidth]{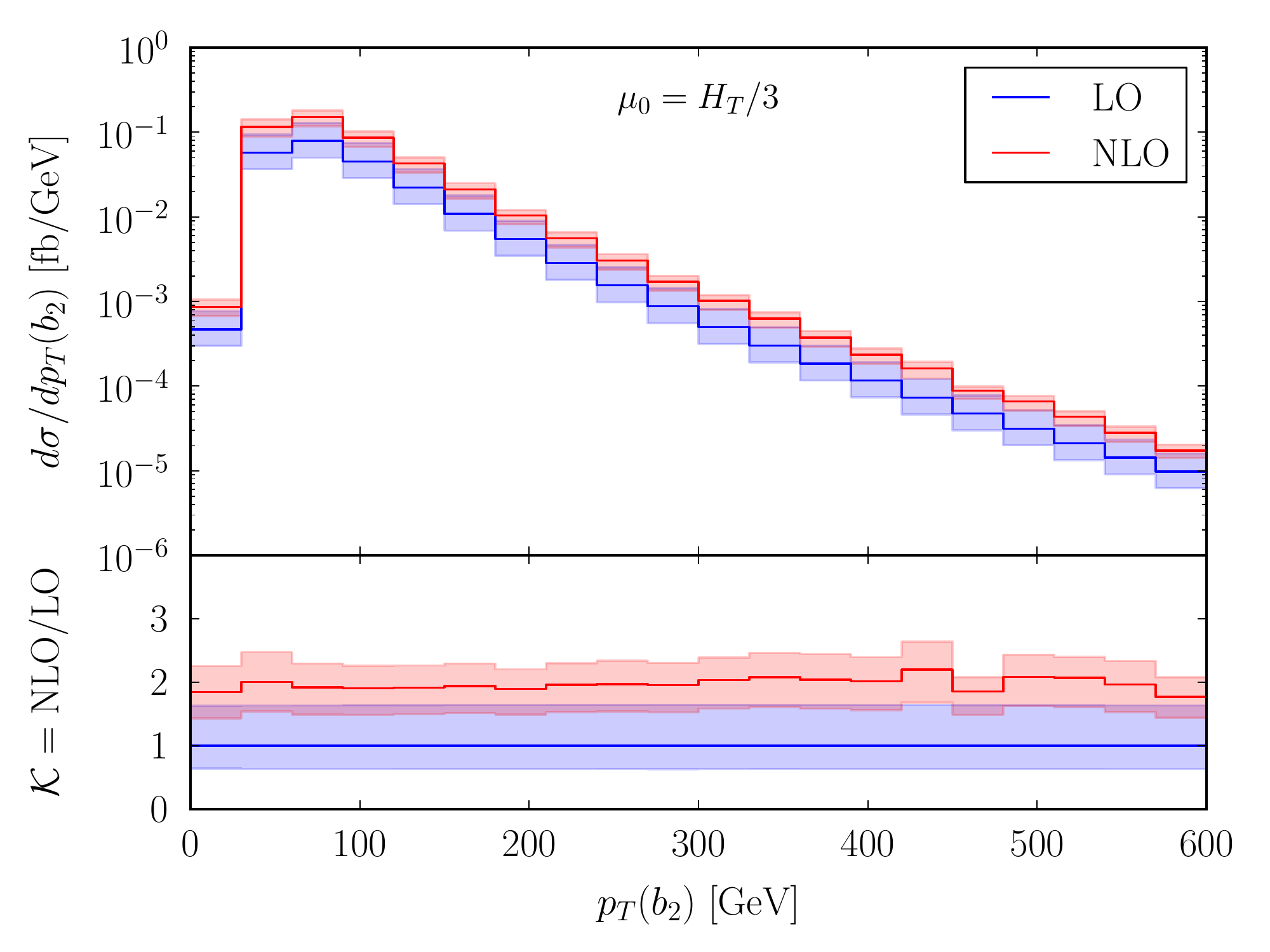}
\includegraphics[width=0.49\textwidth]{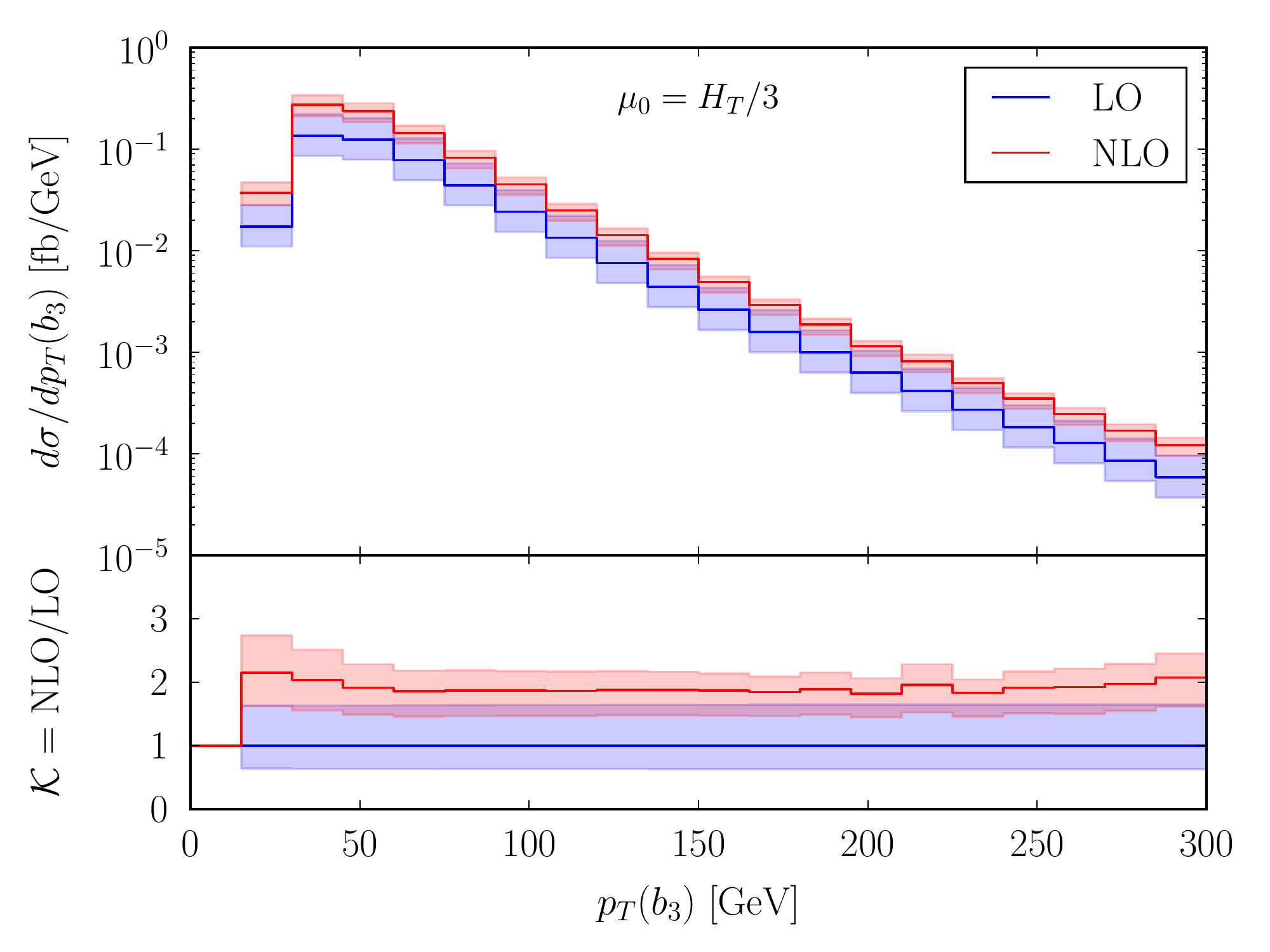}
\includegraphics[width=0.49\textwidth]{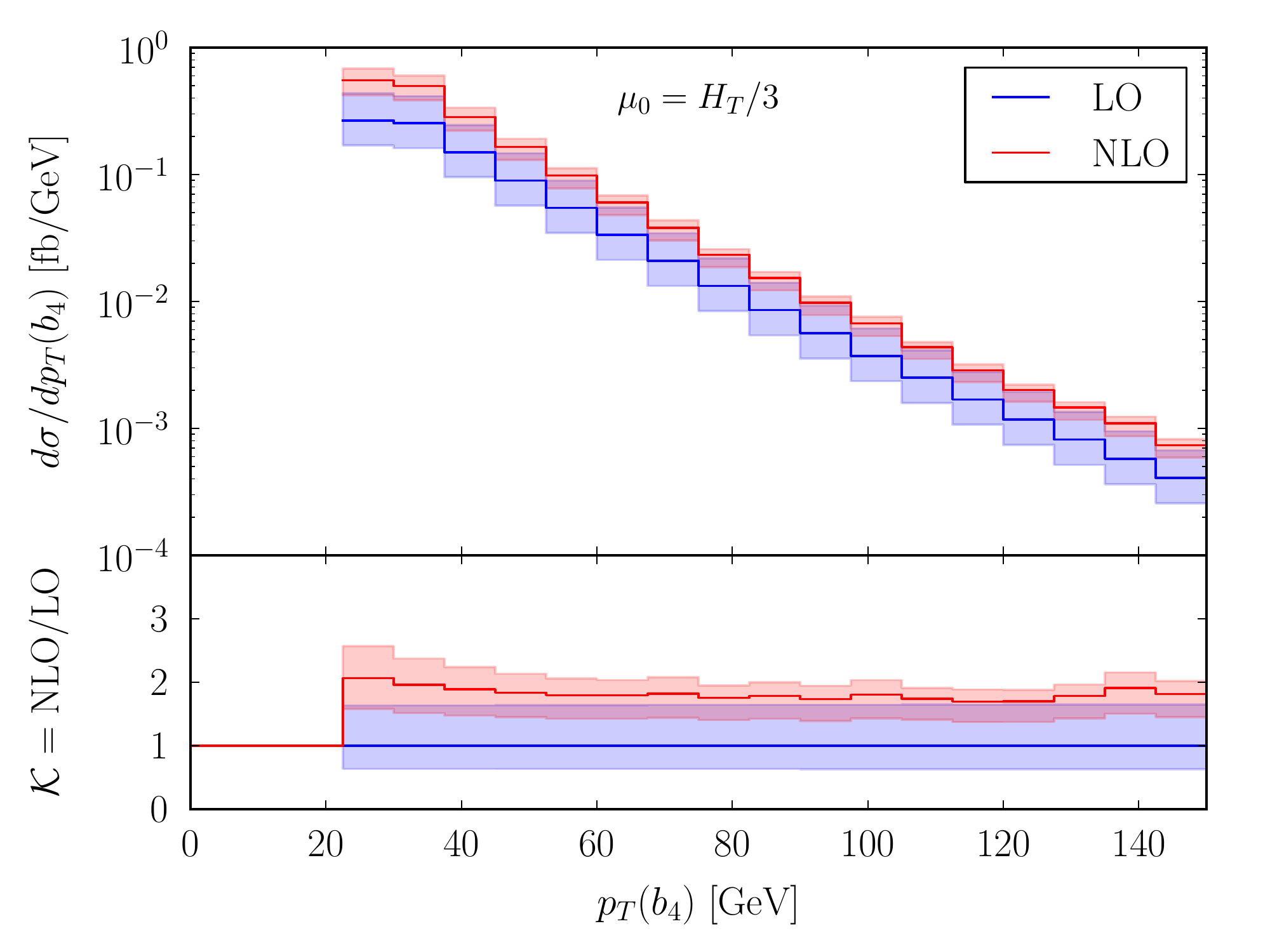}
  \end{center}
  \caption{\label{fig:ptb} \it
Differential cross section distributions as a function of the
transverse momentum of the $1^{st}$, $2^{nd}$, $3^{rd}$ and the
$4^{th}$ hardest $b$-jet
at LO and NLO for the $pp \to e^+ \nu_e \, \mu^- \bar{\nu}_\mu \,
b\bar{b}\, b\bar{b} +X$ process at the LHC with $\sqrt{s}=13$
TeV. The heavy-flavour  jets are ordered in $p_T$. The upper plots
show absolute LO and NLO QCD predictions together with corresponding
uncertainty bands. The
lower panels display the differential ${\cal K}$-factor together with
the uncertainty band and the relative scale uncertainties of the LO
cross section.  Results are provided for
$\mu_R=\mu_F=\mu_0=H_T/3$. The LO and the NLO NNPDF3.1 PDF sets are
employed. }
  \end{figure}

We start with the transverse momentum of the four $b$-jets. They are
depicted in Figure \ref{fig:ptb}.  The heavy-flavour jets are ordered
in $p_T$, thus, $p_T(b_1)$ corresponds to the leading (hardest)
$b$-jet while $p_T(b_4)$ is the fourth-leading (softest)
$b$-jet. The differential cross sections are plagued by the same large
higher order QCD effects as the integrated
fiducial cross sections. Indeed, for example for $p_T(b_1)$ we observe
$90\%-135\%$ NLO QCD corrections, that introduce shape distortions of
the order of $45\%$. The theoretical uncertainties due to scale
dependence are, on the other hand, in the range of $20\%-30\%$.  For
$p_T(b_2)$, $p_T(b_3)$ and $p_T(b_4)$ higher order QCD corrections are
of the order of $70\%-120\%$ while uncertainties are between $10\%$
and $25\%$ depending on the $p_T$ of the $b$-jet.  Had we used 
$\mu_0=m_t$ instead we would rather obtain  shape
distortions up to even $150\%$. This shows the advantage of the
dynamical scale over the fixed one.
%
% =============================================
% =============================================
\begin{figure}[t!]
  \begin{center}
        \includegraphics[width=0.49\textwidth]{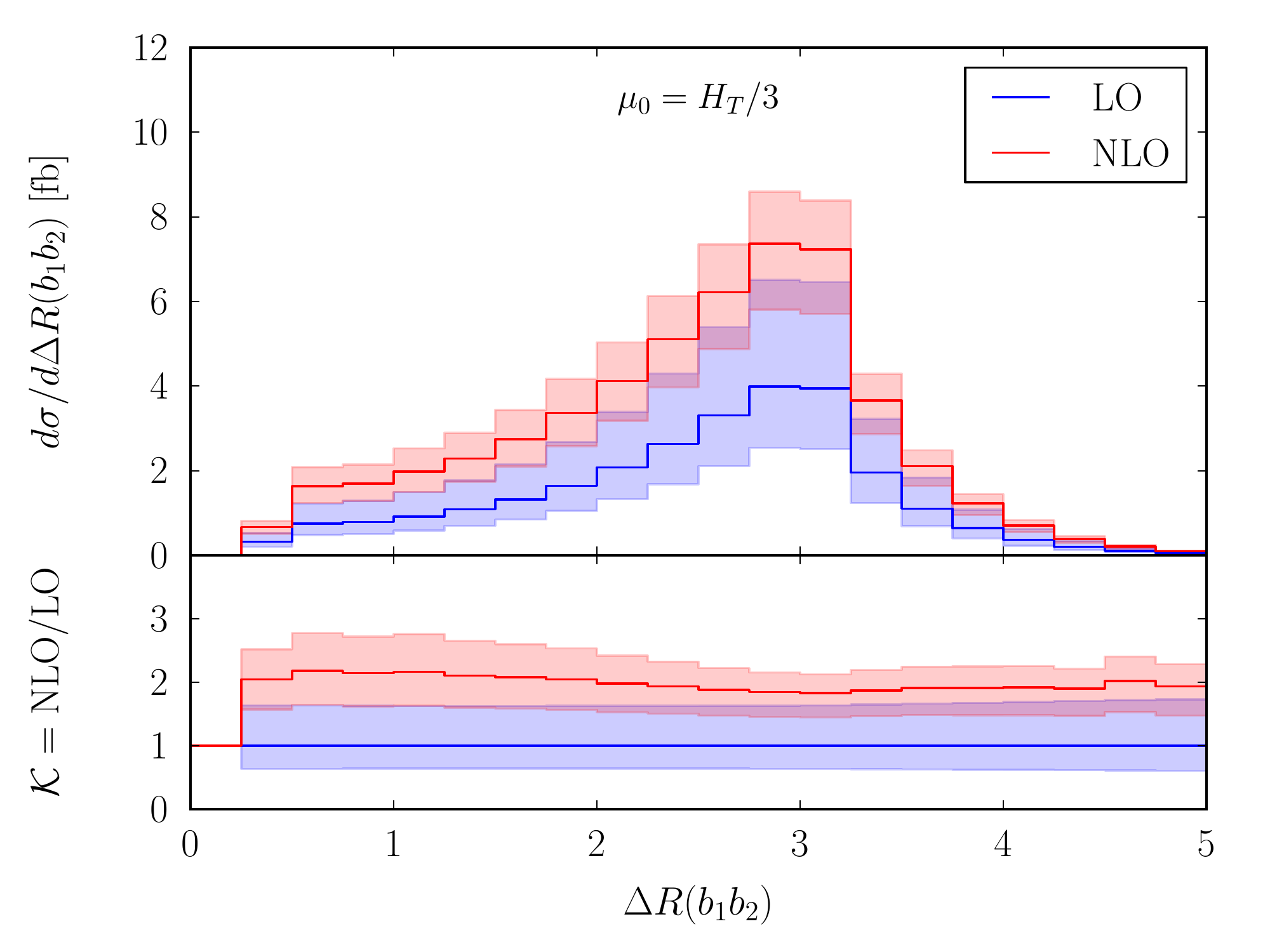}
\includegraphics[width=0.49\textwidth]{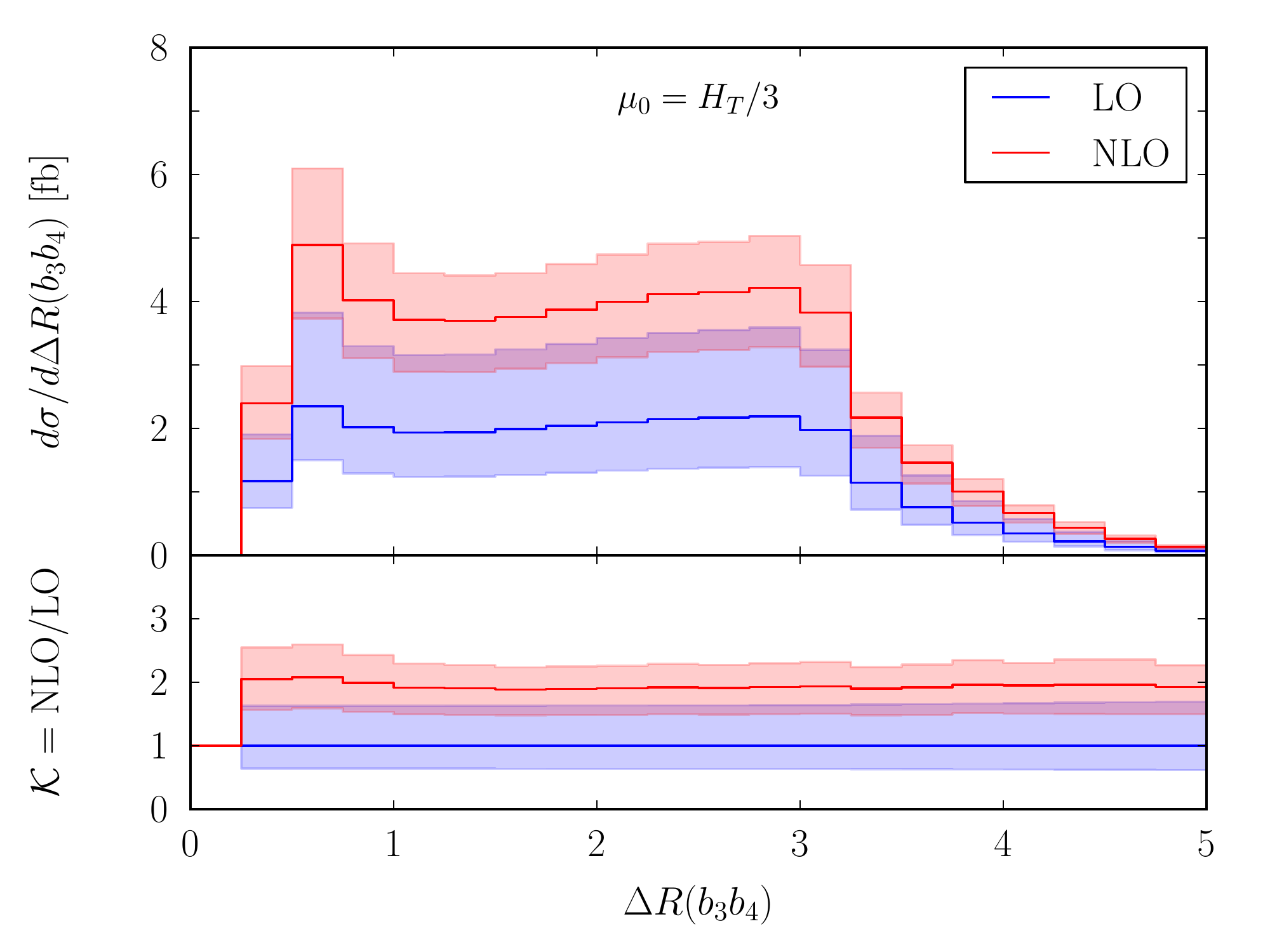}
\includegraphics[width=0.49\textwidth]{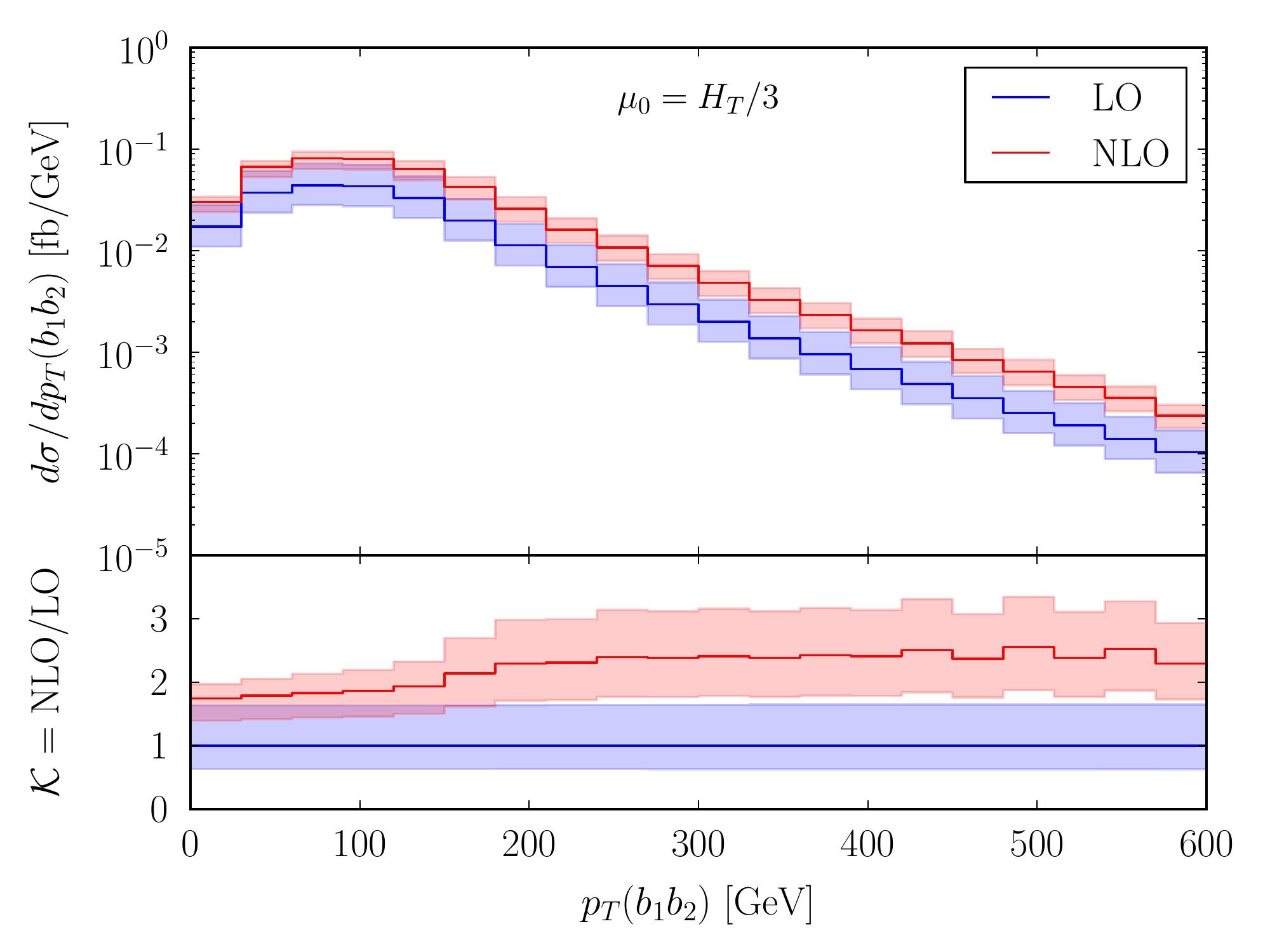}
\includegraphics[width=0.49\textwidth]{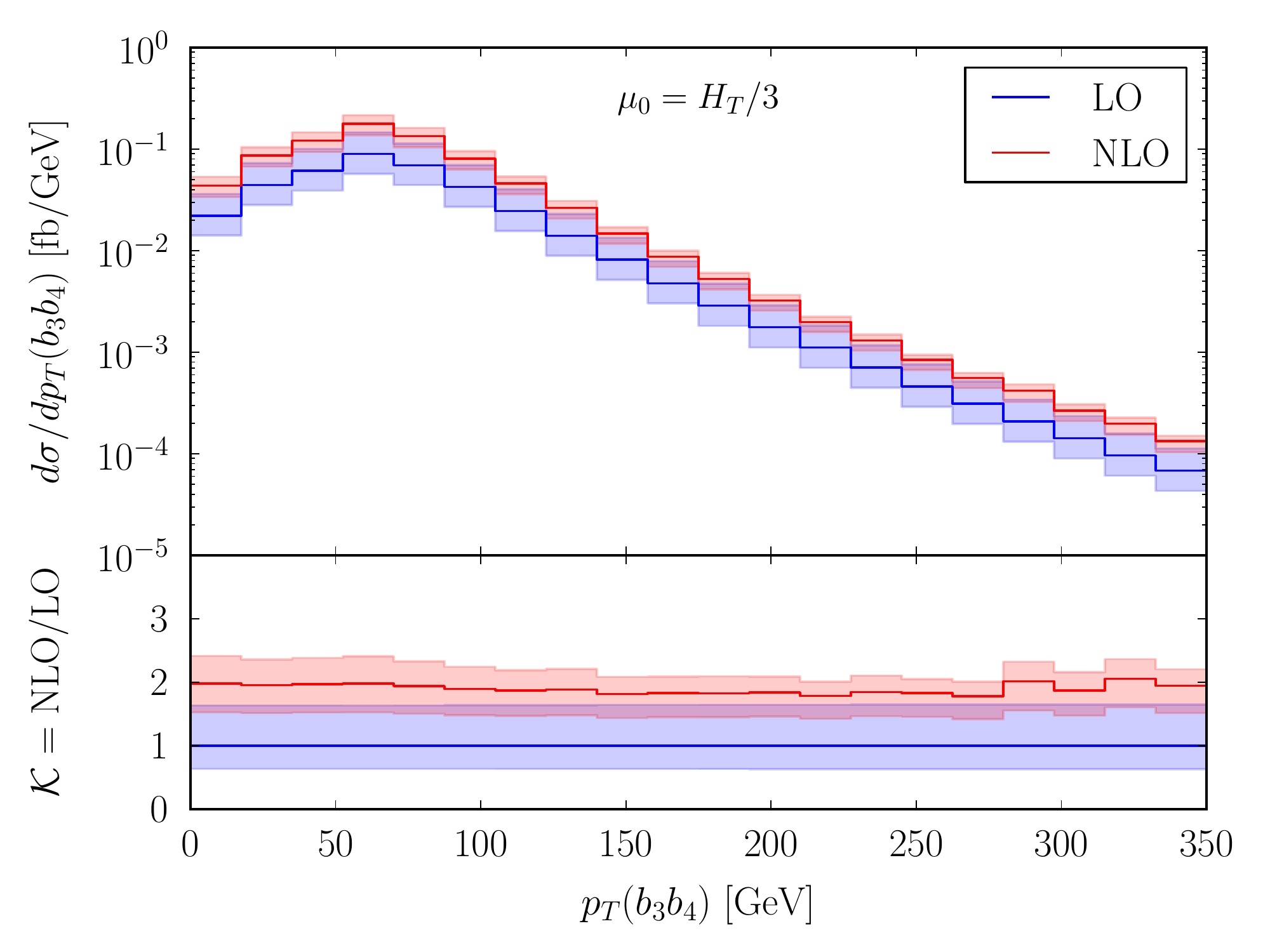}
\includegraphics[width=0.49\textwidth]{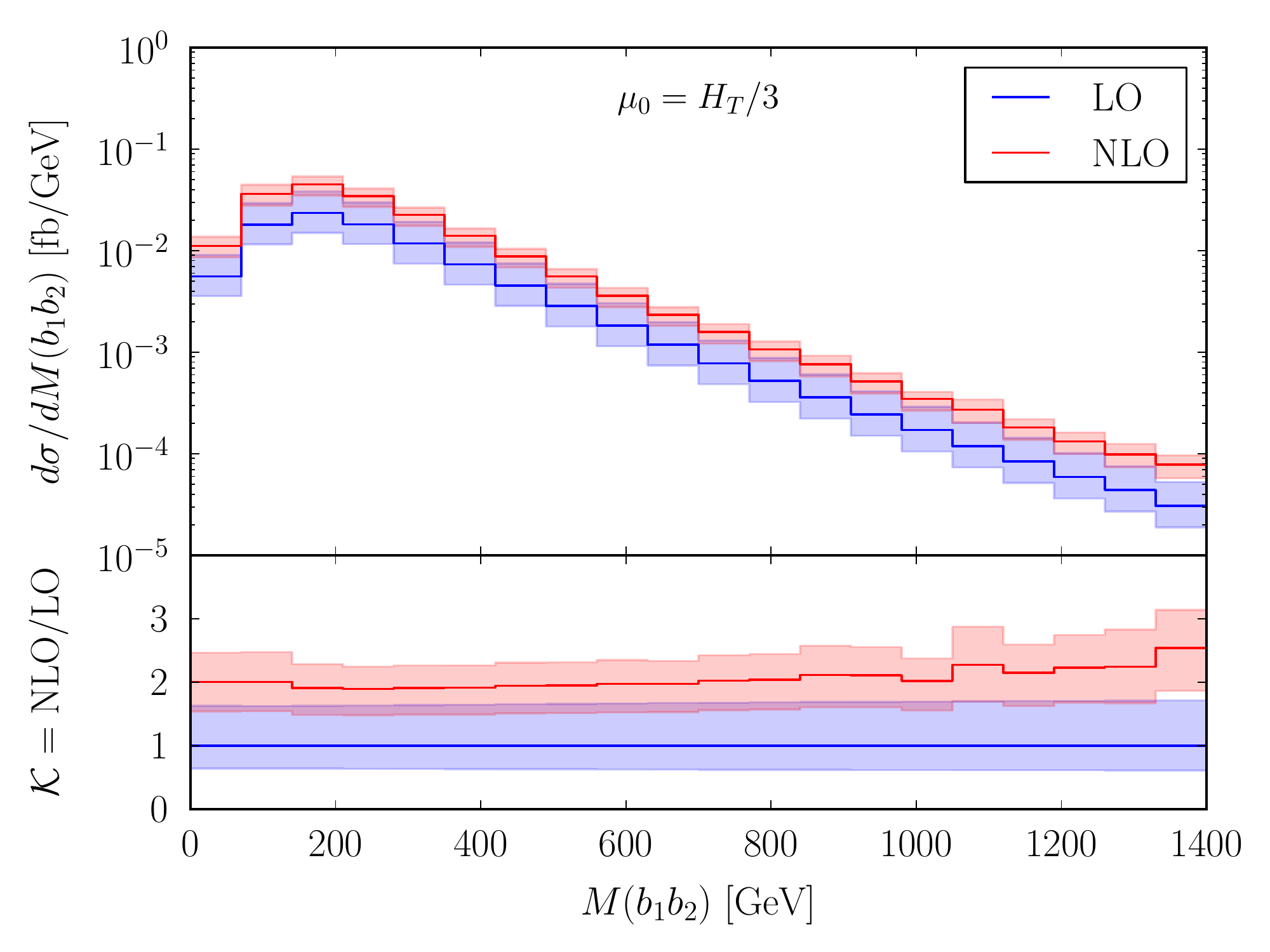}
\includegraphics[width=0.49\textwidth]{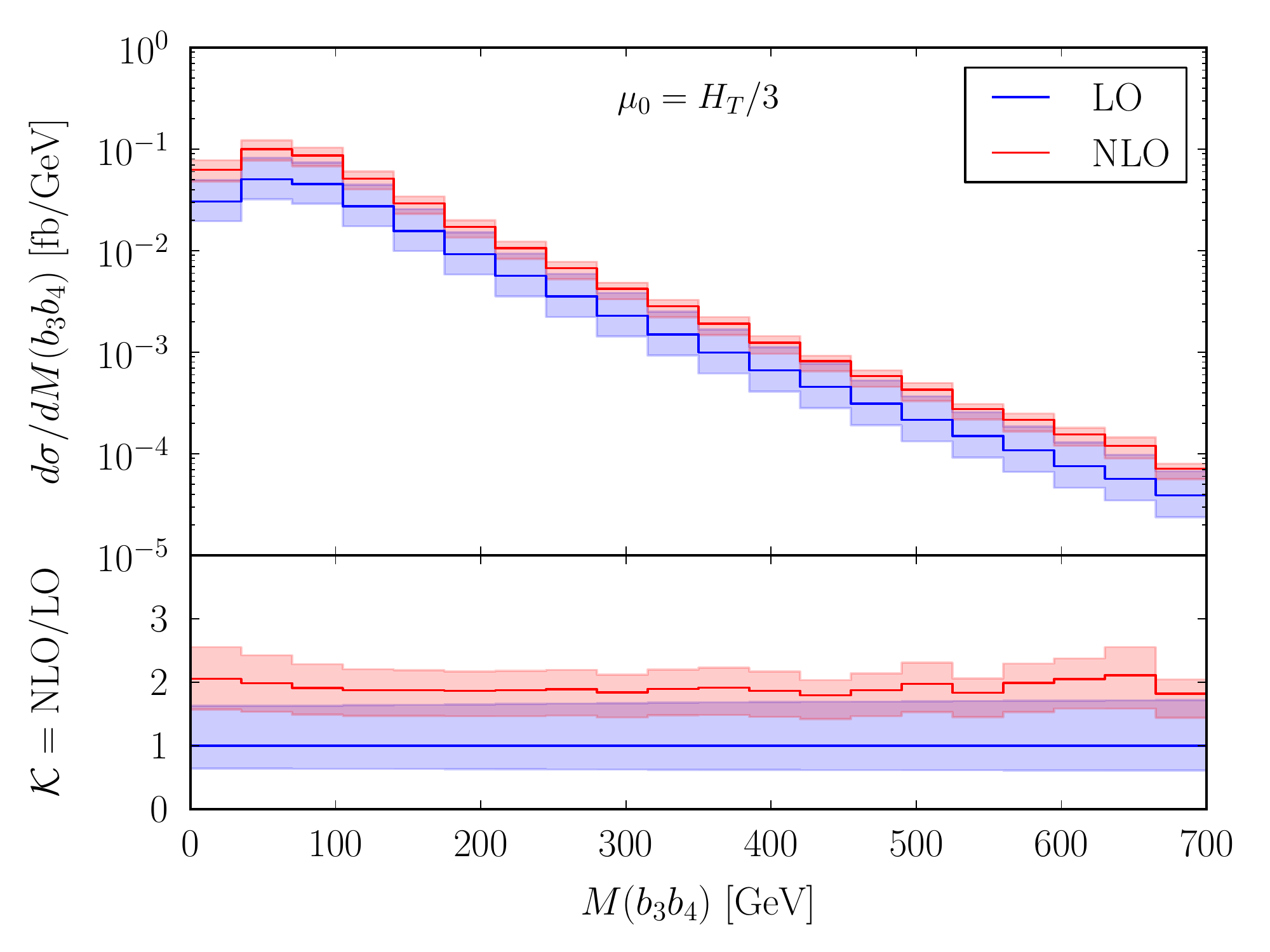}
  \end{center}
\caption{\label{fig:b12-b34} \it  As in Figure \ref{fig:ptb} but for the
  $\Delta R(b_1b_2)$, $\Delta R(b_3b_4)$, $p_T(b_1b_2)$,
  $p_T(b_3b_4)$, $M(b_1b_2)$ and $M(b_3b_4)$ distributions.}
\end{figure}
% =============================================
\begin{figure}[t!]
  \begin{center}
     \includegraphics[width=0.49\textwidth]{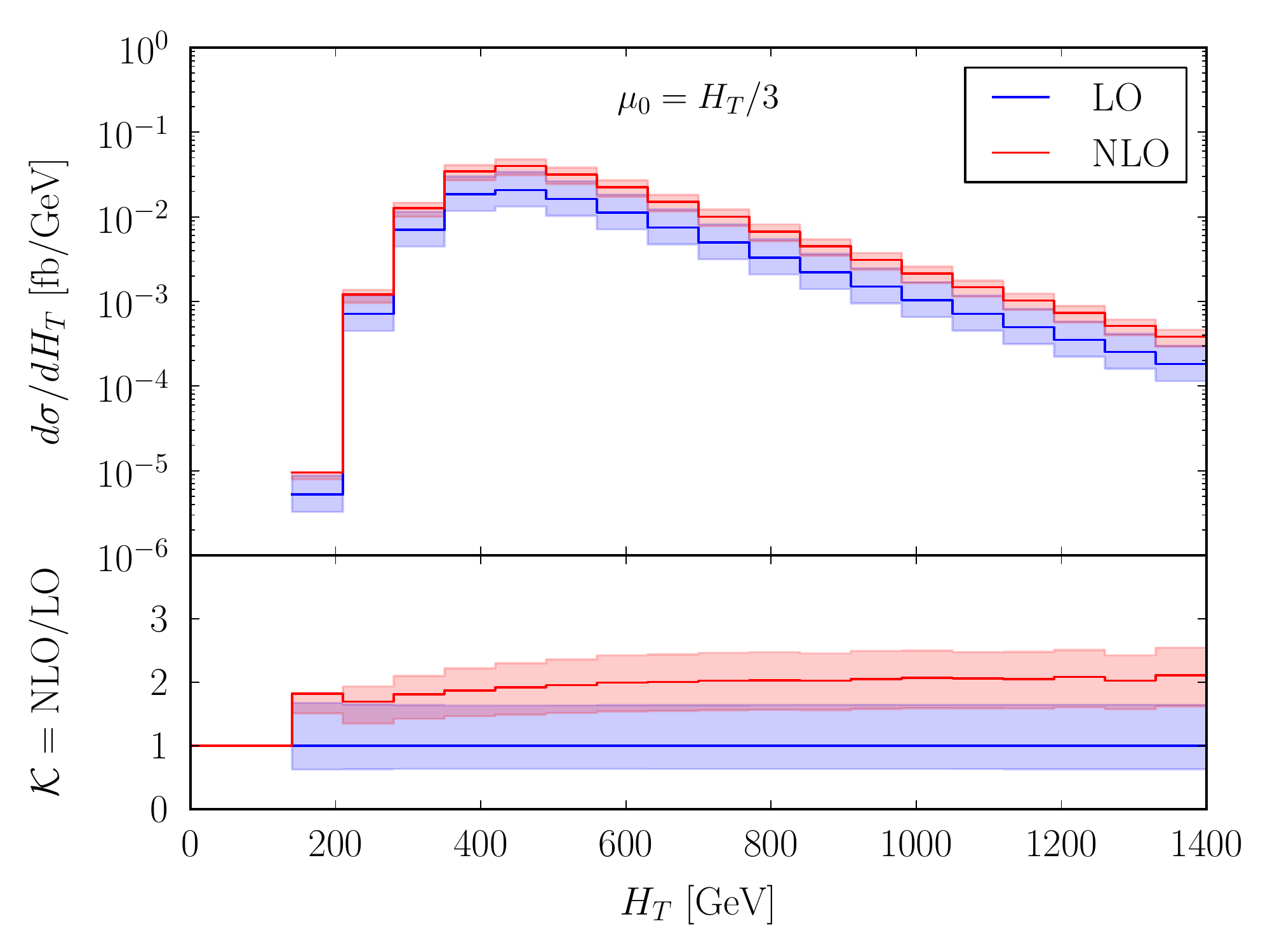}
    \includegraphics[width=0.49\textwidth]{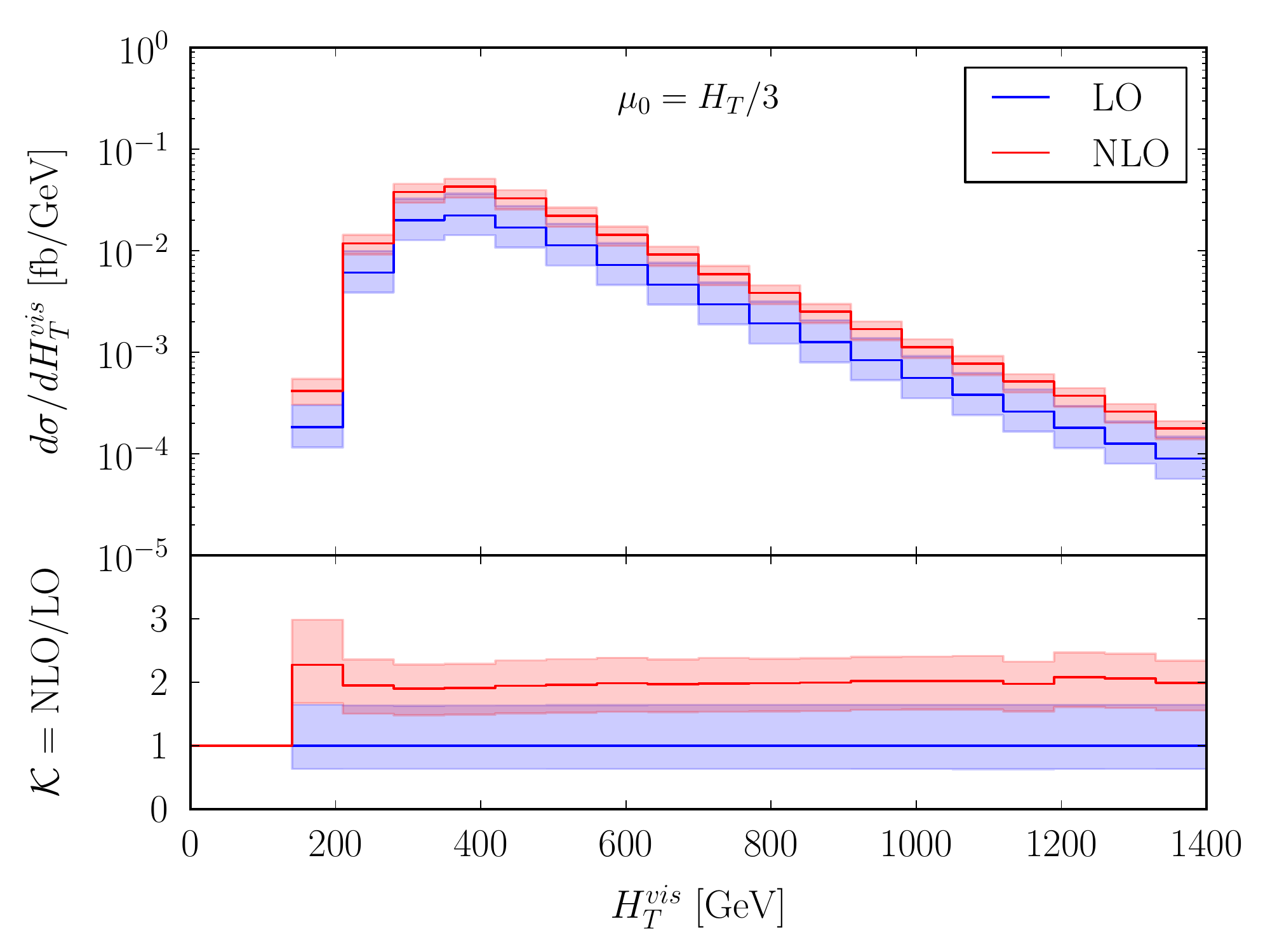}
    \includegraphics[width=0.49\textwidth]{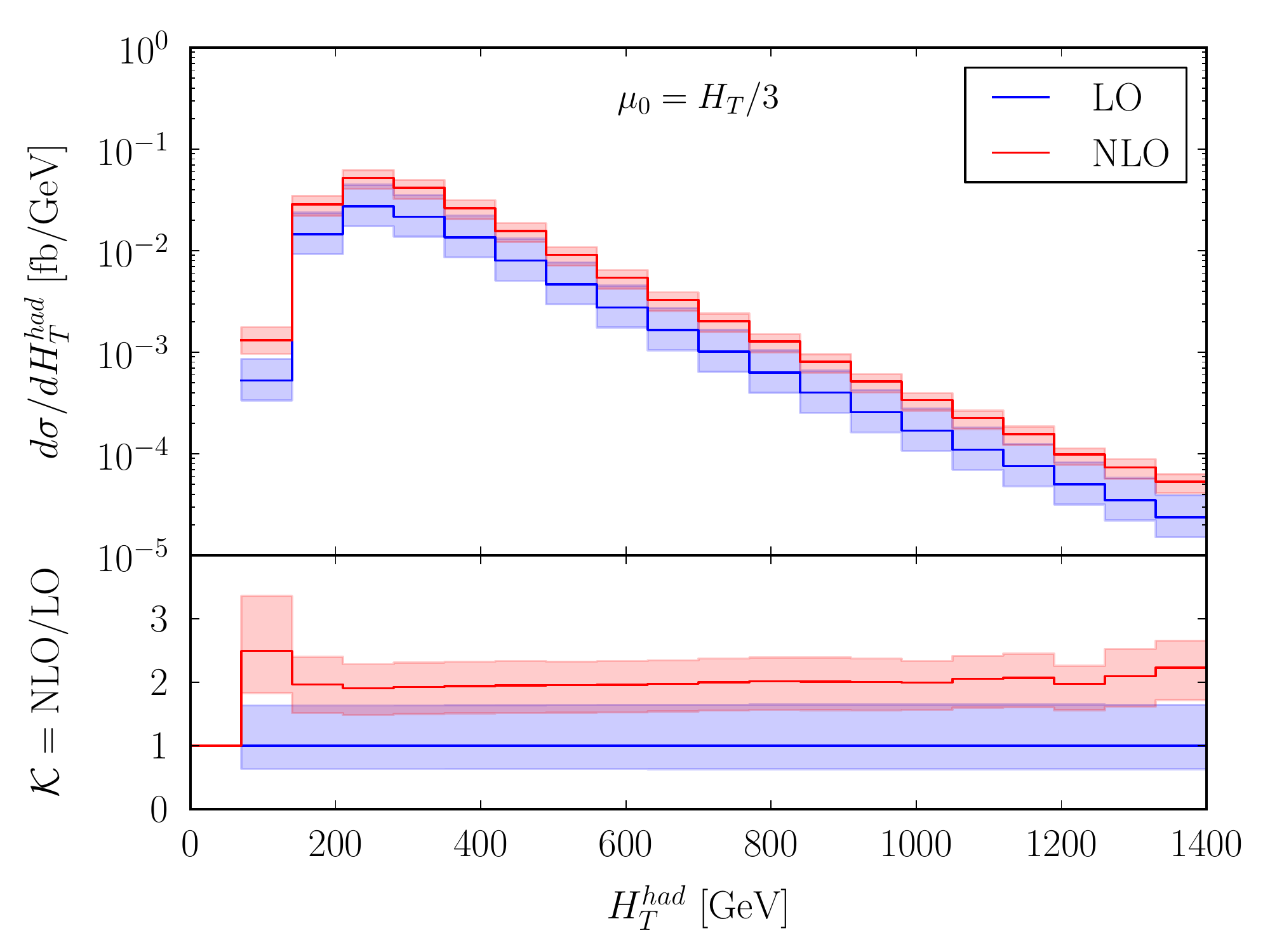}
    \includegraphics[width=0.49\textwidth]{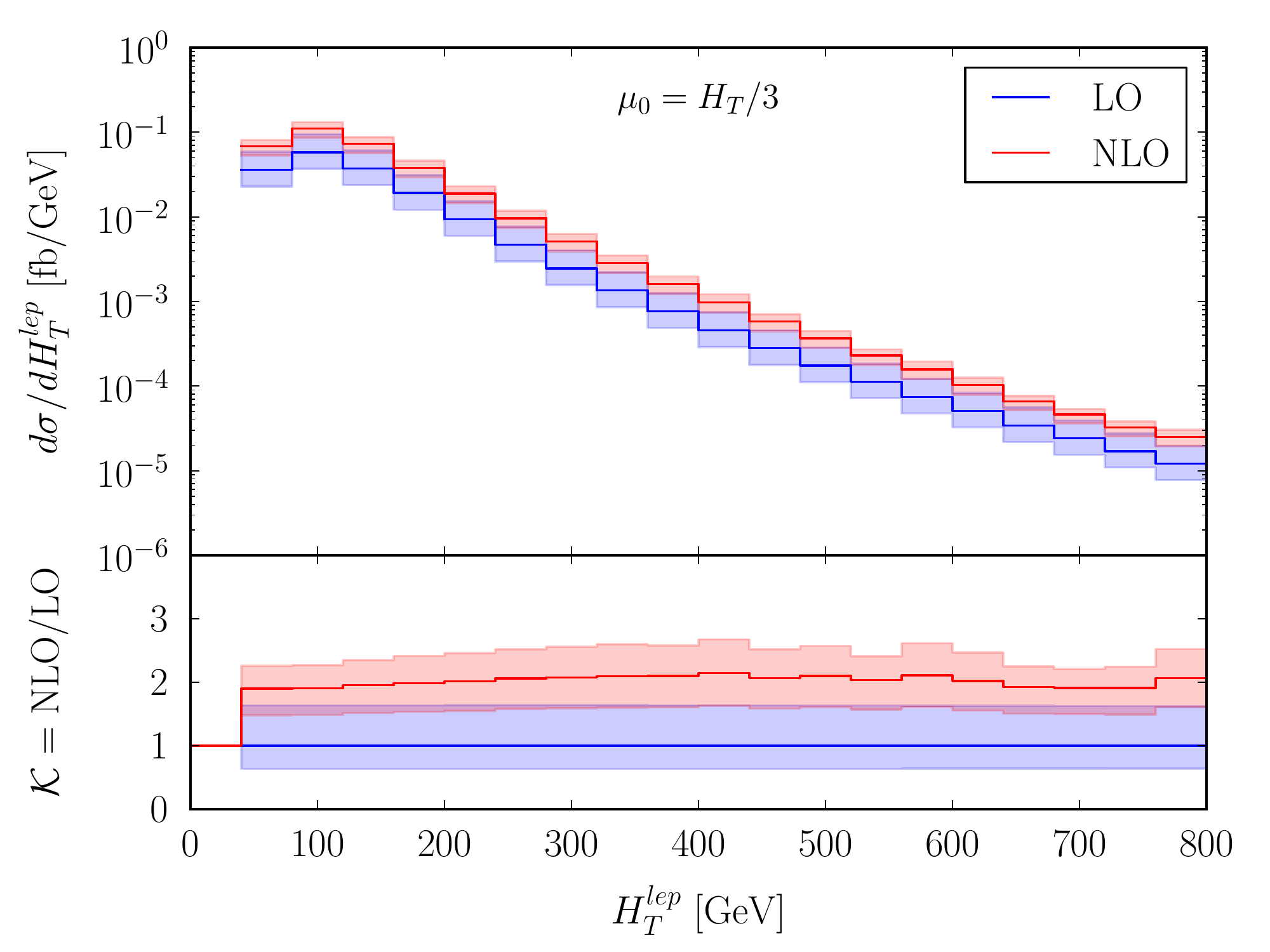}
  \end{center}
\caption{\label{fig:ht} \it As in Figure \ref{fig:ptb} but for the
  $H_T$, $H_T^{vis}$, $H_T^{had}$ and $H_T^{lep}$ distributions.}
\end{figure}
% =============================================
\begin{figure}[t!]
  \begin{center}
 \includegraphics[width=0.49\textwidth]{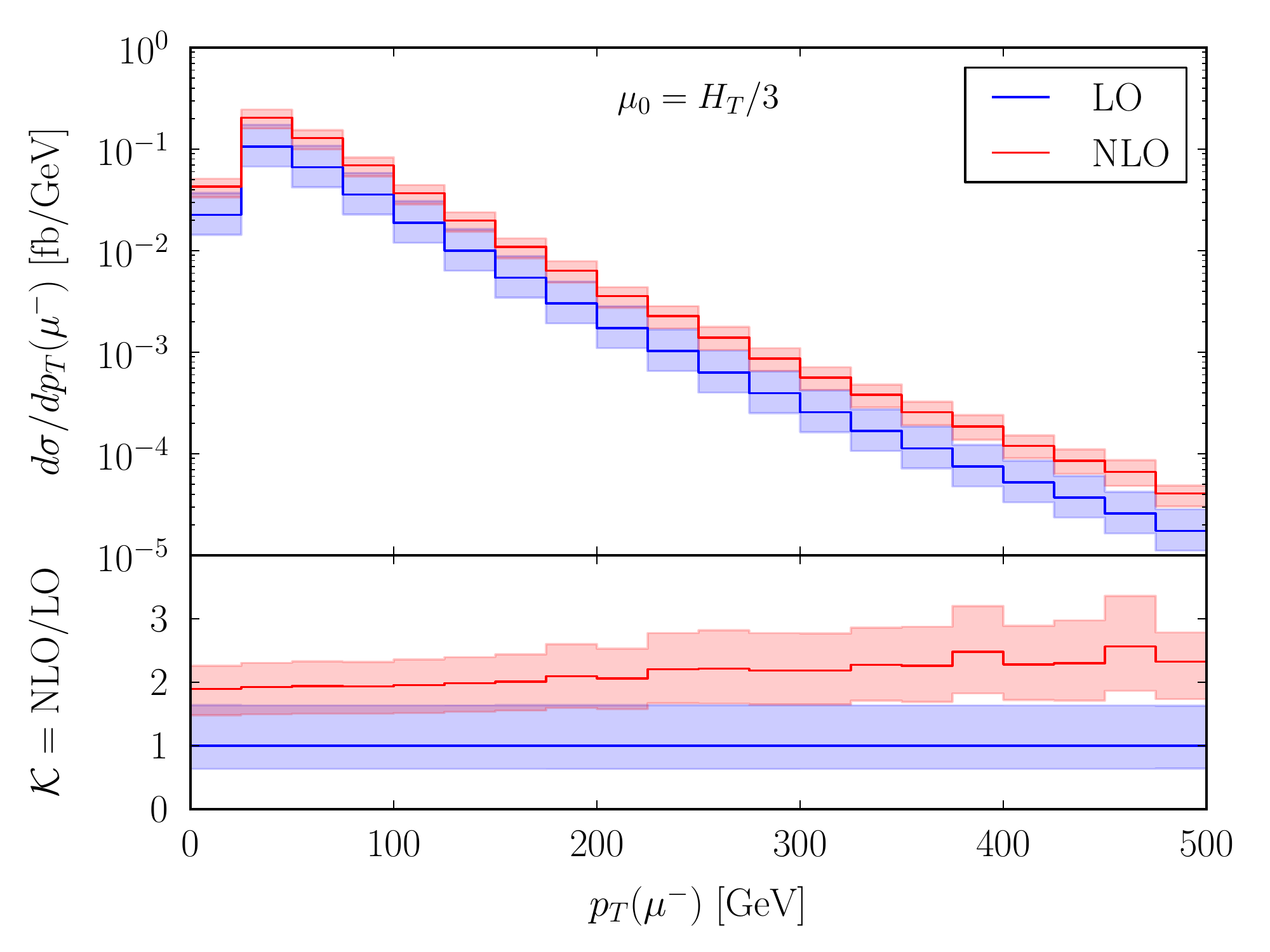}
 \includegraphics[width=0.49\textwidth]{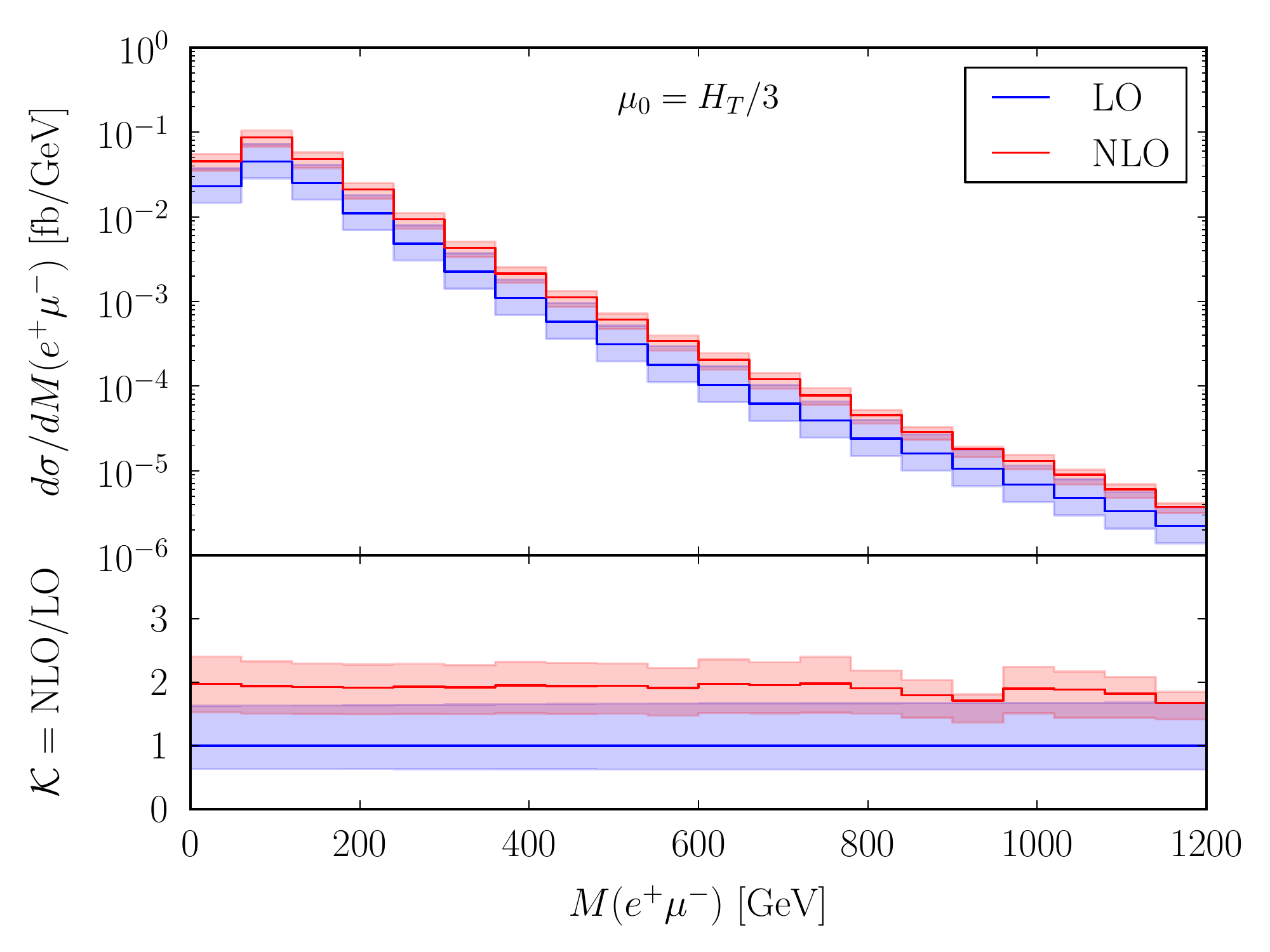}
  \includegraphics[width=0.49\textwidth]{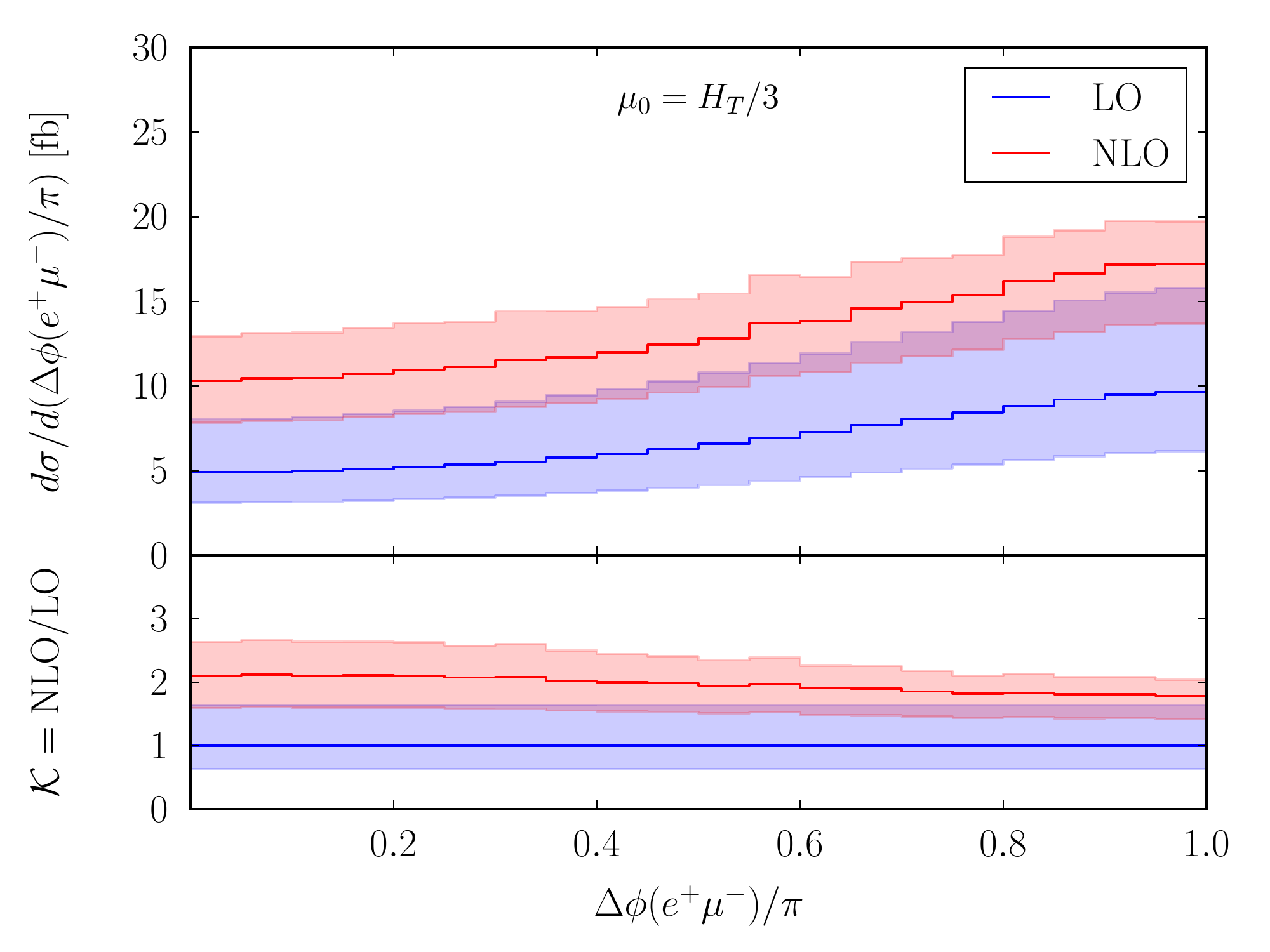}
  \includegraphics[width=0.49\textwidth]{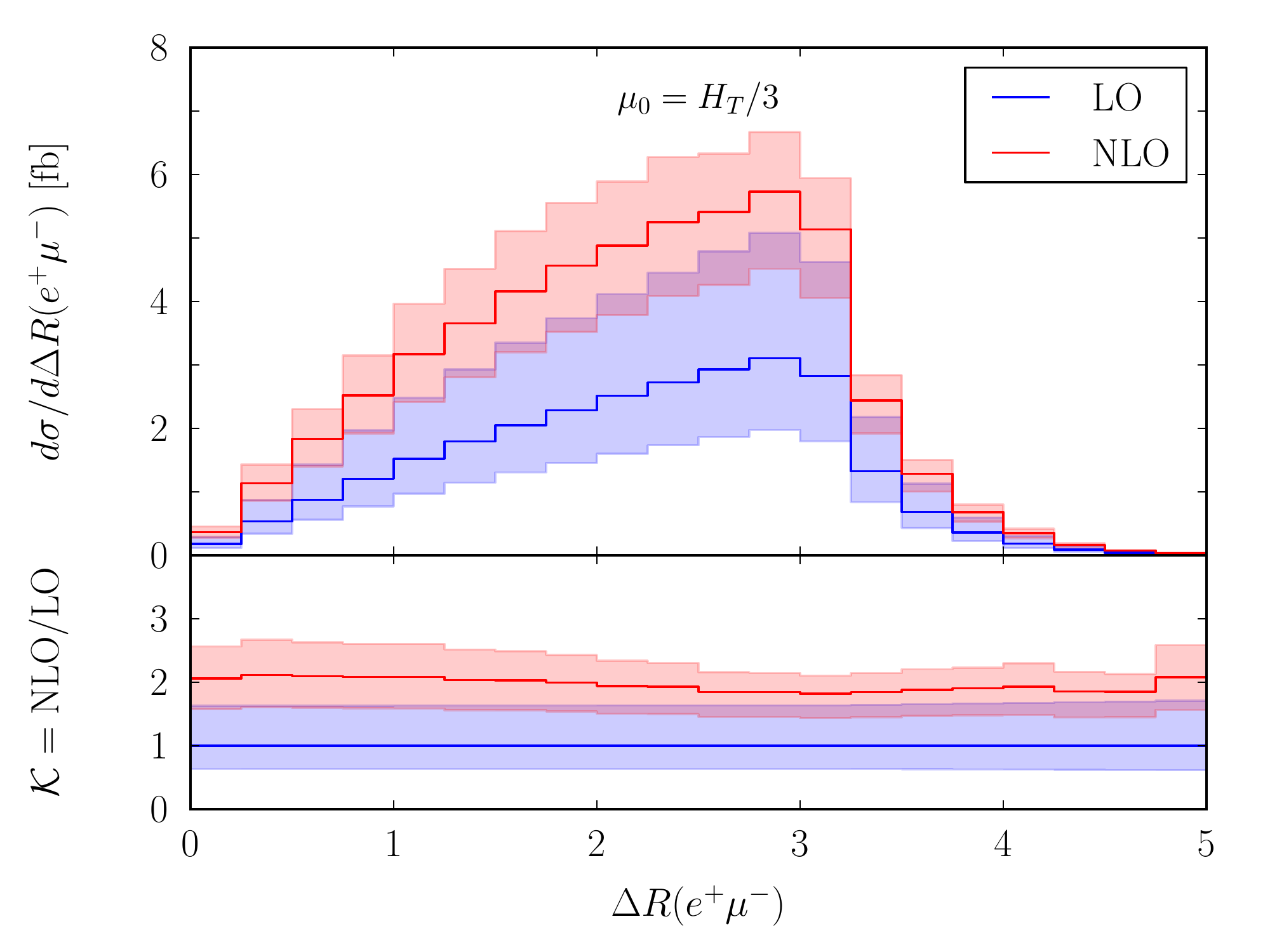}
  \end{center}
  \caption{\label{fig:lep} \it As in Figure \ref{fig:ptb} but for the
    $p_T(\mu^-)$, $M(e^+\mu^-)$, $\Delta \phi (e^+\mu^-)$ and  $\Delta
    R(e^+ \mu^-)$ distributions.}
\end{figure}
% =============================================

In Figure \ref{fig:b12-b34} we show the differential cross section
distribution as a function of the angular separation between the
$b$-jets, $\Delta R(bb)$. Also presented are the transverse momentum
and invariant mass of the $bb$ system. They are labelled as $p_T(bb)$
and $M(bb)$ respectively. Specifically, we display the two hardest
$b$-jets, denoted as $b_1b_2$, and the two softest $b$-jets, denoted
as $b_3b_4$. Looking at $\Delta R(b_1b_2)$ we can notice that the
$b_1b_2$ system originates predominately from top-quark pair
production as $b_1$ and $b_2$ are generated mostly in back-to-back
configurations. This is additionally confirmed by the $p_T(b_1b_2)$
and $M(b_1b_2)$ distributions, that have harder spectra in comparison
to the $b_3b_4$ system. The latter system is expected to receive large
contributions from gluon splittings as manifested by the enhancement at the
beginning of the $\Delta R(b_3b_4)$ distribution. However, we can also
notice rather large contributions in the back-to-back configurations
for $\Delta R(b_3b_4)$. This suggest that the simple picture that the
two high $p_T$ $b$-jets are from top-quark decays while the two low
$p_T$ $b$-jets, which are closest in $\Delta R(bb)$, are $b$-jets from
the $g\to b\bar{b}$ splitting may not apply. The reconstruction of the
production mechanisms for all final states is rather cumbersome when
multiple $b$-jets are present. It requires good reconstruction
techniques and excellent understanding of the modelling of top-quark
decays. The presence of the additional contributions either from
off-shell top quarks or from additional resolved light and/or $b$-jets
makes this picture even more complicated.  As we have mentioned
earlier we leave such studies on the identification of the origin of the
$b$-jets for the future. Instead, in the following we will focus on
the size of NLO QCD corrections to various observables constructed for
the $b_1b_2$ and $b_3b_4$ system. We underline here the fact that such
spectra are measured experimentally, see
e.g. Ref. \cite{Aaboud:2018eki}. When examining dimensionful
observables we notice that the $b_1b_2$ system receives larger
corrections up to even $150\%$. For $b_3b_4$, on the other hand,
corrections are more constant and between $80\%$ and $110\%$. For
dimensionless angular distributions NLO QCD effects are within the
$80\%-100\%$ range. All observables, except for $p_T(b_1b_2)$, have
moderate shape distortions. For the $p_T(b_1b_2)$ distribution,
however, they are even up to $80\%$. Had we used the fixed scale
choice also in this case we would rather obtain more than $100\%$
changes in the shape of the observables due to QCD higher order
corrections. Furthermore, for all observables moderate theoretical
uncertainties up to $25\%-30\%$ are observed.

An interesting set of observables is depicted in Figure
\ref{fig:ht}. They are the scalar sums built out of the transverse
momenta of the  various final states from the $pp \to e^+ \nu_e \,
\mu^- \bar{\nu}_\mu \, b\bar{b}\, b\bar{b} +X$ process. In particular,
we present $H_T$, $H_T^{vis}$, $H_T^{had}$ and $H_T^{lep}$. The first
one has already been defined as
\begin{equation}
  H_T = \sum_{i=1}^{4} p_T(b_i) + \sum_{i=1}^{2}
  p_T(\ell_i)+p_T^{miss}\,,
\end{equation}
where $\ell_{1,2}=e^+,\mu^-$ and
$p_T^{miss}$ is the missing transverse momentum from the two
neutrinos. We also have the visible, hadronic and leptonic versions of
$H_T$. The three observables  are measured experimentally and defined
as follows 
\begin{equation}
 H_T^{vis} = \sum_{i=1}^{4} p_T(b_i) +
 \sum_{i=1}^{2}p_T(\ell_i)\,, \quad \quad \quad \quad 
 H_T^{had} = \sum_{i=1}^{4} p_T(b_i) \,, \quad \quad \quad \quad 
 H_T^{lep} = \sum_{i=1}^{2}
  p_T(\ell_i)\,.
\end{equation}  
These observables have various kinematical thresholds. For example for
$H_{T}$ and $H_{T}^{vis}$ we have $H_{T, \, min}=H_{T,\,
min}^{vis}=140$ GeV as there is no restriction on the kinematics of
the missing transverse momentum. For the remaining two we can write
$H_{T,\, min}^{had}=100$ GeV and $H_{T,\, min}^{lep}=40$ GeV. 
$H_{T}$ observables receive rather constant NLO QCD corrections that are
of the order of $100\%$. Had we changed $\mu_0=H_T/3$ to the fixed
scale setting shape distortions up to almost $200\%$ would rather be
noticed. For all four versions of $H_T$ theoretical uncertainties are
of the order of $20\%-30\%$.

Although the main emphasis is on the understanding of the $b$-jet
kinematics, we can calculate higher order QCD corrections to any
IR-safe observable, which can be constructed from the available final
states, in the $pp \to e^+ \nu_e \, \mu^- \bar{\nu}_\mu \,
b\bar{b}\, b\bar{b} +X$ process.  Therefore, we can examine in detail
for example the two charged leptons. The advantage of the leptonic
observables over hadronic ones lies in the fact that measurements of
lepton observables are particularly precise at the LHC due to the
excellent lepton energy resolution of the ATLAS and CMS detectors.  In
Figure \ref{fig:lep} we present the following leptonic observables:
the transverse momentum of the muon $(p_T(\mu^-))$, the two leptons'
invariant mass $(M(e^+\mu^-))$, angular difference in the transverse
plane $(\Delta \phi(e^+\mu^-))$, and angular separation $(\Delta
R(e^+\mu^-))$. NLO QCD corrections to leptonic observables are also
substantial in the range of $80\%-150\%$. Theoretical uncertainties,
on the other hand, are again up to $20\%-25\%$ only.
%
% =============================================
\begin{figure}[t!]
  \begin{center}
     \includegraphics[width=0.49\textwidth]{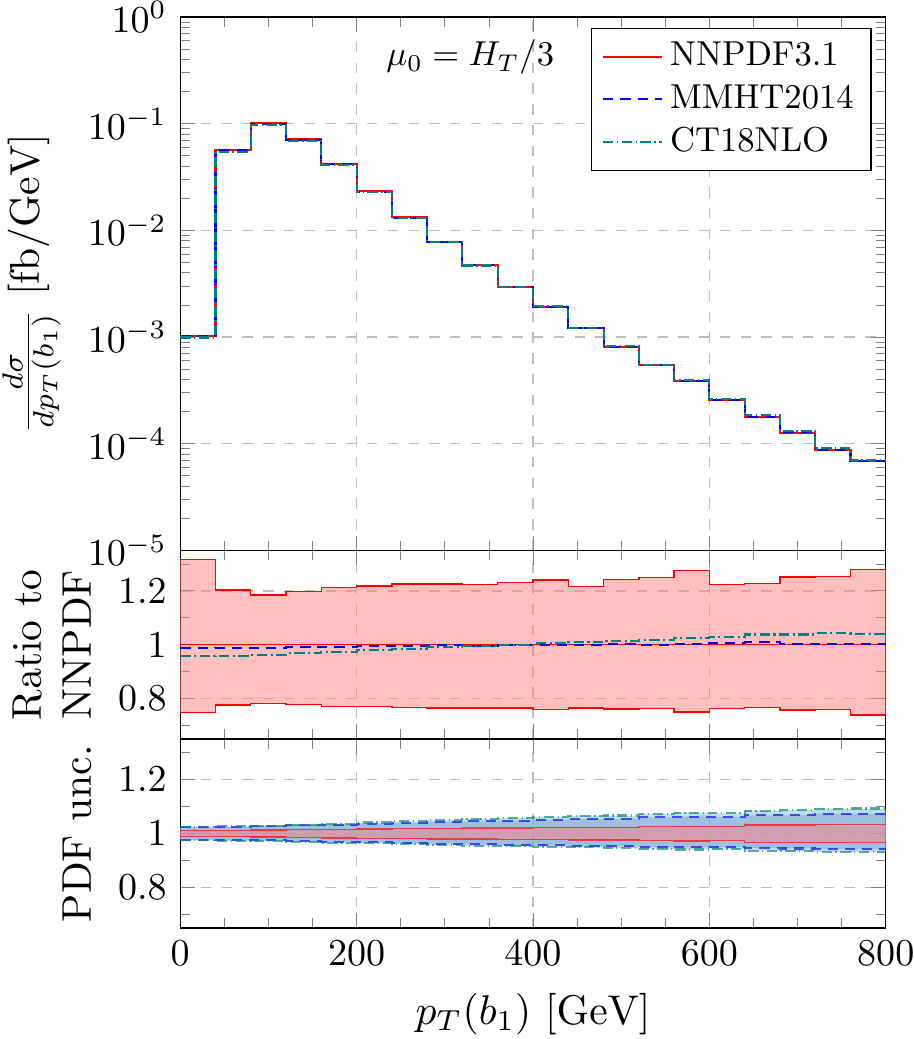}
    \includegraphics[width=0.49\textwidth]{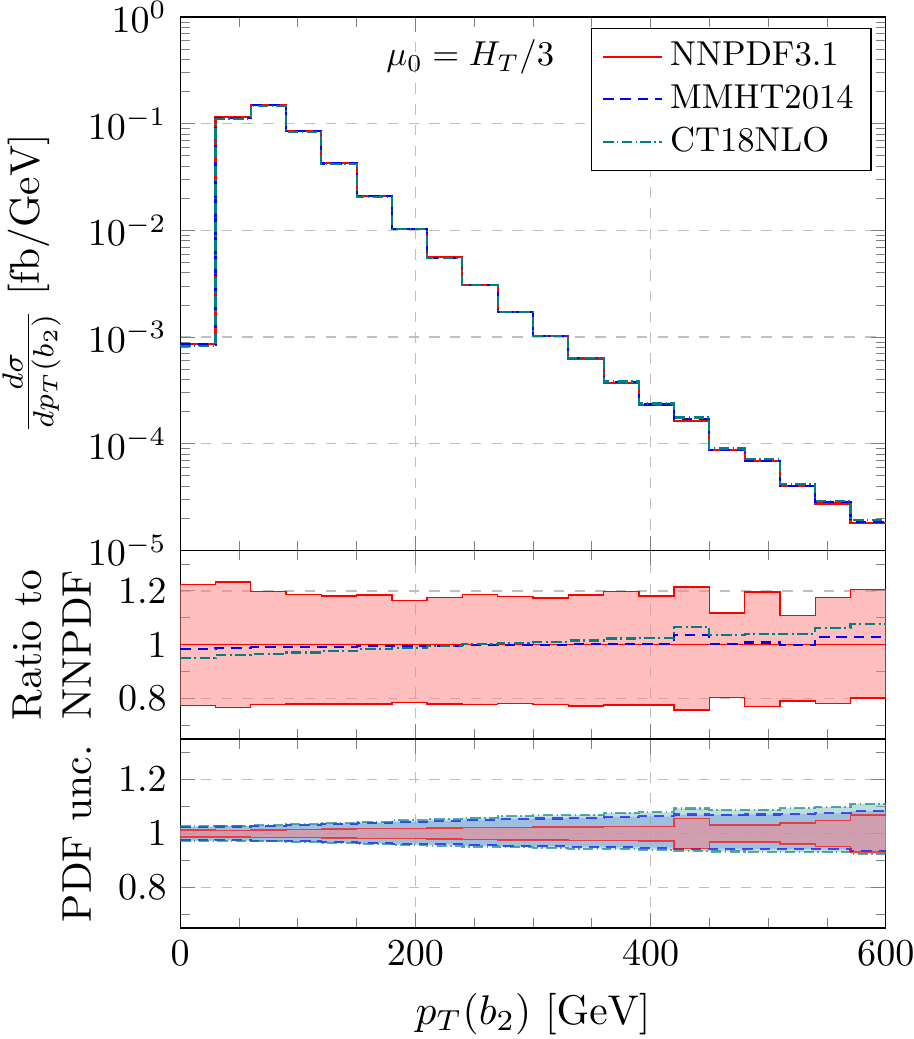}
    \includegraphics[width=0.49\textwidth]{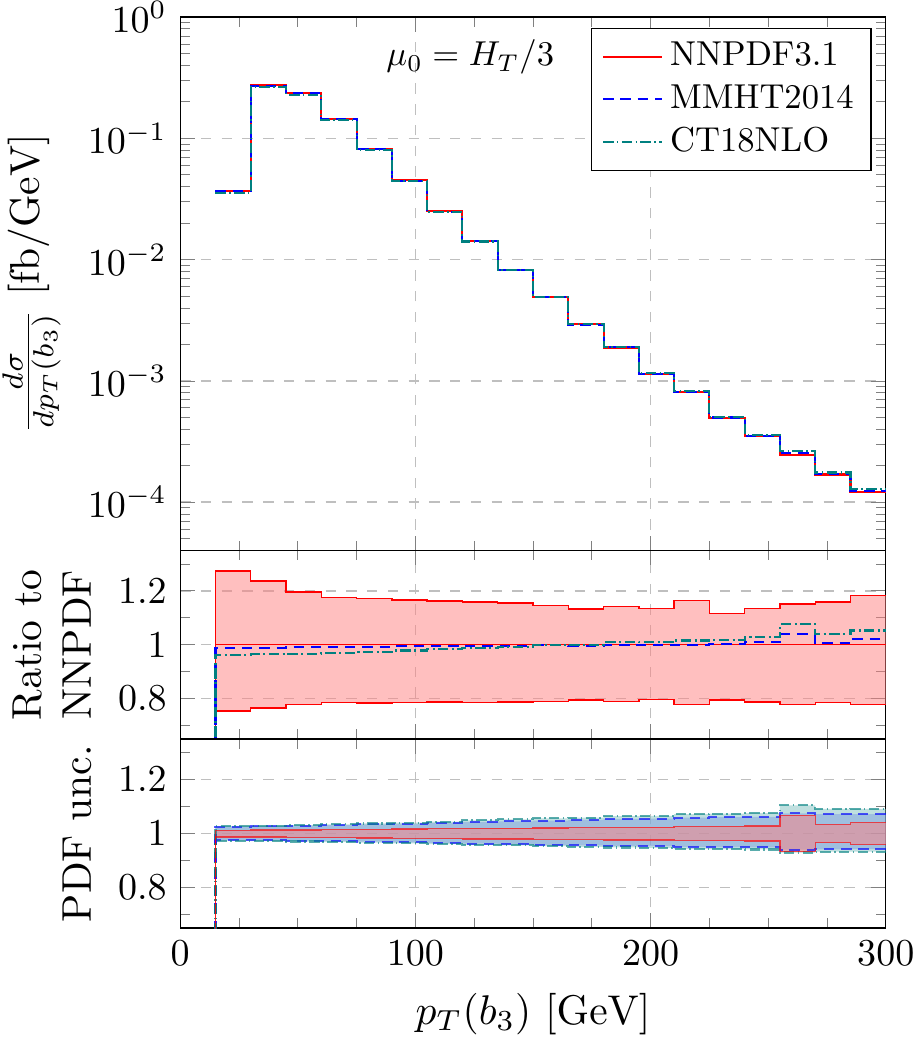}
    \includegraphics[width=0.49\textwidth]{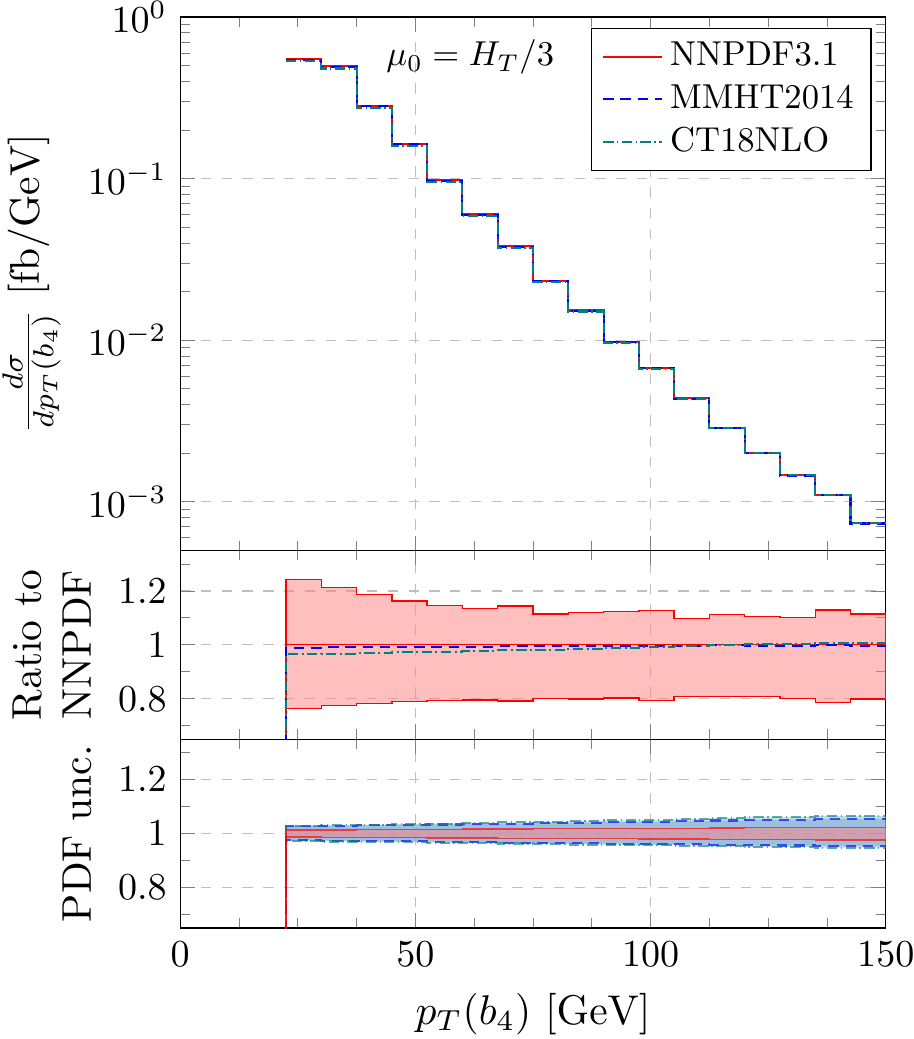}
  \end{center}
  \caption{\label{fig:pdfs} \it NLO differential cross section
    distributions as a function of $p_T(b_1)$, $p_T(b_2)$, $p_T(b_3)$
    and $p_T(b_4)$.  Results are shown for the $pp \to e^+ \nu_e \,
    \mu^- \bar{\nu}_\mu \, b\bar{b}\, b\bar{b} +X$ process at the LHC
    with $\sqrt{s}=13$ TeV. Three different PDF sets are
    employed. Lower panels display the theoretical uncertainties from
    scales and PDFs. All predictions are normalised to the central NLO
    prediction with the NNPDF3.1 PDF set.}
\end{figure}
% =============================================

As already pointed out the theoretical uncertainty related to the lack
of next-to-next-to-leading-order QCD corrections is only a fraction of
the overall theoretical systematics. To complete our analysis we study
differential cross section uncertainties due to the PDF
parameterisation including the latest fits by several groups,
i.e. NNPDF3.1, MMHT2014 and CT18NLO. We have comparatively examined
the impact of PDFs and scale variations on the overall theoretical
uncertainty for differential cross section distributions for the $pp
\to e^+ \nu_e \, \mu^- \bar{\nu}_\mu \, b\bar{b}\, b\bar{b} +X$
process. For all observables that we have examined we could confirm
that the PDF uncertainties are small and well below the theoretical
uncertainties predicted by scale variation.  As an example in Figure
\ref{fig:pdfs} we present NLO differential cross section distributions
as a function of the transverse momentum of the $1^{st}$, $2^{nd}$,
$3^{rd}$ and $4^{th}$ hardest $b$-jet. Upper panels show the absolute
NLO predictions for three different PDF sets as obtained with the
dynamical scale setting.  Middle panels display the relative scale
uncertainties of NLO predictions for the NNPDF3.1 PDF set. Also
presented are the relative predictions for the MMHT2014 and CT18NLO
PDF set. They are obtained by normalisation to the central NLO
prediction with the NNPDF3.1 PDF set. Finally, lower panels display
the relative internal PDF uncertainties of the NLO cross section for
each of the three PDF sets. Also here all NLO predictions are
normalised to the central NLO prediction with the NNPDF3.1 PDF set. A
few comments are in order. To start with, at the differential level
the internal PDF uncertainties are up to $11\%$ for CT18NLO and of the
order of $8\%$ for MMHT2014.  For the default NNPDF3.1 PDF set they
are in the $1\%-7\%$ range. When comparing the NLO QCD results as
obtained with either CT18NLO or MMHT2014 to the findings generated for
the default NNPDF3.1 PDF set the relative differences are up to $8\%$
and $3\%$ respectively. Thus, also at the differential level the
internal NNPDF3.1 PDF uncertainties are similar in size to differences
between the PDF sets that are recommended for applications at the LHC
Run II \cite{Butterworth:2015oua}.

To summarise this part, for the $pp \to e^+ \nu_e \, \mu^-
\bar{\nu}_\mu \, b\bar{b}\, b\bar{b} +X$ production process at the LHC
for a centre-of-mass-system energy of $\sqrt{s}=13$ TeV and with
rather inclusive selection cuts not only big NLO QCD corrections but
also significant shape changes are visible when going from LO to
NLO. This confirms that NLO QCD effects to this process are extremely
important. The theoretical uncertainties due to scale dependence for
$\mu_0=H_T/3$ are rather moderate of the order of $20\%-30\%$. For the
fixed scale setting they are much higher.  The uncertainties due to
the NNPDF3.1 PDF parameterisation are small, i.e. in the $1\%-7\%$ 
range. Overall, when various PDF sets are examined the PDF
uncertainties are maximally up to $11\%$. Consequently, the dominant
component of the final theoretical error for the $pp \to e^+ \nu_e \,
\mu^- \bar{\nu}_\mu \, b\bar{b}\, b\bar{b} +X$ process is determined
by the scale dependence.

% =============================================
%
\section{Contribution of initial state bottom quarks}
\label{sec:ttbb-initial-b}
% 
% =============================================

In the next step we study the contributions that are induced by the
initial state bottom quarks.  To this end, additional subprocesses are
included in the calculation. For the LO part we add $b\bar{b} \to
e^+\nu_e\, \mu^-\bar{\nu}_\mu\, b\bar{b}\,b\bar{b}$, $ bb \to
e^+\nu_e\, \mu^-\bar{\nu}_\mu\, b\bar{b}\,bb$ and $\bar{b}\bar{b}\to
e^+\nu_e\, \mu^-\bar{\nu}_\mu\, b\bar{b}\, \bar{b}\bar{b}$. The three
reactions are related by crossing symmetry, thus, each comprises
$2790$ Feynman diagrams. The last two subprocesses must be taken into
account as they might not necessarily be suppressed other than by
PDFs. Indeed, they are already part of the double-resonant
contribution to the $t\bar{t}$ process with additional two $b$-jets. For
the real emission part, on the other hand, the following subprocesses
are included $ b\bar{b} \to e^+\nu_e\, \mu^-\bar{\nu}_\mu\,
b\bar{b}\,b\bar{b}\, g$, $gb \to e^+\nu_e\, \mu^-\bar{\nu}_\mu\,
b\bar{b}\,b\bar{b}\, b $, $g\bar{b} \to e^+\nu_e\,
\mu^-\bar{\nu}_\mu\, b\bar{b}\,b\bar{b}\, \bar{b}$, $bb\to e^+\nu_e\,
\mu^-\bar{\nu}_\mu\, b\bar{b}\,bb \, g$ and $\bar{b}\bar{b} \to
e^+\nu_e\, \mu^-\bar{\nu}_\mu\,b\bar{b}\, \bar{b}\bar{b} \,g$. Each
subprocess comprises $28728$ Feynman diagrams. We continue to use the
$anti-k_T$ jet algorithm. However a small modification is needed for
the jet flavour assignment. As we are dealing with massless bottom
quarks from the theoretical point of view the important parton
recombination rules, that are required to guarantee the IR-safety of
the jet algorithm, are $b g \to b$, $\bar{b}g\to \bar{b}$ and
$b\bar{b}\to g$. We need, however, additional recombination rules for
$bb$ and $\bar{b}\bar{b}$. We employ two variants that are IR-safe at
NLO. Beyond NLO IR-safety requires the algorithms of
Ref. \cite{Banfi:2006hf,Buckley:2015gua}.

\subsubsection*{Charge-blind $\boldsymbol{b}$-jet tagging}

In the first case we use the
charge-blind $b$-jet tagging. From the experimental point of view
$b$-jet tagging algorithms are sensitive mostly to the absolute flavour
and they do not additionally tag the charge of the $b$-jet. In the
absence of charge tagging any combination that contains an even
number of $b$ and/or $\bar{b}$ quarks should also be considered to
carry zero flavour as from the experimental point of view such
signatures will not be distinguishable from $b\bar{b}\to g$. In this
case the complete  set of  recombination rules for heavy-flavour  jets 
is  given by
\begin{equation}
  bg\to b\,, \quad \quad \quad \quad
  \bar{b} g \to \bar{b}\,, \quad \quad \quad \quad
  b\bar{b} \to g \,, \quad \quad \quad \quad
  bb \to g\,,\quad \quad \quad \quad
  \bar{b}\bar{b}\to g\,.
\end{equation}  
We ask for at least four $b$-jets in the final state and check whether
there is at least one combination of (any) four $b$-jets that fulfils
the required cuts.  We shall refer to this approach as the
charge-blind $b$-jet tagging or in short as the {\it charge-blind
scheme} in the following.

\subsubsection*{Charge-aware $\boldsymbol{b}$-jet tagging}

In the second case, we assume that the charge tagging of $b$-jets is
possible, see e.g. Ref.
\cite{Krohn:2012fg,TheATLAScollaboration:2015ggd,Tokar:2017syr,
ATLAS-CONF-2018-022}. From the experimental point of view the
disadvantage of this approach might lie in the possibility of the
reduction in the $b$-jet tagging efficiency and in smaller event
statistics. In this case the following recombinations rules are
employed:
\begin{equation}
  bg\to b\,, \quad \quad \quad \quad
  \bar{b} g \to \bar{b}\,, \quad \quad \quad \quad
  b\bar{b} \to g \,, \quad \quad \quad \quad
  bb \to b\,,\quad \quad \quad \quad
  \bar{b}\bar{b}\to \bar{b}\,.
\end{equation}  
We ask for at least four $b$-jets in the final state, however, this
time we check whether any combination with zero total charge, i.e. any
$b\bar{b}b\bar{b}$ combination, passes the cuts. In this scheme there
is no need to consider the $bb$ and $\bar{b}\bar{b}$ initiated
subprocess.  We shall refer to this approach as the charge-aware
$b$-jet tagging or in short as the {\it charge-aware scheme}.
%
% =============================================
\begin{table}[t!]
\begin{center}
\begin{tabular}{ccccc}
  \hline \hline\\[-0.4cm]
  \textsc{Scale} & $\sigma^{\rm LO}_{gg}$ [fb]
  & $\sigma^{\rm LO}_{q\bar{q}+\bar{q}q}$ [fb]
  & $\sigma^{\rm LO}_{b\bar{b}+\bar{b}b}$ [fb]
  & $\sigma^{\rm LO}_{bb + \bar{b}\bar{b}}$ [fb]
 \\[0.2cm]
   \hline
  \hline\\[-0.4cm]
  $\mu_0=m_t$      & 6.561(2) & 0.4367(1)
  & 0.008607(7) & 0.006184(8)\\[0.2cm]
  $\mu_0=H_T/3$ & 6.404(3) & 0.4092(1)
  & 0.008428(7) & 0.006005(7)\\[0.2cm]
    \hline
  \hline
\end{tabular}
\end{center}
\caption{\label{tab:b_contribution_lo} \it LO
integrated fiducial cross section for the $pp\to e^+ \nu_e \,\mu^-
\bar{\nu}_\mu\, b\bar{b}\,b\bar{b}+X$ production process at the LHC
with $\sqrt{s} = 13$ TeV for $\mu_0= m_t$ and $\mu_0=H_T/3$.
Theoretical results with and without the initial state $b$
contributions are provided for the LO NNPDF3.1 PDF set.  Also given are
Monte Carlo integration errors (in parenthesis).}
\end{table}
% =============================================
\begin{table}[t!]
\begin{center}
  \begin{tabular}{cccccccc}
    \hline\hline\\[-0.4cm]
 $p_{T}^{\rm veto}(j)$ &$p_{T}(b)$ & $\sigma^{\rm NLO}_{\rm no\,b}$  [fb]
      & $\sigma^{\rm NLO}_{\rm ch-aware}$  [fb]
 & $\sigma^{\rm NLO}_{\rm  ch-blind}$ [fb]
  &  $\delta_{\rm scale}$ &
 $\delta_{\rm PDF}$ \\[0.2cm]
   \hline \hline\\[-0.4cm]
\multicolumn{7}{c}{$\mu_R=\mu_F=\mu_0=m_t$} \\[0.2cm]
    \hline \hline\\[-0.4cm] 
    & $25$ & $13.24(3)$ & $13.33(3)$ & $13.41(3)$ 
 & $^{\,+18\%}_{\,-22\%}$ & $^{\,+1\%}_{\,-1\%}$  \\[0.2cm]
  & $30$ & $9.25(2)$ & $9.32(2)$ & $9.37(2)$ & $^{\,+14\%}_{\,-21\%}$ &
 $^{\,+2\%}_{\,+2\%}$\\[0.2cm]
  & $35$ & $6.57(1)$ & $6.62(1)$ & $6.66(1)$ & $^{\,+12\%}_{\,-20\%}$
  & $^{\,+2\%}_{\,-2\%}$ \\[0.2cm]
  & $40$ & $4.70(1)$ & $4.74(1)$ & $4.77(1)$ & $^{\,+10\%}_{\,-19\%}$
  & $^{\,+2\%}_{\,-2\%}$ \\[0.2cm]
  \hline 
  \hline\\[-0.4cm]
  $100$ & $25$ & $10.37(3)$ & $10.46(3)$ & $10.53(3)$ &
  $^{\,+3\%}_{\,-16\%}$
  & $^{\,+1\%}_{\,-1\%}$  \\[0.2cm]
  \hline
  \hline\\[-0.4cm]
  $50$& $25$ & $7.77(3)$ & $7.85(3)$ & $7.93(3)$ &
  $^{\,+3\%}_{\,-33\%}$
  & $^{\,+1\%}_{\,-1\%}$ \\[0.2cm]
    \hline
  \hline\\[-0.4cm]
  \multicolumn{7}{c}{$\mu_R=\mu_F=\mu_0=H_T/3$} \\[0.2cm]
     \hline \hline \\[-0.4cm]
   & $25$ & $13.22(3)$ & $13.31(3)$ & $13.38(3)$ & $^{\,+20\%}_{\,-22\%}$
   & $^{\,+1\%}_{\,-1\%}$  \\[0.2cm]
   & $30$ & $9.09(2)$ & $9.16(2)$ & $9.21(2)$ & $^{\,+18\%}_{\,-22\%}$
   &  $^{\,+2\%}_{\,-2\%}$  \\[0.2cm]
   & $35$ & $6.37(1)$ & $6.42(1)$ & $6.46(1)$ & $^{\,+17\%}_{\,-21\%}$
   & $^{\,+2\%}_{\,-2\%}$  \\[0.2cm]
   & $40$ & $4.51(1)$ & $4.54(1)$ & $4.57(1)$ & $^{\,+16\%}_{\,-21\%}$
   & $^{\,+2\%}_{\,-2\%}$  \\[0.2cm]
   \hline 
  \hline\\[-0.4cm]
   $100$ & $25$ & $10.77(3)$ & $10.86(3)$ & $10.94(3)$
  & $^{\,+8\%}_{\,-18\%}$
  & $^{\,+1\%}_{\,-1\%}$  \\[0.2cm]
   \hline
  \hline\\[-0.4cm]
   $50$& $25$ & $8.35(3)$ & $8.43(3)$ & $8.51(3)$ & $^{\,+1\%}_{\,-18\%}$
  & $^{\,+1\%}_{\,-1\%}$  \\[0.2cm]
   \hline
  \hline
\end{tabular}
\end{center}
\caption{\label{tab:b_contribution} \it NLO integrated fiducial cross
section for the $pp\to e^+ \nu_e \,\mu^- \bar{\nu}_\mu\,
b\bar{b}\,b\bar{b}+X$ production process at the LHC with $\sqrt{s} =
13$ TeV for $\mu_0= m_t$ and $\mu_0=H_T/3$.  Theoretical results with
and without the initial state $b$ contributions are provided for the
NLO NNPDF3.1 PDF set. Results for four different values
of the $p_T(b)$ cut and for a jet veto with $p_T^{\rm
veto}(j)=50,100$ GeV are additionally provided. Also given are the
theoretical uncertainties due to scale variation $(\delta_{scale})$ and
PDFs $(\delta_{\rm PDF})$. Finally,   Monte Carlo integration
errors are shown (in parenthesis). }
\end{table}
% =============================================

In the following the difference between the two approaches is
examined. We investigate their impact on the full $pp$ cross sections
at LO and NLO in QCD. Specifically, we compare the two schemes to the
case when the initial state bottom-quark contribution is omitted.  At
LO the two approaches are equivalent because we always have
exactly the four $b$-jets that need to pass the cuts. Furthermore,
at this order both schemes differ only by the initial state subprocesses
that must be taken into account. Simply, in the charge-aware scheme
the $bb$ and $\bar{b}\bar{b}$ subprocesses are not considered.

In Table \ref{tab:b_contribution_lo} we present contributions to the LO
$t\bar{t}b\bar{b}$ cross section subprocess by subprocess. As already
stipulated $\sigma^{\rm LO}_{b\bar{b}+\bar{b}b}$ and $\sigma^{\rm
LO}_{bb+\bar{b}\bar{b}}$ are similar in size. For
$\mu_0=m_t$ the full $pp$ LO integrated cross section with the
$b$-initiated contributions included is given by
\begin{equation}
    \sigma^{\rm LO}_{\rm charge-aware} ({\rm
      NNPDF3.1},
    \mu_0=m_t) = 7.006(2)\, {\rm fb}\,,
\end{equation} 
for the charge-aware scheme and reads 
\begin{equation}
    \sigma^{\rm LO}_{\rm charge-blind} ({\rm
      NNPDF3.1},
    \mu_0=m_t) = 7.012(2)\, {\rm fb}\,,
\end{equation} 
for the charge-blind one. The two findings can be compared to the
result where the initial state $b$ contributions are neglected
\begin{equation}
    \sigma^{\rm LO}_{\rm no-b} ({\rm
      NNPDF3.1},
    \mu_0=m_t) = 6.998(2)\, {\rm fb}\,.
\end{equation} 
Similar results are obtained for $\mu_0=H_T/3$. We can conclude that
at LO the initial state $b$-quark contributions are at the
$0.1\%-0.2\%$ level independently of the scale choice and the $b$-jet
tagging scheme. Thus, they can be safely neglected.

In Table \ref{tab:b_contribution} we display our findings at NLO in
QCD. The theoretical results are given for both scale choices
$\mu_0=m_t$ and $\mu_0=H_T/3$. For $\mu_0=m_t$ the initial state
bottom-quark contributions are of the order of $1.3\%$. The variation
in the $p_T(b)$ cut in the range of $25-40$ GeV has increased these
contributions up to $1.5\%$. A similar increase can be obtained with the
imposition of a jet veto on the fifth jet (upper bound on the allowed
transverse momentum) of $100$ GeV. On the other hand, for $p_T^{\rm
veto} (j) =50$ GeV these contributions have reached $2\%$. Results for
the dynamical scale choice are very similar. Consequently, the
bottom-quark initiated contributions can be safely omitted when
compared to the theoretical error for this process. The latter is at
the $20\%-22\%$ level.

A few comments are in order. If we look at the bottom-quark
initiated subprocesses only, their combined contribution at NLO is
dominated by the $bg/\bar{b}g$ initial states. Furthermore, at this
level there are substantial differences between the two
schemes. Specifically, for $\mu_0=m_t$ we have for the charge-aware
and charge-blind scheme the following results
\begin{equation}
  \begin{split}
    \sigma^{\rm NLO}_{\rm b-initated, \, charge-aware} ({\rm
      NNPDF3.1},
    \mu_0=m_t)&= 0.0954(2) \, {\rm fb}\\[0.2cm]
    \sigma^{\rm NLO}_{\rm b-initated, \, charge-blind}  ({\rm
      NNPDF3.1},
    \mu_0=m_t)&= 0.1684(4)\,
    {\rm fb}\,,
    \end{split}
\end{equation} 
where in bracket the MC error is additionally provided. It is only
when the dominant $gg$ partonic subprocess is taken into account that
the differences between the two schemes become
insignificant. Furthermore, having at hand the results with an
additional jet veto on the fifth jet we can make the following remarks
on the size of the higher order QCD effects.  We observe that the jet
veto can reduce the ${\cal K}$-factor for the default set of
cuts. Specifically, for $\mu_0=m_t$ $(\mu_0=H_T/3)$ when $p_T^{\rm
veto} (j) =100$ GeV is used we obtain ${\cal K}=1.48$ $({\cal
K}=1.58)$, whereas for $p_T^{\rm veto} (j) =50$ GeV we have ${\cal
K}=1.11$ $({\cal K}=1.23)$.
%
% =============================================
\begin{figure}[t!]
  \begin{center}
  \includegraphics[width=0.49\textwidth]{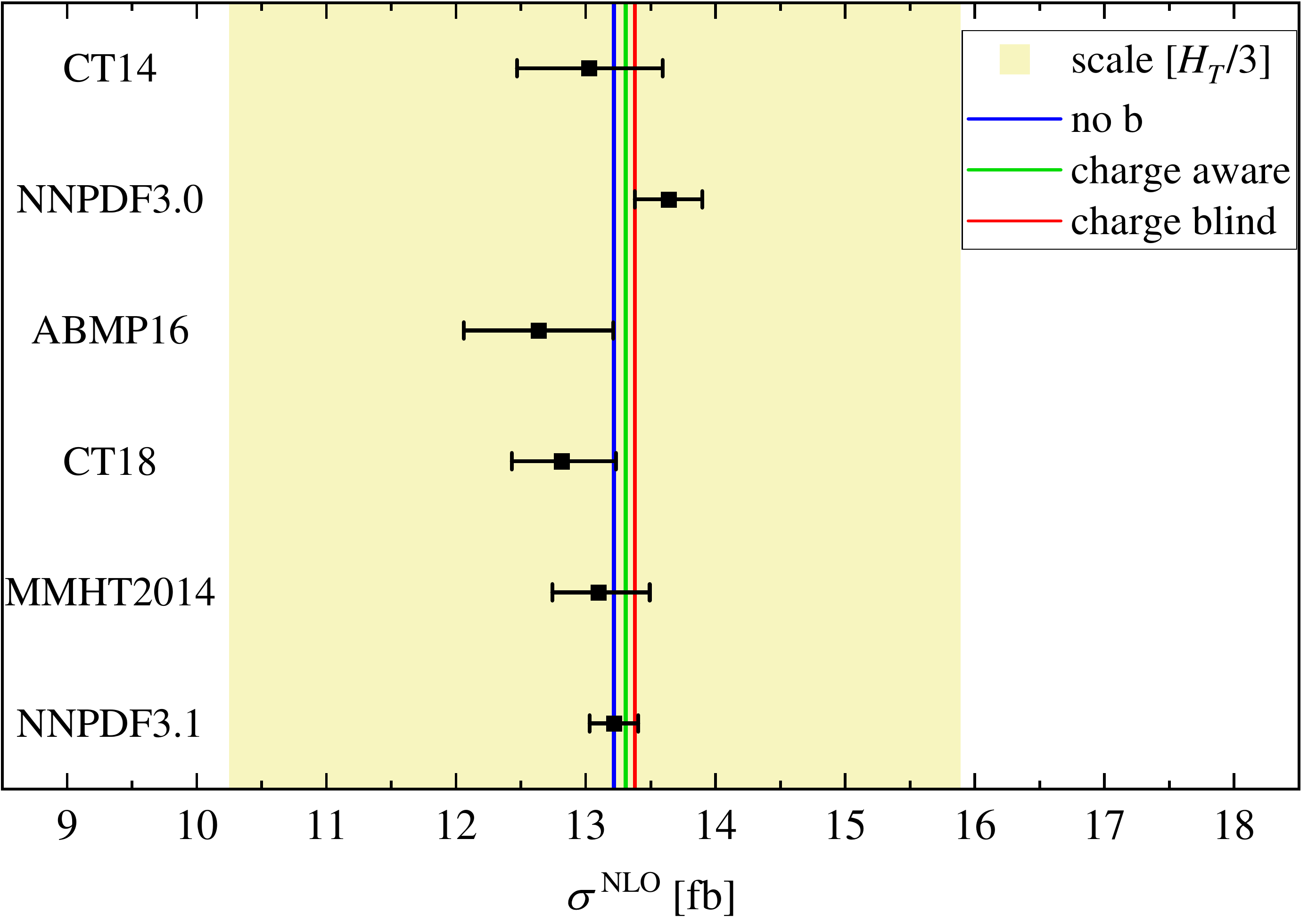}
  \end{center}
\caption{\label{fig:b-contribution} \it NLO
integrated fiducial cross section for the $pp\to e^+ \nu_e \,\mu^-
\bar{\nu}_\mu\, b\bar{b}\,b\bar{b}+X$ production process at the LHC
with $\sqrt{s} = 13$ TeV for $\mu_0= H_T /3$. Theoretical results
with and without the initial state $b$-quark contributions are
provided for the NLO NNPDF3.1 PDF set. Also given are results with
other PDF sets for the case with no initial state bottom-quark
contributions. }
\end{figure}
% =============================================
\begin{figure}[t!]
  \begin{center}
     \includegraphics[width=0.49\textwidth]{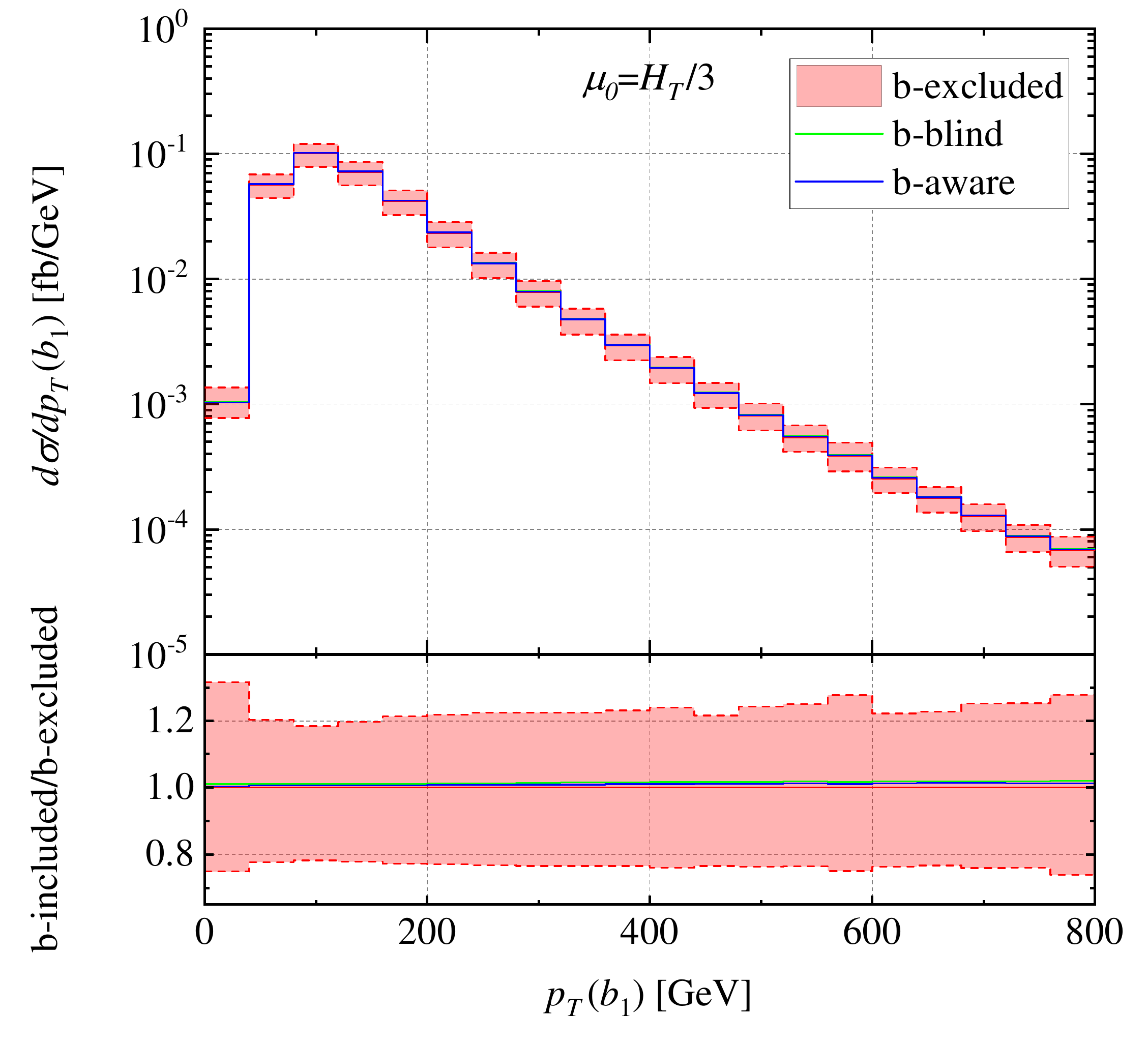}
    \includegraphics[width=0.49\textwidth]{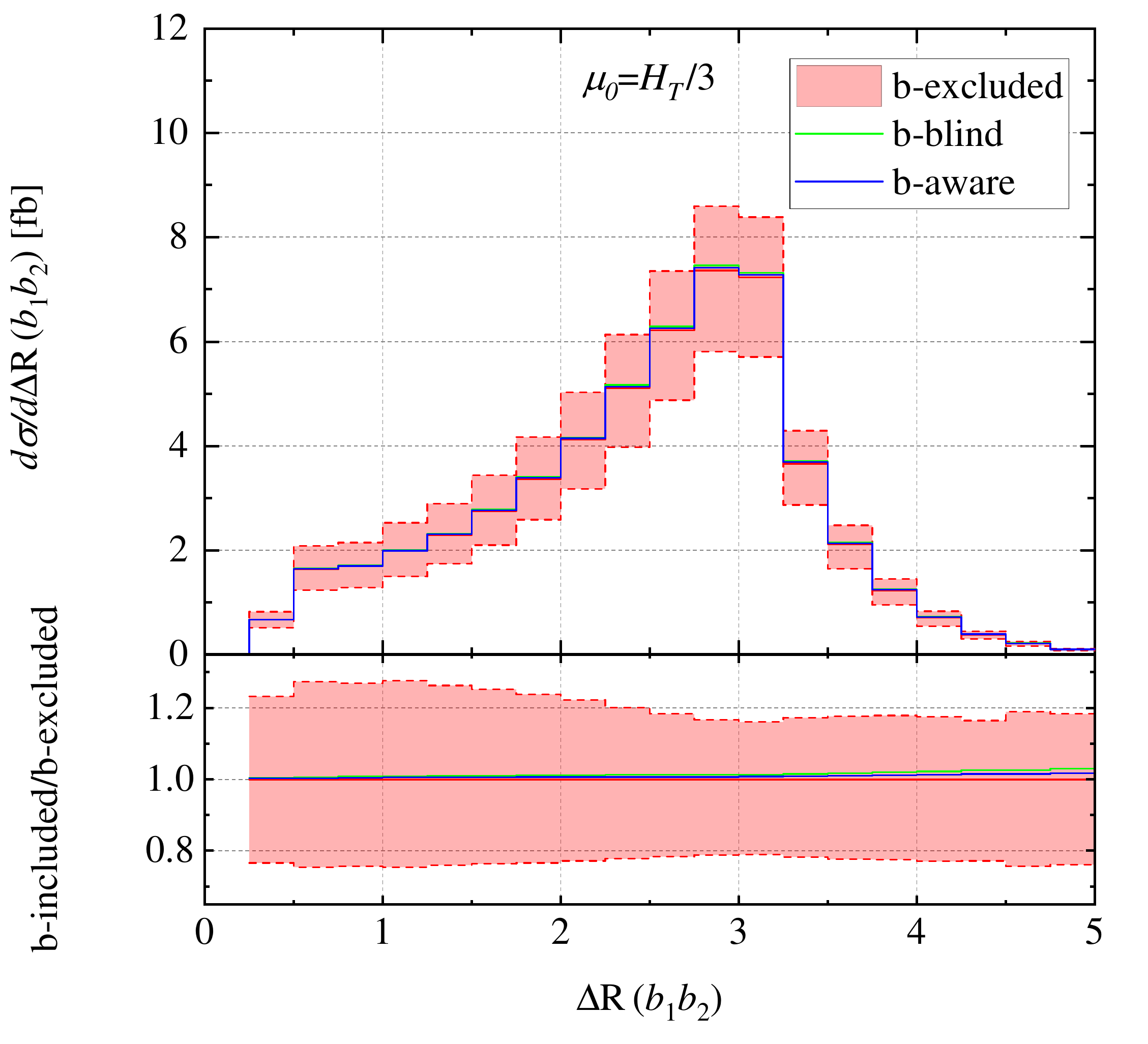}
    \includegraphics[width=0.49\textwidth]{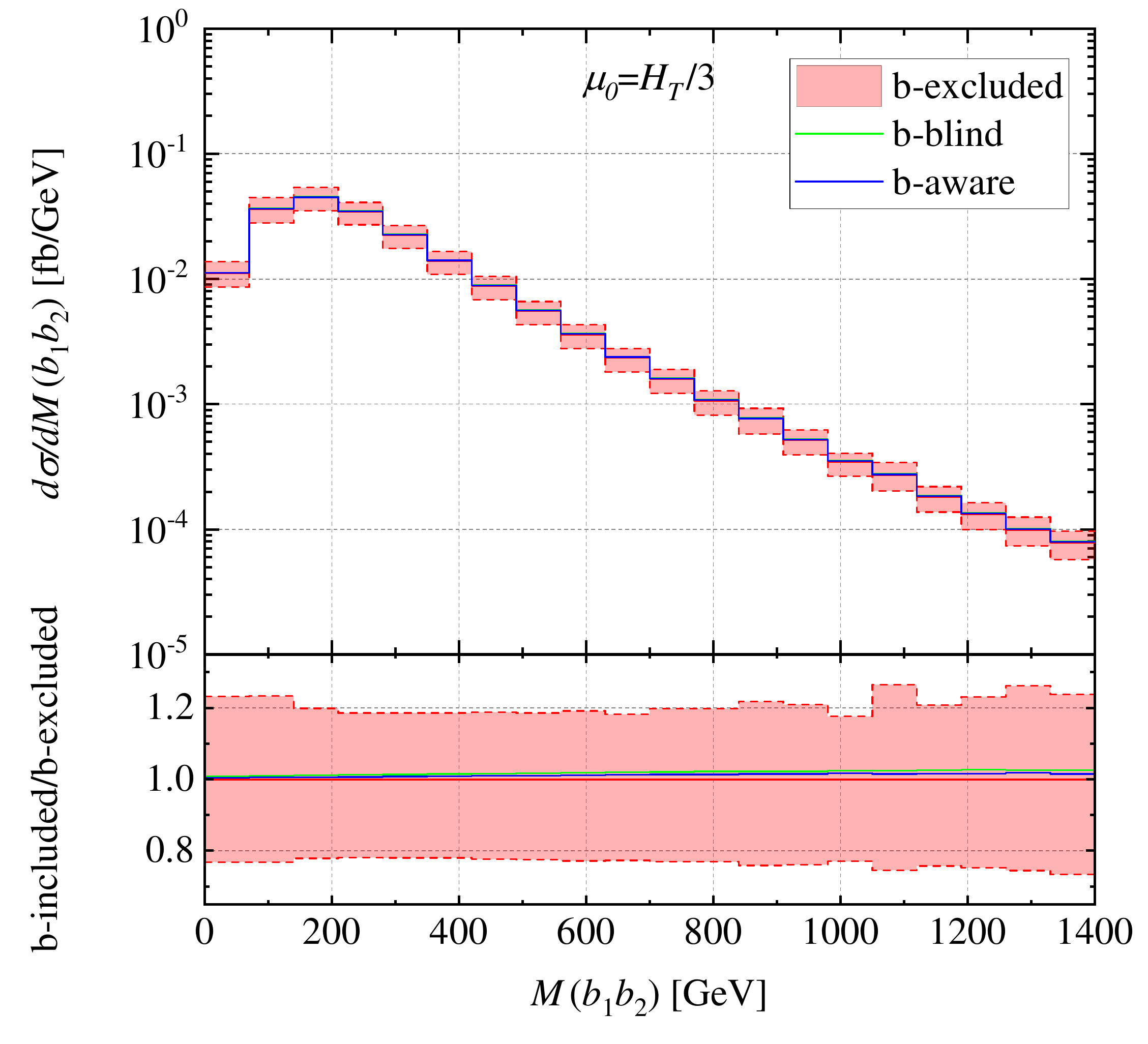}
    \includegraphics[width=0.49\textwidth]{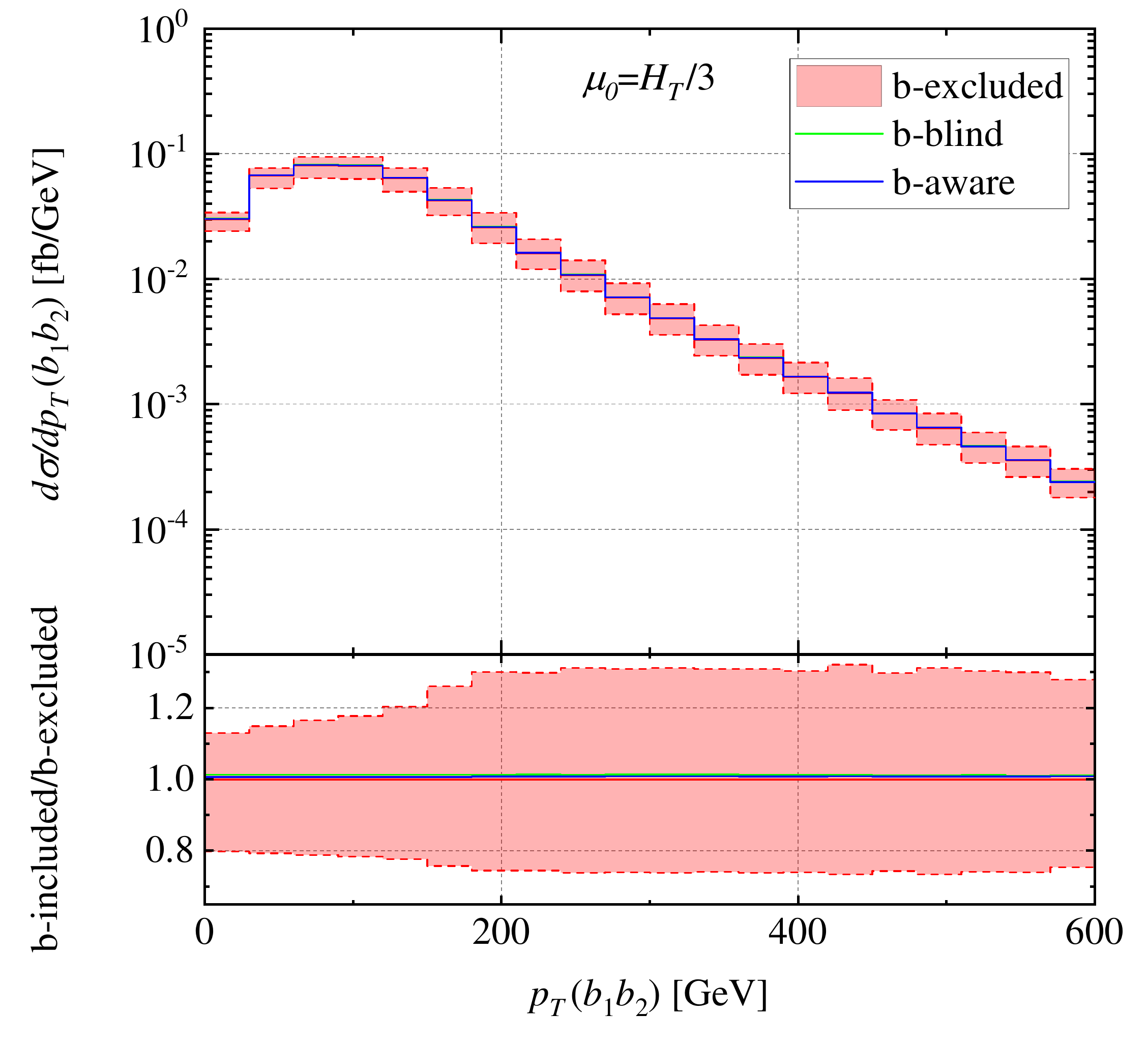}
  \end{center}
  \caption{\label{fig:b_contribution} \it
    Differential cross section distributions as a function of
$p_T(b_1)$, $\Delta R(b_1b_2)$, $M(b_1b_2)$ and $p_T(b_1b_2)$ for the
$pp\to e^+ \nu_e \,\mu^- \bar{\nu}_\mu\, b\bar{b}\,b\bar{b}+X$
production process at the LHC with $\sqrt{s} = 13$ TeV. The upper
plots show absolute NLO QCD predictions without and with the initial
state bottom quark contributions. In the latter case results are shown
for the charge blind and charge aware $b$-jet tagging. The lower
panels display the differential ratios of these contributions. Also
shown is the NLO scale dependence for the case without the initial state
bottom quark contributions. The NLO NNPDF3.1 PDF set is employed and
$\mu_0= H_T /3$ is used.}
\end{figure}
% =============================================

As a bonus of our study in Figure \ref{fig:b-contribution} we show
graphically NLO integrated fiducial cross section for the $pp\to
t\bar{t} b\bar{b}$ process with and without the initial state $b$
contributions. Results are given for the NLO NNPDF3.1 PDF
set. However, we also provide results with other PDF sets for the case
when the initial state $b$-quark contributions are not included.  We
display the result for CT14, CT18, NNPDF3.0, MMHT2014 and for the NLO
ABMP16 PDF set \cite{Alekhin:2018pai}. In the latter case the internal
PDF uncertainties are evaluated using the symmhessian method, see
e.g. Ref. \cite{Buckley:2014ana}. To put all the results in
perspective the theoretical uncertainty resulting from scale
dependence are also shown. All predictions are calculated for
$\mu_0=H_T/3$. We notice that, all theoretical predictions agree
within their internal PDF errors with the default result for the
NNPDF3.1 PDF set. The ABMP16 PDF uncertainties are, however, the
largest, i.e. at the $5\%$ level. Moreover, the earlier version of CT
and NNPDF PDF sets, i.e. CT14 and NNPDF3.0, show larger uncertainties
than those given by CT18 and NNPDF3.1. For CT14 we obtained $4\%$
while for NNPDF3.0 $2\%$. This should be compared to  $3\%$ and $1\%$
for CT18 and NNPDF3.1  respectively.

Finally, we perform a similar comparative analysis at the differential
level.  In Figure \ref{fig:b_contribution} we display NLO differential
cross section distributions as a function of $p_T(b_1)$, $\Delta
R(b_1b_2)$, $M(b_1b_2)$ and $p_T(b_1b_2)$. The upper plots show
absolute NLO QCD predictions with and without the initial state
$b$-quark contributions. Also shown is the NLO scale dependence for the
default case. The lower panels display the differential ratios of the
results with the initial state bottom-quark contributions to the one
without such contributions. Overall, for the process at hand when
comparing to the differential theoretical errors due to scale
dependence the inclusion of the bottom-quark induced subprocesses is
not important also at the differential level. Nonetheless, a consistent
treatment of heavy-flavour jets is necessary to obtain IR-finite
results.

% =============================================
%
\section{Comparison with previous results}
\label{sec:comparison}
%
% =============================================

In what follows we compare our predictions for the $pp\to e^+ \nu_e
\,\mu^- \bar{\nu}_\mu\,b\bar{b}\,b\bar{b} +X$ process at the LHC
running at $\sqrt{s}=13$ TeV to the theoretical predictions from
Ref. \cite{Denner:2020orv}, which we dub as ${\rm DLP}$ computation.
The comparison is carried out at the integrated and differential level
at LO and NLO in QCD. We adopt the dynamical scale setting used in
Ref. \cite{Denner:2020orv}. Specifically, we employ
$\mu_R=\mu_F=\mu_0$ where $\mu_0=\mu_{\rm DLP}$ is given by
  \begin{equation}
    \mu_0=\mu_{\rm DLP}=\frac{1}{2} \left[
        \left(
          p_T^{miss} + \sum_{i =e^+,\, \mu^-, \, {b_1},   {b_2},
            {b_3},    {b_4}} E_T(i) 
\right) + 2m_t
\right]^{1/2}
\left(
\sum_{i= {b_1},   {b_2},
            {b_3},    {b_4}} E_T(i)
  \right)^{1/2}\,.
\end{equation}
Furthermore, $E_T(i) =\sqrt{p^2_T(i)+M^2(i)}$ and $M^2(i)$ is the
invariant mass squared of the object considered. Similarly to our
dynamical scale setting also in this case the top-quark reconstruction
is  not attempted. In Ref. \cite{Denner:2020orv} the bottom-quark induced
contributions are included at LO and NLO in QCD.  The scheme adopted
there  is that of the charge-blind $b$-jet tagging, where the $bb$ and
$\bar{b}\bar{b}$ subprocesses are neglected. We start
with the LO results for the integrated fiducial cross section. When
the $bb$ and $\bar{b}\bar{b}$ contributions are neglected the LO
integrated fiducial cross section as obtained with the help of
\textsc{Helac-Nlo} is given by 
\begin{equation}
    \sigma^{\rm LO}_{\rm HELAC-NLO}({\rm NNPDF3.1},\mu_0
    =\mu_{\rm DLP}) 
    = 5.201(2)^{\, +60\%}_{\, -35\%}  \, {\rm fb} \,.
\end{equation}
It can be directly compared to the result form
Ref. \cite{Denner:2020orv}
\begin{equation}
    \sigma^{\rm LO}_{\rm DLP} ({\rm NNPDF3.1},\mu_0
    =\mu_{\rm DLP}) 
    = 5.198(4)^{\, +60\%}_{\, -35\%}  \, {\rm fb}\,.
\end{equation}
The two results agree perfectly within the given MC errors. Had we
additionally included the two missing subprocesses $bb$ and
$\bar{b}\bar{b}$ the result would rather be $\sigma^{\rm LO}_{\rm
HELAC-NLO}=5.206(2)$ fb. The latter is $1.8\sigma$ away from
$\sigma^{\rm LO}_{\rm DLP}$.

In order to
reproduce the NLO QCD result from Ref. \cite{Denner:2020orv} the
additional fifth jet (if resolved) must be added to the definition of
$\mu_0=\mu_{\rm DLP}$. The $\mu_0=\mu_{\rm DLP}$  scale is, thus,
given by
  \begin{equation}
    \mu_0=\mu_{\rm DLP}=\frac{1}{2} \left[
        \left(
          p_T^{miss} + \sum_{i =e^+ ,\, \mu^- ,\,  {b_1},   {b_2},
            {b_3},    {b_4},\, j} E_T(i) 
\right) + 2m_t
\right]^{1/2}
\left(
\sum_{i= {b_1},   {b_2},
            {b_3},    {b_4}, \, j} E_T(i)
  \right)^{1/2}\,.
\end{equation}
The \textsc{Helac-Nlo} NLO result and the $\sigma^{\rm NLO}_{\rm DLP}$
one, both  without the $bb$ and $\bar{b}\bar{b}$ contributions, read
\begin{equation}
  \begin{split}
    \sigma^{\rm NLO}_{\rm HELAC-NLO}({\rm NNPDF3.1},\mu_0
    =\mu_{\rm DLP}) &
    = 10.28(1)^{+18\%}_{-21\%} \, {\rm fb} \,,\\[0.2cm]
    \sigma^{\rm NLO}_{\rm DLP} ({\rm NNPDF3.1},\mu_0
    =\mu_{\rm DLP}) &
    = 10.28(8)^{+18\%}_{-21\%}  \, {\rm fb}\,.
\end{split}
\end{equation}
Theoretical predictions are in perfect agreement.  Had 
we also included the $bb$ and $\bar{b}\bar{b}$ contributions at NLO our
result would rather be $\sigma^{\rm NLO}_{\rm HELAC-NLO}=10.29(1)$
fb, so still in agreement with $\sigma^{\rm NLO}_{\rm
  DLP}$. 
%
% =============================================
\begin{figure}[t!]
  \begin{center}
 \includegraphics[width=0.49\textwidth]{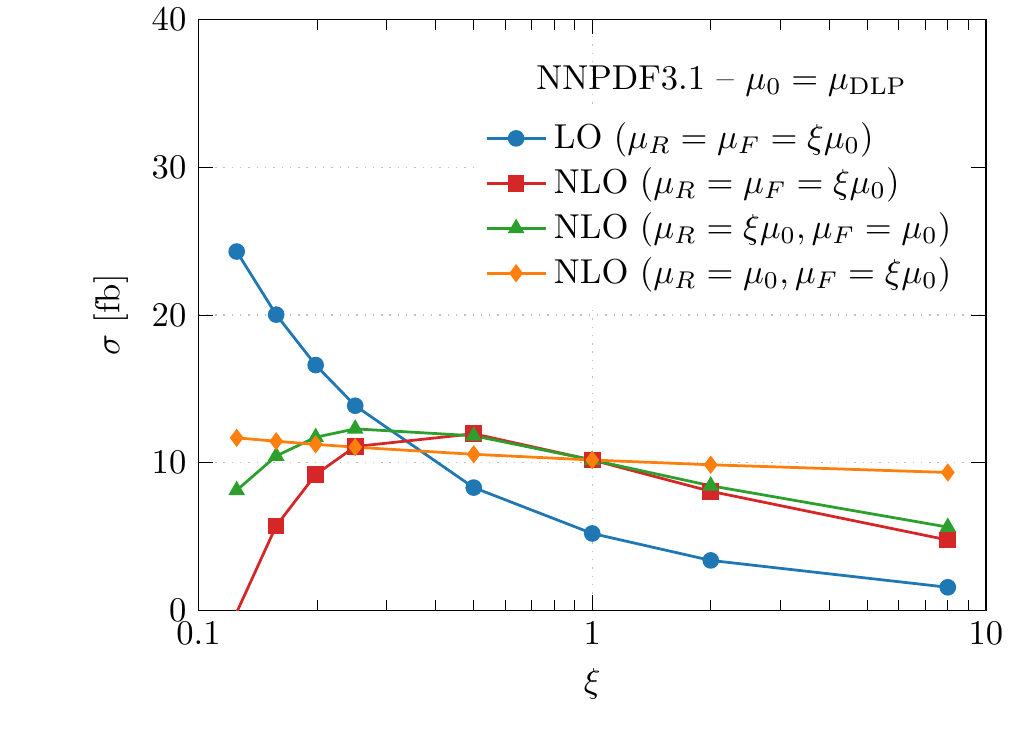}
 \includegraphics[width=0.49\textwidth]{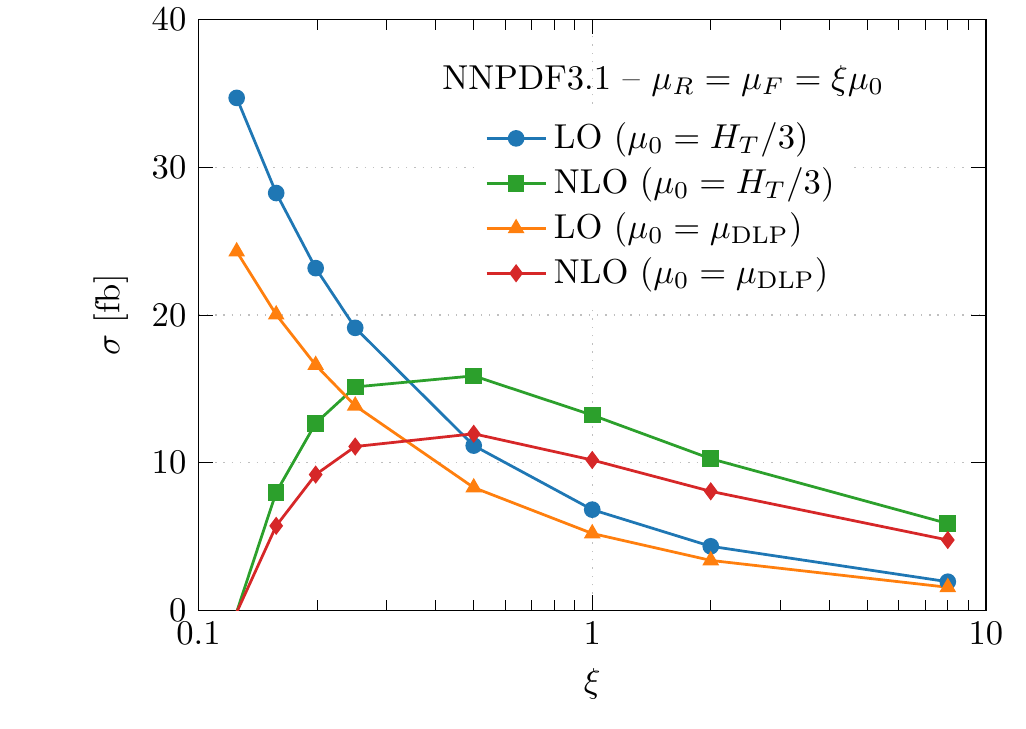}
  \end{center}
\caption{\label{fig:scale_dep3} \it Scale dependence of the LO and NLO
integrated fiducial cross sections for the $pp\to e^+ \nu_e \,\mu^-
\bar{\nu}_\mu\, b\bar{b}\,b\bar{b}+X$ production process at the LHC
with $\sqrt{s} = 13$ TeV for $\mu_0=\mu_{\rm DLP}$. The LO and NLO
NNPDF3.1 PDF sets are employed. Also shown is
the variation of $\mu_R$ with fixed $\mu_F$ and the variation of
$\mu_F$ with fixed $\mu_R$ as well as the comparison to our default 
scale choice $\mu_0=H_T/3$.}
\end{figure}
% =============================================
\begin{figure}[t!]
  \begin{center}
     \includegraphics[width=0.49\textwidth]{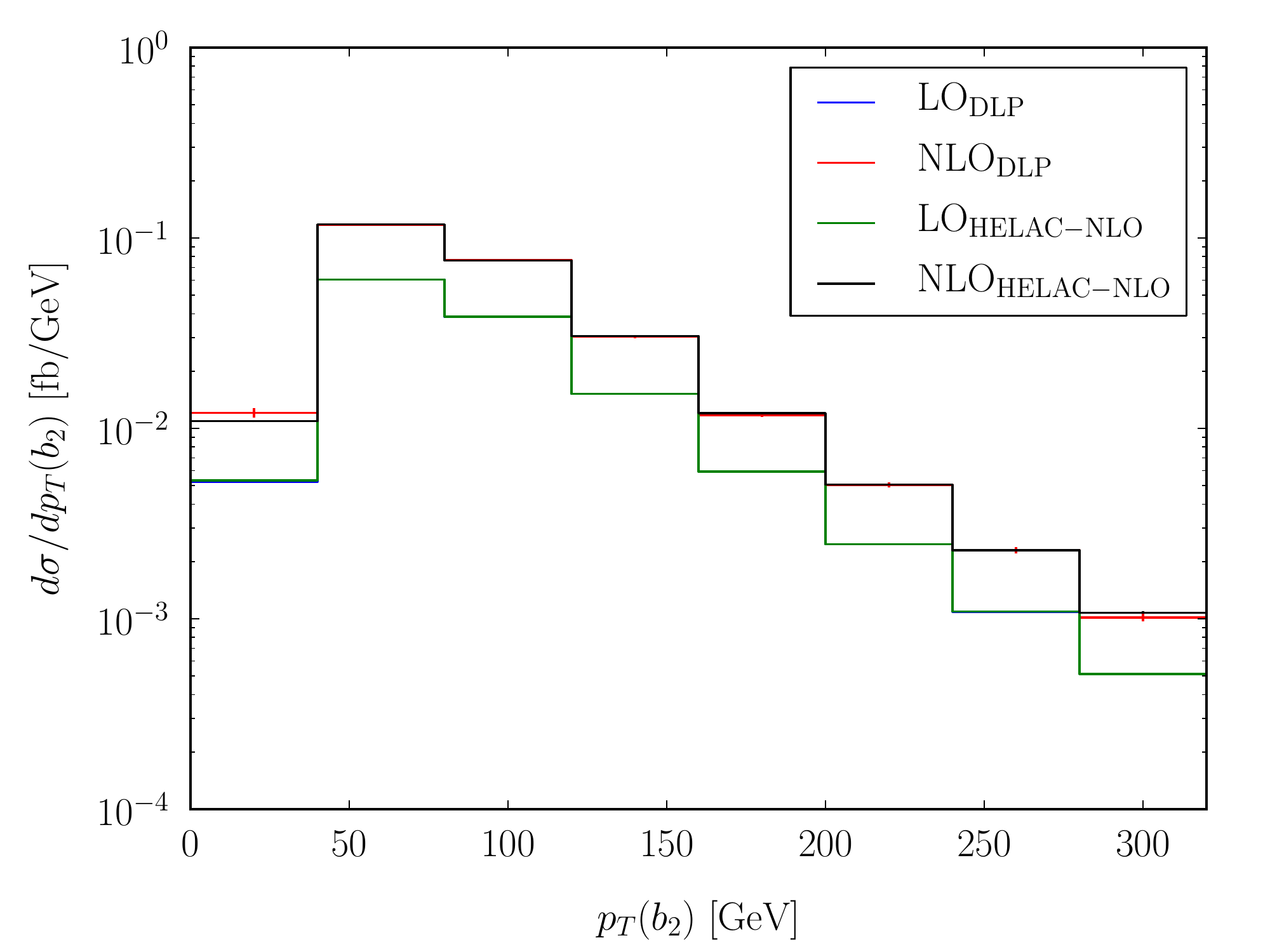}
    \includegraphics[width=0.49\textwidth]{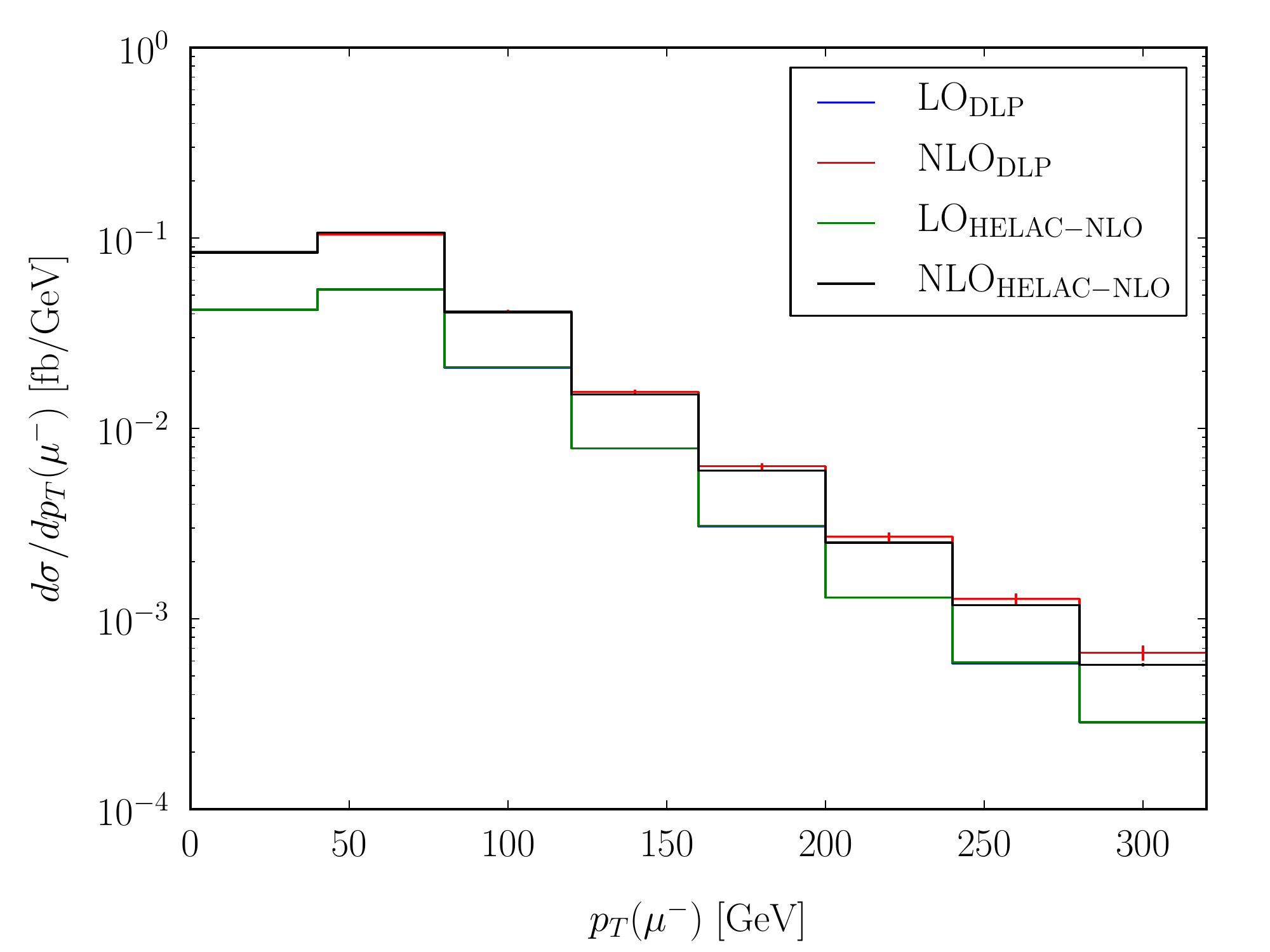}
    \includegraphics[width=0.49\textwidth]{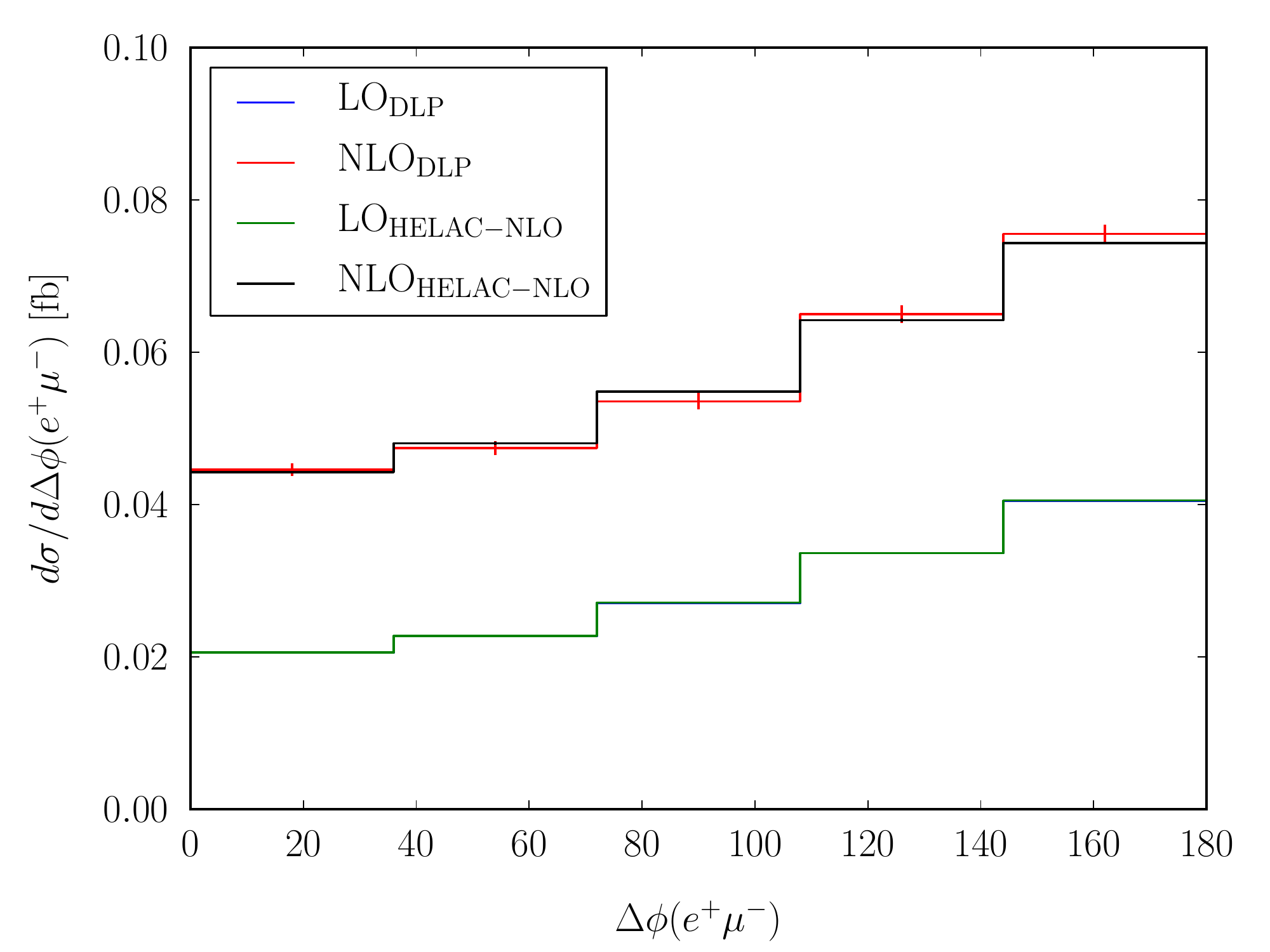}
    \includegraphics[width=0.49\textwidth]{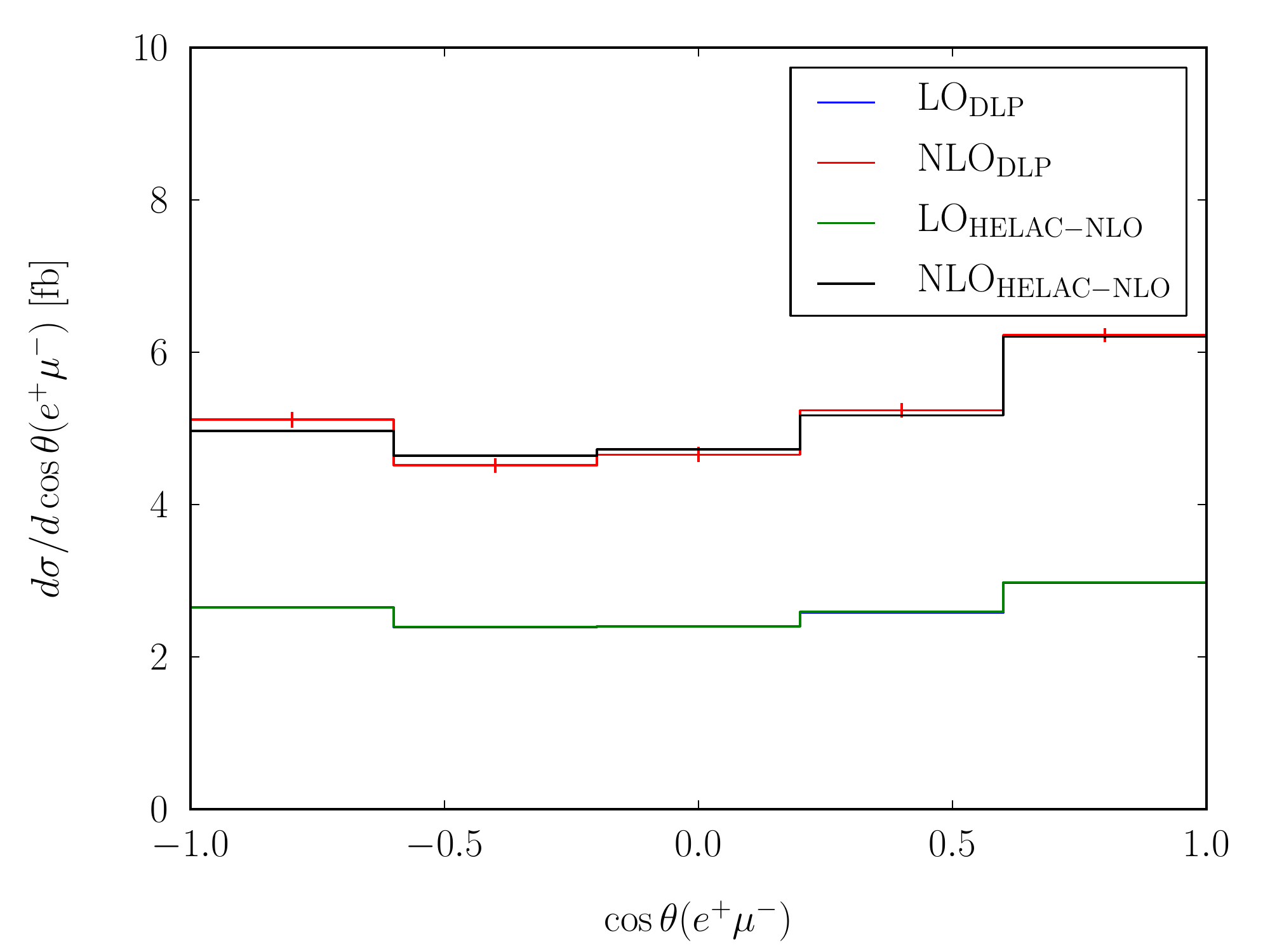}
  \end{center}
  \caption{\label{fig:comparison} \it
 Differential cross section distributions as a function of $p_T(b_2)$,
$p_T(\mu^-)$, $\Delta \phi(e^+\mu^-)$ and $\cos \theta(e^+\mu^-)$ for
the $pp\to e^+ \nu_e \,\mu^- \bar{\nu}_\mu\, b\bar{b}\,b\bar{b}+X$
production process at the LHC with $\sqrt{s} = 13$ TeV. Results are
presented at LO and NLO for $\mu_0=\mu_{\rm DLP}$.  Comparison between
\textsc{Helac-Nlo} predictions and results from
Ref. \cite{Denner:2020orv} (given with the subscript {\rm DLP}) is
shown. The NNPDF3.1 PDF sets are employed. Also displayed are Monte
Carlo integration errors.}
\end{figure}
% =============================================

A few comments are in order. Both LO and NLO QCD results for
$\mu_0=\mu_{\rm DLP}$ are much smaller than these obtained either for
$\mu_0=m_t$ or $\mu_0=H_T/3$. In the case of the NLO predictions the
differences exceed the theoretical uncertainties due to scale
dependence. The dynamical scale $\mu_0=\mu_{\rm DLP}$ includes the
dependence on the $p_T$ of the additional jet, thus, on the average
its value is larger.  On the other hand, the asymptotic freedom
guarantees that the value of $\alpha_s$ becomes smaller for larger
$\mu_0$, resulting in lower NLO and LO cross sections. Had we removed
the additional jet from the definition of $\mu_0=\mu_{\rm DLP}$ the
resulting NLO cross section would increase by about
$13\%$. Consequently, it would be in agreement with the NLO findings
both for $\mu_0=m_t$ and $\mu_0=H_T/3$ within  the quoted theoretical
uncertainties. 

In Figure
\ref{fig:scale_dep3} we provide the graphical representation of the scale
dependence of the LO and NLO integrated fiducial cross sections for the
$pp\to e^+ \nu_e \,\mu^- \bar{\nu}_\mu\,b\bar{b}\,b\bar{b} +X$
production process as obtained with $\mu_0=\mu_{\rm DLP}$. Also shown
are the variation of $\mu_R$ with fixed $\mu_F$ and the variation of
$\mu_F$ with fixed $\mu_R$. Finally, the  comparison to our dynamical
scale setting of $\mu_R=\mu_F=\mu_0=H_T/3$ is depicted.

In Figure \ref{fig:comparison} we depict the  comparison
between the \textsc{Helac-Nlo} predictions and the theoretical results from
Ref. \cite{Denner:2020orv} at the differential level. Specifically, we
show LO and NLO differential cross section distributions as a function
of $p_T(b_2)$, $p_T(\mu^-)$, $\Delta \phi(e^+\mu^-)$ and $\cos
\theta(e^+\mu^-)$.  Also here  heavy-flavour induced subprocesses are
included except from $bb$ and $\bar{b}\bar{b}$. For each observable  good
agreement has been found.

% =============================================
%
\section{Summary}
\label{sec:sum}
%
% =============================================
%

In this paper we provided state-of-the-art theoretical predictions for
the $pp\to e^+ \nu_e \,\mu^- \bar{\nu}_\mu\, b\bar{b}\,b\bar{b}+X$
process at the LHC with $\sqrt{s}=13$ TeV. In the computation
off-shell top quarks were described by Breit-Wigner
distributions. Furthermore double-, single- as well as non-resonant
top-quark contributions and interference effects were consistently
incorporated at the matrix element level. Non-resonant and off-shell
effects due to the finite $W$-boson width were also consistently taken
into account. All results were obtained with the help of the
\textsc{Helac-Nlo} MC package. We have shown LO and NLO predictions
for the integrated and differential cross sections, that are
phenomenologically relevant for LHC physics.  We assessed the
theoretical uncertainties of our high-precision theoretical
predictions stemming from scale dependence and PDF parameterisation
while using fixed and dynamical scale settings,
i.e. $\mu_R=\mu_F=\mu_0$, where $\mu_0=m_t$ or
$\mu_0=H_T/3$. Furthermore, the following PDF sets were examined
NNPDF3.1, CT18 and MMHT14.

The full $pp$ cross section receives positive and large NLO QCD
corrections of $89\%$. The theoretical uncertainties resulting from
scale variation are $65\%$ at LO and $22\%$ at NLO. The internal PDF
uncertainties are very small, at the level of $1\%-3\%$ only. Thus,
the NLO theoretical error is completely dominated by the scale
dependence. We provide LO and NLO results for other PDF sets and
observe that the ${\cal K}$-factor has a very large spread from $1.81$
down to $1.23$ depending on the LO PDF set employed and specifically
on the value of the strong coupling constant $\alpha_s(m_Z)$ used. On the other
hand, we observed a very stable behaviour of the systematics when varying
the $p_T (b)$ cut or adding additional cuts.

The differential cross sections have been plagued by the same large
higher order QCD effects as the integrated fiducial cross
sections. Not only big NLO QCD corrections but also significant shape
changes were visible when going from LO to NLO. This confirms that NLO
QCD effects to the $pp\to e^+ \nu_e \,\mu^- \bar{\nu}_\mu\,
b\bar{b}\,b\bar{b}+X$ process are extremely important. The theoretical
uncertainties due to scale dependence for $\mu_0 = H_T /3$ are rather
moderate of the order of $20\% - 30\%$. For the fixed scale setting
they are much higher. The uncertainties due to the NNPDF3.1 PDF
parameterisation are small, i.e. in the $1\% - 7\%$ range. When other
PDF sets have been examined the PDF uncertainties increased maximally
up to $11\%$. Consequently, the final theoretical error for the
process at hand remains dominated by the scale dependence.

In the next step we have studied the contributions that are induced by
the initial state bottom quarks. To this end, additional subprocesses
were included in the calculation.  Additionally, we needed new
recombination rules for partons to construct light- and heavy-flavour
jets. We employed two variants that are IR-safe at NLO: charge-blind
and charge-aware $b$-jet tagging. The differences between the two
approaches were examined in detail. Overall, for the process at hand
when comparing to the theoretical errors due to scale dependence the
size of the contributions generated by the bottom-quark induced
subprocesses is negligible both at the integrated and differential
level. Nonetheless, a consistent treatment of heavy-flavour jets is
necessary to obtain IR-finite results.

Finally, we compared our predictions for the $pp\to e^+ \nu_e \,\mu^-
\bar{\nu}_\mu\, b\bar{b}\,b\bar{b}+X$ process to the previous results
obtained in Ref. \cite{Denner:2020orv}. After clarifying the scale
choice used in Ref. \cite{Denner:2020orv} with the authors and the
status for the two $bb$ and $\bar{b}\bar{b}$ subprocesses perfect
agreement has been found both at the integrated and differential
level.

\acknowledgments{

We would like to thank Jasmina Nasufi for her contributions at early
stages of this project.  We would like to thank  Mathieu
Pellen for clarifications concerning the dynamical scale setting and
the setup used in Ref. \cite{Denner:2020orv}. Furthermore, we would
like to thank him for providing the data files for differential cross
sections distributions that we have used for comparisons in Section
\ref{sec:comparison}.
  
The work of H.B. and M.W.  was supported by the Deutsche
Forschungsgemeinschaft (DFG) under grant 396021762 $-$ TRR 257: {\it
P3H - Particle Physics Phenomenology after the Higgs
Discovery}. Support by a grant of the Bundesministerium f\"ur Bildung
und Forschung (BMBF) is additionally acknowledged.

The research of M.L. was supported by the DFG under grant 400140256 - GRK
2497: {\it The physics of the heaviest particles at the Large Hardon
  Collider}.

The work of G.B. was supported by grant K 125105 of the National
Research, Development and Innovation Office in Hungary.

H.B.H. has received funding from the European Research Council (ERC)
under the European Union's Horizon 2020 Research and Innovation
Programme (grant agreement no. 683211). Furthermore, the work of H.B.H
has been partially supported by STFC consolidated HEP theory grant
ST/T000694/1

The work of M.K. was supported in part by the U.S. Department of Energy
under grant DE-SC0010102.

Simulations were performed with computing resources granted by RWTH
Aachen University under projects {\tt rwth0414} and  {\tt rwth0531}.

%\bibliography{References} 

%\bibliographystyle{JHEP}

%\printbibliography[heading=bibintoc, title={References}]

\providecommand{\href}[2]{#2}\begingroup\raggedright\endgroup

\end{document}